\documentclass[fleqn,usenatbib]{mnras}

\usepackage{mathptmx}
\usepackage{txfonts}

\usepackage[T1]{fontenc}
\usepackage{ae,aecompl}

\DeclareRobustCommand{\VAN}[3]{#2}
\let\VANthebibliography\thebibliography
\def\thebibliography{\DeclareRobustCommand{\VAN}[3]{##3}\VANthebibliography}

\usepackage{graphicx}	
\usepackage{hyperref}

\newcommand{\gaia}{\textit{Gaia}}
\newcommand{\Teff}{\mbox{$T_{\rm eff}$}}
\newcommand{\Msun}{\mbox{$M_\odot$}}

\title[\gaia\ white dwarfs within 40\,pc III]{\gaia\ white dwarfs within 40\,pc III: spectroscopic observations of new candidates in the southern hemisphere}

\author[O'Brien et al]{Mairi W. O'Brien,$^{1}$\thanks{E-mail: Mairi.O-Brien@warwick.ac.uk} P.-E. Tremblay,$^{1}$ N.~P.~Gentile Fusillo,$^{2}$  M.~A.~Hollands,$^{3}$ 
\newauthor B.~T.~G\"ansicke,$^{1}$ D.~Koester,$^{4}$ I. Pelisoli,$^{1}$ E.~Cukanovaite,$^{1}$ T.~Cunningham,$^{1}$ 
\newauthor A.~E.~Doyle,$^{5}$ A.~Elms,$^{1}$ J.~Farihi,$^{6}$ J.~J.~Hermes,$^{7}$ J.~Holberg,$^{8}$ S.~Jordan,$^{9}$   
\newauthor B.~L.~Klein,$^{10}$ S.~J.~Kleinman,$^{11}$ C.~J.~Manser,$^{1, 12}$ D.~De Martino,$^{13}$ T.~R.~Marsh,$^{1}$ 
\newauthor J.~McCleery,$^{1}$ C.~Melis,$^{14}$ A.~Nitta,$^{15}$ S.~G.~Parsons,$^{3}$ R.~Raddi,$^{16}$
\newauthor A.~Rebassa-Mansergas,$^{16, 17}$  M.~R.~Schreiber,$^{18, 19}$ R.~Silvotti,$^{20}$ D.~Steeghs,$^{1}$ 
\newauthor O.~Toloza,$^{18,19}$  S.~Toonen,$^{21}$  S.~Torres$^{16, 17}$, A.~J.~Weinberger,$^{22}$ and B.~Zuckerman$^{10}$
\newauthor (affiliations can be found after the references)
}

\date{Accepted 2022 November 8. Received 2022 November 8; in original form 2022 October 4}

\pubyear{2022}
\begin{document}
\label{firstpage}
\pagerange{\pageref{firstpage}--\pageref{lastpage}}
\maketitle

\begin{abstract}
We present a spectroscopic survey of 248 white dwarf candidates within 40 pc of the Sun; of these 244 are in the southern hemisphere. Observations were performed mostly with the Very Large Telescope (X-Shooter) and Southern Astrophysical Research Telescope. Almost all candidates were selected from \textit{Gaia} Data Release 3 (DR3). We find a total of 246 confirmed white dwarfs, 209 of which had no previously published spectra, and two main-sequence star contaminants. Of these, 100 white dwarfs display hydrogen Balmer lines, 69 have featureless spectra, and two show only neutral helium lines. Additionally, 14 white dwarfs display traces of carbon, while 37 have traces of other elements that are heavier than helium. We observe 35 magnetic white dwarfs through the detection of Zeeman splitting of their hydrogen Balmer or metal spectral lines. High spectroscopic completeness ($>$\,97\,per\,cent) has now been reached, such that we have 1058 confirmed \textit{Gaia} DR3 white dwarfs out of 1083 candidates within 40 pc of the Sun at all declinations.

\end{abstract}



\section{Introduction}
\label{intro}
Approximately 97\,per\,cent of stars will end their lives as white dwarfs \citep{Fontaine2001}. As stars with masses below $\approx$ 10\,M$_{\odot}$ leave the main-sequence they become red giants, eventually shedding their outer layers as a planetary nebula, revealing the remaining core --- a dense white dwarf held up by electron degeneracy pressure. 
Once the star is a white dwarf, it cools down for the remainder of its lifetime, a process that is accurately modelled. Photometry and spectroscopy are used to estimate the cooling age of a white dwarf. An initial-to-final mass relation (IFMR; e.g. \citealt{El-Badry2018,Cummings2018,Barrientos2021,Barnett2021}) is employed to estimate the progenitor mass of the white dwarf, and evolutionary models are used to determine the main-sequence lifetime. From large samples of white dwarfs with known ages and Galactic kinematics, the stellar formation history at different look-back times in the Milky Way's past can be mapped \citep[][and references therein]{Fantin2019}. 

Studies of white dwarf spectral types \citep{Sion1983} reveal the chemical composition of the atmosphere and non-degenerate convectively mixed envelope, which has far-reaching implications. White dwarfs typically only show spectral lines from either hydrogen or helium, depending on their temperature and atmospheric composition. \citet{vanMaanen1917} discovered the first white dwarf spectrum that displays elements heavier than helium, a spectral class that is now indicative of accreted planetary debris \citep{Zuckerman2007,Farihi2016,Veras2021}. These metal-polluted systems are used to understand how planets evolve along with  their host stars. Ongoing accretion of planetary debris has been observed directly through the detection of X-rays from a metal-polluted white dwarf \citep{Cunningham2022}. In contrast, the presence of trace carbon in the atmosphere of the classical DQ stars below 10\,000\,K is currently explained by convective dredge-up from the interior \citep{Coutu2019, Koester2020,Bedard2022}. High-mass DQ white dwarfs (and possibly some lower mass DQ) are likely explained by stellar mergers (\citealt{Dunlap2015, Cheng2019, Coutu2019, Hollands2020, Farihi2022}). 

Degenerate stars provide a unique opportunity to probe extreme astrophysical environments, due to their large surface gravities. White dwarfs can have very strong magnetic fields and there are many proposed channels currently in use to explain their origin (see e.g. \citealt{Schreiber2021a,Schreiber2021b,Bagnulo2022}). Measured field strengths range from 10$^4$ to 10$^9$ Gauss, although the lower observational limit depends on spectral type and the availability of spectro-polarimetric observations \citep{Ferrario2020,Bagnulo2021}.

The highly accurate astrometry and photometry of nearby stars measured from the \textit{Gaia} spacecraft have enabled rapid progress in the definition of white dwarf samples. \citet{Gentile2021} have created a catalogue of $\approx$ 360\,000 high-confidence white dwarf candidates present in \textit{Gaia} Early Data Release 3 (EDR3) based on the positions of the candidates on the Hertzsprung-Russell (HR) diagram. No new $G$, $BP$ or $RP$ magnitudes or astrometry have been released in \textit{Gaia} DR3. Therefore, we reference DR3 as our source in this paper \citep{Gaia2021}. 

Cooling white dwarfs have a relatively large range of absolute \textit{Gaia} magnitudes (8 $\lesssim M_{\rm G} \lesssim$ 18 mag). In particular, the very faint end of the white dwarf luminosity function, which includes ultra-cool white dwarfs from old disc and halo stars \citep{Hollands2021,Kaiser2021,Bergeron2022,Elms2022}, can only be observed up to a distance of 40--100\,pc given a \textit{Gaia} limiting magnitude of $G \approx 20$--$21$. A sample which includes all ages and types of white dwarfs can only be achieved for 40--100\,pc, therefore a volume-limited sample out to these distances is needed.

Spectroscopic follow-up observations of \textit{Gaia} candidates are needed to confirm their classification as white dwarfs. Fortunately this work can build upon two decades of observations to define volume-limited samples of white dwarfs within 13\,pc, 20\,pc or 40\,pc \citep{Holberg2002,Giammichele2012,Limoges2015,Holberg2016}. Additional spectroscopic campaigns in the northern hemisphere have targeted 40\,pc white dwarfs \citep[][hereafter Paper I]{Tremblay2020} using the \textit{Gaia} DR2 white dwarf candidate catalogue from \citet{Gentile2019}. This resulted in a high level of spectroscopic completeness in the northern hemisphere within 40\,pc \citep[][hereafter Paper II]{Mccleery2020}. 

As of now, \citet{Gentile2021} have identified 542 white dwarf candidates in the northern hemisphere within 40\,pc, 531 of which are spectroscopically confirmed from the literature (e.g. \citealt[][Paper I]{Gianninas2011, Kawka2012,Limoges2015, Subasavage2017}). In Paper II, the 40\,pc northern sample was analysed based on a DR2 catalogue, which contained 521 confirmed white dwarfs \citep{Gentile2019}. 

In the southern hemisphere, \citet{Gentile2021} have identified 541 white dwarf candidates within 40\,pc, of which 304 are spectroscopically confirmed from the literature. There is a significant gap in the southern hemisphere observations that needs to be filled before meaningful analysis of the volume-limited 40\,pc sample can occur. 

In this Paper III on \textit{Gaia} white dwarfs in 40\,pc, we present spectroscopic follow-up observations of white dwarf candidates from DR3 within 40\,pc, the vast majority of which are in the southern hemisphere.

We present 220 updated or confirmed spectral types in the southern hemisphere, and three in the northern hemisphere. We observe two DR3 candidates in the south that are main-sequence stars. We also find two white dwarfs not in the DR3 catalogue, and four white dwarfs within 1$\sigma_\varpi$ of 40\,pc. Following the results from the present work, the full \textit{Gaia} 40\,pc sample of white dwarf candidates has 1058 confirmed white dwarfs out of 1083 initial DR3 candidates (97\,per\,cent spectroscopic completeness). Of the 25 remaining white dwarf candidates in DR3, two are confirmed as main-sequence stars in this paper, and 23 are unobserved. A detailed statistical analysis of the full 40\,pc white dwarf sample, including a list of all spectral types and references, will appear in the upcoming Paper IV. 

In this work, we discuss the nature of 246 \textit{Gaia} white dwarf candidates, 34 of which have previous spectral type classifications in the literature (see Table~\ref{tab:final_all} for citations). Four of these sources lie outside of 40\,pc but are within 1$\sigma_\varpi$ of that distance. The majority of targets, 242, are located in the southern hemisphere ($\delta < 0$\,deg), while the remaining four are in the northern hemisphere. 

\begin{table*}
	\centering
        \caption{Log of spectroscopic observations, where wavelength ranges are those used for analysis in this work.}
        \label{tab:log}
        \begin{tabular}{llllll}
                \hline
                Telescope/ & Programme IDs & No. of objects & Wavelength & Spectral Resolution (R) \\
                Instrument &  & in this work & Coverage [\AA] & \\
                \hline
                VLT/X-Shooter & 0102.C-0351 & 181 & 3600~--~10\,200 & UVB: 5400, VIS: 8900 \\ 
                 & 1103.D-0763 & & & \\
                 & 105.20ET.001 & & & \\
                SOAR/Goodman & SO2017B-009 & 49 & 3850~--~5550 & 1100 \\
                 & SO2018A-013 & & & \\
                 & SO2018B-015 & & & \\
                Shane/Kast & -- & 11 & 3600~--~7800 & 1900 \\
                GTC/OSIRIS & GTC103-21A & 3 & 3950~--~5700 & 2200 \\
                WHT/ISIS & ITP08 & 2 & 3730~--~7290 & Blue: 2000, Red: 3900 \\
                Tillinghast/FAST & -- & 2 & 3600~--~5500 & 1500 \\
                \hline
        \end{tabular}\\
\end{table*}

\section{Observations}

\subsection{Catalogue photometry and astrometry}
\citet{Gentile2021} used spectroscopically confirmed white dwarfs from the Sloan Digital Sky Survey (SDSS) \citep{SDSS_dr16} to select regions of the \textit{Gaia} DR3 HR diagram in which white dwarfs are likely to be present. We selected white dwarf candidates from the catalogue of \citet{Gentile2021} with a parallax $\varpi - \sigma_\varpi > 25$\,mas such that all sources are within $1\sigma_\varpi$ of 40\,pc. For each source, \citet{Gentile2021} provide a parameter, the probability of being a white dwarf ($P_{\rm WD}$). \citet{Gentile2021} suggest using $P_{\rm WD} > 0.75$ as a cut for the best compromise between completeness and contamination, and within 40\,pc only eight candidates out of 1083 do not meet this cut, so we therefore include all 1083 candidates in our sample. We prioritised observations of high-confidence candidates within the southern hemisphere that had no previously published spectral type, or an ambiguous classification, as our goal is to increase the spectroscopic completeness of the overall 40\,pc white dwarf sample. We use the WD\,Jhhmmss.ss\,$\pm$\,ddmmss.ss naming convention introduced by \citet{Gentile2019} in Table~\ref{tab:final_all} and figures throughout the Appendix of this paper. For simplicity, we shorten their WD\,J names to WD\,Jhhmm\,$\pm$\,ddmm in all other tables and text in this paper.

The \citet{Gentile2021} catalogue does not include white dwarfs in unresolved binaries with brighter main-sequence companions. \citet{Toonen2017} predicts that 0.5--1\,per\,cent of white dwarfs are part of an unresolved WD+MS binary, therefore in 40\,pc we would expect that only 5--10 of these systems would be excluded from the \citet{Gentile2021} DR3 catalogue.

\subsection{Spectroscopy}
We observed a total of 248 white dwarf candidates with parallaxes $\varpi - \sigma_\varpi > 25$\,mas as presented in Table~\ref{tab:log}. The majority of targets (181) were observed from the VLT with the X-Shooter spectrograph \citep{Vernet2011}, where we employed slit widths of $1.0$, $0.9$ and $0.9$\,arcsec in the UVB ($3000$~--~$5600$\,\AA, $R=5400$), VIS ($5500$~--$~10\,200$\,\AA, $R=8900$) and NIR ($10\,200$~--~$24\,800$\,\AA, $R=5600$) arms, respectively. 

The data were reduced following a standard procedure employing the \texttt{Reflex} pipeline \citep{Freudling2013}. The flux calibration used observations of hot DA white dwarfs obtained with the same instrument setup as the science spectroscopy, while telluric correction was performed using \texttt{molecfit} \citep{Kausch2015,Smette2015}. We extracted and inspected X-Shooter NIR spectra, and concluded that the signal-to-noise ratio was insufficient for meaningful analysis. Therefore we do not present any NIR spectra in this work.

We also observed 49 white dwarfs using the Goodman spectrograph \citep{Clemens2004} mounted on the Southern Astrophysical Research telescope (SOAR). We used the 930 line mm$^{-1}$ grating in the M2 mode (3850~--~5550\,\AA) and a 1.5\,arcsec slit. The data were reduced using the \texttt{iraf} package \texttt{ccdproc}, and extracted using \texttt{noao.twodspec.apextract}. Flux calibration was carried out using spectrophotometric standard stars observed on the same night and with the same setup. The 930--M2 mode does not cover any skylines, and since arcs were not taken close in time to the observations, radial velocities (RVs) from these observations are not reliable.

We also present two observations using the Intermediate-dispersion Spectrograph and Imaging System (ISIS) on the William Herschel Telescope (WHT) and three observations using the Optical System for Imaging and low-Resolution Integrated Spectroscopy (OSIRIS) on the Gran Telescopio Canarias (GTC) \citep{Cepa2000,Cepa2003}, which have the same setup as the observations reported in Paper I. 

We also present eleven observations from the Kast Double Spectrograph mounted on the Shane 3\,m telescope at the Lick Observatory. We used the 600/4310 grism for the blue, and either 830/8460 or 600/7500 gratings for the red, and we used slit widths of 1, 1.5, or 2\,arcsec. We also present two observations from the FAst Spectrograph for the Tillinghast Telescope (FAST) at the F. L. Whipple Observatory. Instrument details for FAST are found in \citet{Fabricant1998}.

We have used spectroscopic and photometric data to determine spectral types by human inspection for all 248 observed white dwarf candidates, which are listed in Table~\ref{tab:final_all}.

\section{Atmosphere and Evolution Models}
\label{sec:models}

All white dwarfs in this work are classified into one of the spectral types (SpT) described in Table~\ref{tab:definitions} \citep{Sion1983}. Spectral types are allocated visually according to the relative strength of absorption lines in the spectrum, with `H' representing Zeeman splitting from the presence of a magnetic field. We have derived atmospheric parameters and chemical abundances using photometric and spectroscopic fitting where appropriate. The notation $\log(\rm{X/Y})$ used in Table~\ref{tab:definitions} and throughout this work refers to the logarithm of the number abundance ratio of any two chemical elements, X and Y.

\begin{table*}
	\centering
        \caption{Definitions of all white dwarf spectral types discussed in this work, where photometric model composition refers to composition-selected \citet{Gentile2021} parameters. Adopted parameters for DZ and DQ white dwarfs in this work use the hybrid photometric/spectroscopic methods and are shown instead in Tables~\ref{table:dazparams}--\ref{table:dqparams}.}
        \label{tab:definitions}
        \begin{tabular}{llll}
                \hline
                Spectral type & Number in & Spectral features in order & Photometric model composition \\
                (SpT) & this work & of strength & \\
                \hline
                DA & 100 & Hydrogen Balmer & pure-H \\ 
                DAH & 28 & Hydrogen Balmer + magnetic & pure-H \\
                DB & 2 & Neutral helium & $\log({\rm H/He})$ $= -$5 \\
                DC & 69 & Featureless & $\log({\rm H/He})$ $= -$5, pure-He below 7000\,K, \\
                 & & & assumed pure-H below 5000\,K \\
                DAZ & 10 & Hydrogen Balmer + metal & pure-H \\
                DZ & 12 & Metal & $\log({\rm H/He})$ $= -$5, pure-He below 7000\,K \\
                DZH & 5 & Metal + magnetic & $\log({\rm H/He})$ $= -$5, pure-He below 7000\,K \\
                DZA & 4 & Metal + hydrogen Balmer & $\log({\rm H/He})$ $= -$5, pure-He below 7000\,K  \\
                DZAH & 2 & Metal + hydrogen Balmer + magnetic & $\log({\rm H/He})$ $= -$5, pure-He below 7000\,K  \\
                DQ & 7 & Carbon (molecular bands) & $\log({\rm H/He})$ $= -$5, pure-He below 7000\,K \\
                warm DQ & 1 & Carbon (atomic lines) &  pure-He \\
                DQpec & 2 & Carbon (molecular bands, shifted wavelengths) & $\log({\rm H/He})$ $= -$5, pure-He below 7000\,K \\
                DQZ & 2 & Carbon + metal & $\log({\rm H/He})$ $= -$5, pure-He below 7000\,K \\
                DZQ & 1 & Metal + carbon & $\log({\rm H/He})$ $= -$5, pure-He below 7000\,K  \\
                DZQH & 1 & Metal + carbon + magnetic & $\log({\rm H/He})$ $= -$5, pure-He below 7000\,K  \\
                \hline
        \end{tabular}\\
\end{table*}

\subsection{Photometric parameters}

Effective temperatures (\Teff) and stellar radii can be derived for most white dwarfs using photometric and parallax fits to model atmospheres, providing the composition of the white dwarf atmosphere is known \citep{koester1979, bergeron2001, Gentile2021}. 

In this work, we rely on the photometric parameters already made available in \citet{Gentile2021}. In brief, either pure-hydrogen \citep{Tremblay2011b}, pure-helium \citep{Bergeron2011}, or mixed hydrogen and helium \citep{Tremblay2014} model atmospheres are used, depending on the spectral type (see Table~\ref{tab:definitions}), to fit the \textit{Gaia} DR3 photometry to determine \Teff\ and radii of all white dwarfs in the sample. Mixed atmosphere models use the ratio $\log(\rm{H/He})$ = $-5$ for all photometric fitting of DC white dwarfs above 7000\,K. For DC stars within 5000\,K $< T_{\rm eff} <$ 7000\,K we use pure-helium atmospheres. For DC white dwarfs below 5000\,K it is difficult to constrain the atmospheric composition, as the H~$\alpha$ line would be very difficult to detect with most ground- and space-based current or near-future spectroscopic instruments, so we assume pure-hydrogen atmospheres \citep[Paper II]{Gentile2020}. 

Surface gravities ($\log(g)$), masses and cooling ages are derived using evolutionary models \citep{Bedard2020}. Table~\ref{tab:final_all} shows the derived parameters from a homogeneous set of photometric fits from \citet{Gentile2021} using \textit{Gaia} data only. In this work we also derive independent parameters from hybrid fits using spectroscopy and photometry for DQ and DZ stars (see Section~\ref{sec:hybrid_params} for details).

\subsection{Spectroscopic parameters}
\label{sec:spectro_params}
We derive \Teff\ and $\log(g)$ from spectroscopic fits of Balmer lines in non-magnetic DA white dwarfs using a \texttt{Python} implementation adapted from previous Balmer line fitting procedures described extensively in \citet[Paper I]{Liebert2005,Tremblay2011,Gianninas2011}. This modern fitting code is part of the 4MOST multi-object spectroscopic (MOS) survey consortium pipeline \citep{4MOSTS3,4MOST} and will also be a key resource for other MOS surveys such as WEAVE \citep{WEAVE}. We rely on DA models from \citet{Tremblay2011} with 3D corrections from \citet{tremblay13c}. Table~\ref{tab:final_all} shows spectroscopic parameters determined from this method. 

Only DA spectra with at least two visible Balmer lines are fitted. If there is only one spectral line available, either due to the \Teff\ and $\log(g)$ of the white dwarf or incomplete spectral coverage, the best-fit parameters cannot be well constrained. For DA white dwarfs below $\approx$ 5200\,K observed with X-Shooter, Balmer lines from H~$\beta$ and above become very weak while \Teff\ and $\log(g)$ are degenerate in predicting the equivalent width of the H~$\alpha$ line. It is therefore not possible to fit both parameters. 

For the two DB white dwarfs in our sample, we use the 3D model atmospheres of \citet{Cukanovaite2021} to obtain $\log({\rm H/He})$ and \Teff. We use a fitting procedure similar to that of \citet{Bergeron2011}.

The DC and magnetic white dwarfs in the sample are not fitted spectroscopically but best-fit parameters from \textit{Gaia} photometry are presented in Table~\ref{tab:final_all}. Best-fit parameters for confirmed unresolved binary systems are not given. White dwarf candidates that were found to be main-sequence stars are not analysed further.

\subsection{Combined spectroscopic and photometric parameters}
\label{sec:hybrid_params}

Atmospheres with carbon traces and metal-polluted white dwarfs are fitted using models from \citet{Koester2010} and improvements described therein. Fits are presented in Sections~\ref{sec:DZ} and \ref{sec:DQ}. We adopt an iterative approach of combined photometric and spectroscopic fitting. We start by computing a small grid of models with an initial guess on the metal abundances to fit the photometry for \Teff\ and $\log(g)$. The subsequent step is then to calculate a new grid of models with variable metal abundances at fixed atmospheric parameters in order to fit chemical composition. We repeat these two steps until convergence.

\begin{figure}
    \centering
	\includegraphics[width=\columnwidth]{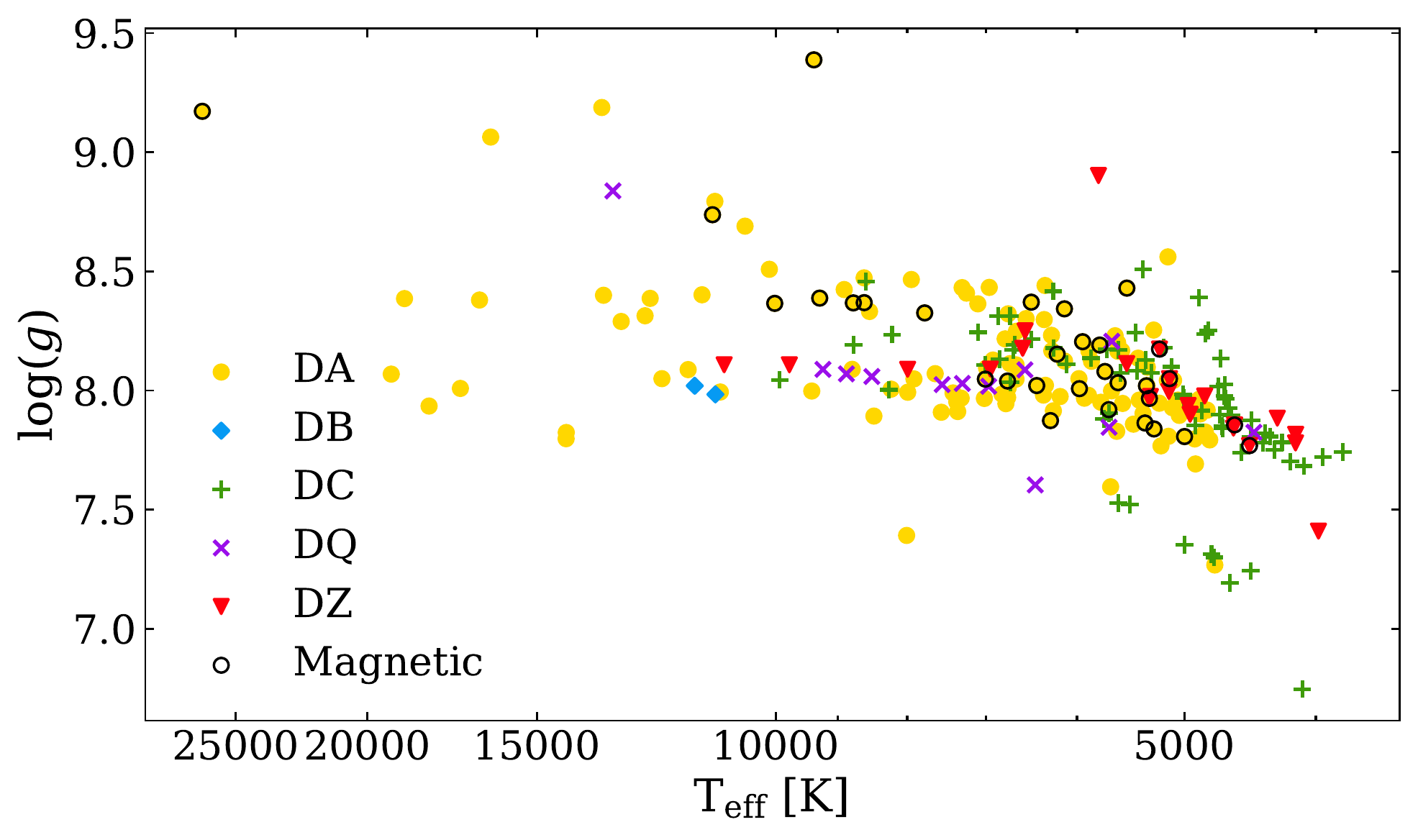}
	\caption{$\log(g)$ against \Teff\ distribution for white dwarfs within 40\,pc that have been spectroscopically observed in this work, where parameters have been determined from fitting of \textit{Gaia} DR3 photometry. Magnetic stellar remnants have black contours. Data are colour- and symbol-coded by their primary spectral type classification only, for simplicity.}
    \label{fig:teff_logg}
\end{figure}

\section{Results}
\label{results}

We confirm the classification of 246 white dwarfs within 1$\sigma_\varpi$ of 40\,pc, 213 of which had no previous observations from literature. The distribution of $\log(g)$ as a function of \Teff\ for all white dwarfs in our sample is shown in Fig.~\ref{fig:teff_logg} based on \textit{Gaia} DR3 photometric parameters \citep{Gentile2021}. In Fig.~\ref{fig:teff_logg}, all sources are fitted as single stars. There is a visible second track at $\log(g) \sim 7.4$, below the main distribution at $\log(g) \sim 8.0$ in Fig.~\ref{fig:teff_logg}, where double degenerate binary candidates with about twice the luminosity of a single white dwarf are located. Their $\log(g)$ values are underestimated as their photometry is fitted here as if they were single stars.

In Fig.~\ref{fig:teff_logg} we observe a downward trend in photometric $\log(g)$ against \Teff\ below around 6000\,K. A similar trend has been discussed following \textit{Gaia} DR2 \citep[Paper I, Paper II]{Hollands2018_Gaia,Bergeron2019}, and could be due to \textit{Gaia} temperatures being too low or luminosities being too large (see Paper I for details).

Only the two DZH white dwarfs \textbf{WD\,J0548$-$7507} and \textbf{WD\,J2147$-$4035}, and the DA \textbf{WD\,J1956$-$5258} do not have atmospheric parameters determined from \textit{Gaia} DR3 photometry in \citet{Gentile2021}. \textbf{WD\,J2147$-$4035} is a very cool IR-faint white dwarf \citep{Apps2021}, and its spectroscopy and photometry has been fitted in \citet{Elms2022}. \textbf{WD\,J0548$-$7507} was selected as a white dwarf candidate by \citet{Gentile2019} in \textit{Gaia} DR2, but it was not selected in the DR3 catalogue due to failing the BP$-$RP excess factor rule, as it is in the Large Magellanic Cloud region \citep{Gentile2021}. \textbf{WD\,J0548$-$7507} has parameters of \Teff\ = 4720\,$\pm$\,170\,K and $\log(g)$ = 7.9\,$\pm$\,0.1 from \textit{Gaia} DR2 photometric fitting. \textbf{WD\,J1956$-$5258} was not selected in either of the DR2 or DR3 white dwarf catalogues, due to its bright, \textit{Gaia} G-band magnitude 10, M-dwarf companion separated by 4.7\,arcsec on the sky.

We have updated the spectral types of five white dwarfs in the sample previously classified as DC, owing to the higher-quality spectroscopy we have obtained as follows: \textbf{WD\,J1821$-$5951} \citep{Subasavage2017} and \textbf{WD\,J1430$-$2403} \citep{Reid2005} are DAs, \textbf{WD\,J0252$-$7522} \citep{Subasavage2007} and \textbf{WD\,J1412$-$1842} \citep{Dupuis1994} are DAHs and \textbf{WD\,J2112$-$2922} \citep{Raddi2017} is a DZQ. These updated spectral types are shown in italics in Table~\ref{tab:final_all}.

While observations focused on southern hemisphere white dwarfs, we also obtained spectroscopy of three northern hemisphere targets omitted from Paper I due to low $P_{\rm WD}$ values in DR2: \textbf{WD\,J1318+7353}, \textbf{WD\,J1815+5532}, and \textbf{WD\,J1919+4527}. In DR3 \citep{Gentile2021}, the $P_{\rm WD}$ values of these white dwarfs increased to 0.96, 0.75, and 0.87 respectively. We also re-observed the highly-polluted northern white dwarf \textbf{WD\,J0358+2157} with X-Shooter.

All objects with a parallax below 25 mas are flagged with an asterisk, these objects may be a member of the 40\,pc sample within 1$\sigma_\varpi$. The best estimates of spectroscopic atmospheric parameters and chemical abundances are displayed in Table~\ref{table:dbparams} for DB white dwarfs, Table~\ref{table:dazparams} for DAZ white dwarfs, Table~\ref{table:dzparams} for DZ and DZA white dwarfs, and Table~\ref{table:dqparams} for all white dwarfs with carbon features. The observations of main-sequence stars that contaminate our sample are discussed in Section~\ref{sec:STAR}.

\setcounter{table}{2}

\begin{table*}
	\centering
     \scriptsize
        \caption{Spectral types and parameters of the white dwarf sample}
        \label{tab:final_all}
        \begin{tabular}{llllllll}
                \hline
                WD\,J name  & SpT & Parallax (mas) & $T_{\rm eff}$ [K] & $\log(g)$ & $T_{\rm eff}$ [K] & $\log(g)$ & Note\\
                &  & & 3D Spectro &  	3D Spectro &  Gaia &  Gaia & \\
                \hline
                001349.89$-$714954.26   & DAH  	& 53.21 (0.02) &   --  	&	-- & 6280 (30)	 	& 7.87 (0.02) &	(a) \\
                001830.36$-$350144.71   & DAH   	& 28.05 (0.06) &   -- 	&	-- & 7010 (60)	 	& 8.05 (0.03) &	\\
                003036.62$-$685458.25 & DA  & 25.46 (0.04) & 	8640 (40) &	7.98 (0.05)	& 8790 (230)	 	& 8.09 (0.06) & \\
                003713.77$-$281449.81 & DC:   	&  26.5 (0.1) &  --	& --	&  5340 (60)	 	& 8.13 (0.04) & \\
                004126.61$-$503258.58 & DC  &  31.84 (0.09) &  --	& --	&   4180 (60)	 	& 7.70 (0.04) & \\
                004434.77$-$114836.05 & DZ   	& 27.1 (0.1) &   -- 	& -- 	& 5300 (70)	 	& 7.98 (0.06) & \\
                005311.22$-$501322.87 & DC   	& 28.72 (0.06) &   --	& -- & 5570 (60)	 	& 8.08 (0.03) & \\
                005411.42$-$394041.53 & DA   	& 37.34 (0.05) &   6580 (20)	& 	8.43 (0.02) 	& 6260 (40)	 	& 8.23 (0.02) & \\
                010338.56$-$052251.96 & DAH   	&  34.4 (0.1) &  --	& --	&   	9380 (290)	 	& 9.39 (0.05) & (b) \\
                012953.18$-$322425.86 & DA   	&  26.10 (0.05) &   6770 (80) &	8.1 (0.1)	& 	6720 (50)	 	& 8.11 (0.03) &\\
                013843.16$-$832532.89 & DA   & 31.92 (0.03) &  7750 (70) &	8.14 (0.09) & 	7630 (60)	 	& 8.07 (0.02) &\\
                $*$ 014240.09$-$171410.85 & DAH   	&  24.97 (0.09) &  -- 	&	-- &    	5560 (50)	 	& 8.00 (0.03) &	\\
                014300.98$-$671830.35 & DAZ     & 102.91 (0.01) & -- &   -- 	& 6350 (30)	 	& 7.98 (0.02) &	(c) \\
                015038.47$-$720716.54 & DC  	& 31.53 (0.04) &  --  	&	-- & 6840 (60)	 	& 8.13 (0.03) & (d) \\
                021228.98$-$080411.00 & DA  	& 59.76 (0.02) &  9020 (20) &    8.14 (0.02)	& 8470 (110)	 	& 7.89 (0.03) &	 \\
                024300.36$-$603414.82 & DA  & 29.86 (0.06) &  5760 (120) & 8.5 (0.3) 	& 5600 (50)	 	& 8.20 (0.03) & \\
                024527.76$-$603858.32 & DA   	& 28.08 (0.04) &   6150 (70) 	& 	8.4 (0.1) &    	5880 (50)	 	& 7.98 (0.03) & \\
                025017.18$-$224130.53 & DA   	& 27.91 (0.08) & -- &   -- 	& 5620 (60)	 	& 8.23 (0.03) & \\
                025245.61$-$752244.56 & \textit{DAH}  	& 32.05 (0.04)  &  --  	&	--	& 6200 (50)	 	& 8.15 (0.02) & (e) \\
                025332.00$-$654559.93 & DA   	&  26.99 (0.05) &  5600 (60) 	& 	8.0 (0.1) & 5450 (50)	 	& 7.86 (0.03) & \\
                025759.87$-$302709.99 & DA   	& 25.95 (0.06) &   6330 (60) 	& 	8.1 (0.1)	& 6170 (40)	 	& 7.98 (0.02) & \\
                030154.44$-$831446.19 & DA   	&  29.89 (0.03) &   6860 (60)	& 	8.0 (0.1)	& 6810 (50)	 	& 7.99 (0.02) & \\
                030407.15$-$782454.62 & DA   	&  25.11 (0.07) &    5500 (30)	& 	7.99 (0.04) & 5360 (60)	 	& 7.90 (0.04) & \\
                031225.70$-$644410.89 & DA   	& 27.33 (0.02) &  -- &  --  	& --	 	& -- & DA+DA (f) \\
                031318.66$-$560734.99 & DA   	& 28.70 (0.02) &    11\,230 (60) 	& 	8.03 (0.03)	& 10\,990 (120)	 	& 7.99 (0.02) & \\
                031646.48$-$801446.19 & DA   	&  28.02 (0.03) &    7510 (50)	& 8.0 (0.1)	& 7360 (60)	 	& 7.95 (0.02) & \\
                031715.85$-$853225.56 & DAH 	& 34.04 (0.03)  &  --  	&	-- & 26470 (1370)	 	& 9.17 (0.05) & (g) \\
                031719.13$-$853231.29 & DA   	&  34.02 (0.02) &  17\,050 (230) 	& 	8.43 (0.03) & 16\,530 (290)	 	& 8.38 (0.02) & (h) \\
                032646.69$-$592700.23 & DA   	&  32.13 (0.05) &  6380 (90) 	& 	8.5 (0.2)	& 6330 (60)	 	& 8.44 (0.02) & \\
                034010.17$-$361038.22 & DA   	&  29.08 (0.05) &   5870 (60) 	& 	8.2 (0.1)	& 5610 (40)	 	& 7.83 (0.03) & (i) \\
                034347.42$-$512516.55 & DAZ  	& 35.83 (0.03) &   -- &  --  	& 6740 (50)	 	& 8.01 (0.02) & \\
                035005.27$-$685307.56 & DA   	&  30.02 (0.05) &  -- 	& 	--	& 4910 (50)	 	& 7.80 (0.03) & \\
                035531.89$-$561128.32 & DAH  	& 30.35 (0.05) &   --  	&	-- & 5770 (50)	 	& 8.19 (0.03) & \\
                035826.49+215726.16 & DAZ  	& 27.67 (0.07) & -- & -- & 6780 (80)	 	& 8.22 (0.03) & (b) \\
                041630.04$-$591757.19 & DA 	& 54.58 (0.03) &  15\,540 (70) & 7.96 (0.01) & 14\,270 (240) & 7.82 (0.02) & (j) \\
                041823.34$-$500424.14 & DC   & 41.93 (0.06) &  --  	&	-- & 4700 (40)	 	& 8.14 (0.03) & \\
                042021.33$-$293426.26 & DAH  	& 32.16 (0.04) &  --  	&	-- & 6420 (40)	 	& 8.02 (0.02) & \\
                042357.67$-$455042.27 & DA   	&  33.40 (0.04) &   5900 (40)	& 8.49 (0.06)	& 5550 (40)	 	& 7.95 (0.02) & (k) \\
                042643.98$-$415341.44 & DAZ  	& 29.06 (0.04) &  -- & -- & 6130 (60)	 	& 8.12 (0.03) & \\
                042731.73$-$070802.80 & DC 	& 25.17 (0.06) &  --  	&	-- & 6720 (60)	 	& 8.04 (0.03) & (b) \\
                044538.42$-$423255.05 & DAZ  	& 36.60 (0.02) &  -- & -- & 6750 (50)	 	& 7.97 (0.02) & \\
                044903.21$-$241239.20 & DA   	&  33.70 (0.07) &  -- 	& 	-- 	& 4870 (50)	 	& 7.96 (0.04) & \\
                045943.21$-$002238.86 & DA   	&  40.46 (0.03) &  11\,060 (100) 	& 8.81 (0.04)	& 11\,090 (120)	 	& 8.79 (0.02) & (l)\\
                050552.46$-$172243.48 & DAH 	&  51.68 (0.03) & --  	&	-- & 5350 (30)	 	& 7.86 (0.02) & (m)\\
                051942.85$-$701401.50 & DC & 25.22 (0.10) &  --  	&	-- & 4540 (70)	 	& 7.74 (0.05) & \\
                052436.27$-$053510.52 & DA   	& 27.98 (0.02) &   17\,330 (120) 	& 	8.08 (0.03)	 & 17\,080 (310)	 	& 8.01 (0.02) & (b) \\
                052844.01$-$430449.21 & DA   	&  26.09 (0.03) &   10\,620 (140)	& 	8.70 (0.04) 	& 10\,540 (140)	 	& 8.69 (0.02) & (n) \\
                053446.50$-$524150.29 & DA   	&  25.21 (0.05) &  6110 (60) 	& 	8.2 (0.1)	& 5980 (70)	 	& 8.05 (0.04) & \\
                054249.69$-$190107.34 & DC 	& 32.79 (0.03) &  --  	&	-- & 8763 (80)	 	& 8.19 (0.02) & \\
                $*$ 054858.25$-$750745.20 & DZH  	& 24.96 (0.09) & --  &  -- 	& 4720 (170) & 7.9 (0.1)  & DR2 Parameters \\
                055118.71$-$260912.89 & DC & 25.28 (0.06) & -- & -- & 4750 (40) & 7.30 (0.03) & \\
                055443.04$-$103521.34 & DZ    & 65.41 (0.02) & --  &  -- 	& 6580 (40)	 	& 8.12 (0.02) & (b) \\
                055802.46$-$722848.43 & DC 	& 25.70 (0.05) &  --  	&	-- & 6720 (80)	 	& 8.31 (0.03) & \\
                055808.89$-$542804.68 & DA   	&  25.24 (0.08) &  -- 	& 	-- & 4850 (60)	 	& 7.92 (0.05) & \\
                061813.08$-$801155.22 & DA   	&  27.98 (0.02) &   14\,800 (240)	& 	8.37 (0.06) 	& 13\,400 (230)	 	& 8.40 (0.01) & (o) \\
                062620.54$-$185006.83 & DAZ  	& 27.94 (0.04) & -- & -- & 7300 (60)	 	& 7.97 (0.02) & \\
                064604.27$-$224633.04 & DC	& 31.26 (0.09) &  --  	&	-- & 4380 (60)	 	& 7.78 (0.04) & \\
                064806.66$-$205839.53 & DA   	&  36.97 (0.06) &  -- 	& --	& 5040 (30)	 	& 7.91 (0.02) & \\
                070551.92$-$083526.76 & DC & 39.42 (0.08) & -- & -- & 4620 (340) & 7.9 (0.3) &  \\
                071550.55$-$370642.20 & DA   	&   29.23 (0.04) &  7260 (90)	& 	8.3 (0.2)	& 7240 (70)	 	& 8.41 (0.02) & \\
                072251.38$-$304234.38 & DA   	&  42.72 (0.07) &   --	& 	--	& 5140 (40)	 	& 8.56 (0.02) & \\
                073326.40$-$445325.34 & DA   	&  25.60 (0.02) &  9500 (40) 	& 	7.98 (0.04) & 9410 (80)	 	& 8.00 (0.02) & \\
                075328.47$-$511436.98 & DAH 	& 30.56 (0.03) &  --  	&	-- & 9280 (100)	 	& 8.39 (0.02) & \\
                075447.40$-$241527.71 & DAH 	& 26.54 (0.07) &  --  	&	-- & 5940 (50)	 	& 8.21 (0.03) & \\
                080151.04$-$282831.73 & DQpec & 28.54 (0.06) & -- & --	& 5680 (40)	 	& 7.85 (0.03) & \\
                \hline
        \end{tabular}
        \\
        Notes: (a) \citet{Landstreet2019}, (b) \citet{Tremblay2020}, (c) \citet{Subasavage2017}, (d) \citet{Subasavage2008}, (e) \citet{Subasavage2007}, (f) \citet{Kulebi2010}, (g) \citet{Kilic2020}, (h) \citet{Barstow1995}, (i) \citet{Reid2005}, (j) \citet{Bedard2017}, (k) \citet{Scholz2000}, (l) \citet{Gianninas2011}, (m) \citet{Blouin2019b}, (n) \citet{ODonoghue2013}, (o) \citet{Kepler2000}, (p) \citet{Dufour2005}, (q) \citet{bergeron2001}, (r) \citet{Coutu2019}, (s) \citet{Hollands2017}, (t) \citet{Dupuis1994}, (u) \citet{Bagnulo2021}, (v) \citet{Kirkpatrick2016}, (w) \citet{Raddi2017}, (x) \citet{Bergeron2021}, (y) \citet{Elms2022}. 
        Objects with an asterisk before their name have a parallax value outside of 40\,pc but may still be within that volume at 1$\sigma$. A spectral type in italics indicates we have updated the classification in this work. A spectral type followed by a colon represents a tentative classification. Table~\ref{tab:definitions} shows which atmospheric composition was used for the photometric fits of each white dwarf. All quoted uncertainties represent the intrinsic fitting errors. The \textit{3D Spectro} column for DA white dwarfs presents fitted Balmer line parameters. 
\end{table*}

\setcounter{table}{2}

\begin{table*}
\centering
\scriptsize
        \caption{Spectral types and parameters of the white dwarf sample (continued)}
        \begin{tabular}{llllllll}
                \hline
                WD\,J name  & SpT &  Parallax (mas) & $T_{\rm eff}$ [K] & $\log(g)$ & $T_{\rm eff}$ [K] & $\log(g)$ &  Note\\
                &  & & 3D Spectro &  	3D Spectro &  Gaia &  Gaia & \\
                \hline
                080833.93$-$530059.48 & DZA  	& 33.29 (0.08) &   -- & -- &  4140 (100)	 	& 7.78 (0.06) & \\
                081200.29$-$610809.79 & DA   	&  25.02 (0.05) &   6340 (60)	& 8.2 (0.1) & 6260 (60)	 	& 8.17 (0.03) & \\
                081227.07$-$352943.32 & DC   	&  89.51 (0.02) & --  	&	-- & 6240 (30)	 	& 8.18 (0.01) & \\
                081630.14$-$464113.24 & DC   	&  43.48 (0.06) & --  	&	-- & 4240 (40)	 	& 7.78 (0.03) & \\
                081716.19$-$680838.31 & DQpec & 25.7 (0.1) & --  	& 	--	& 4440 (100)	 	& 7.83 (0.07) \\
                081843.92$-$151208.31 & DZ     & 30.41 (0.14) & --  &  --  	&  3980 (210)	 	& 7.4 (0.2) & \\
                082533.15$-$510730.83 & DC:   	& 37.42 (0.05) &  --  	&	-- & 5010 (40)	 	& 7.98 (0.03) &  \\
                083759.16$-$501745.76 & DA   	&  31.52 (0.02) &  12\,860 (40) & 	8.33 (0.02) & 12\,490 (160)	 & 8.31 (0.01) & \\
                084635.27$-$362206.68 & DA    & 30.89 (0.07) & -- & -- & 4890 (40)	 	& 7.91 (0.03) & \\
                085021.30$-$584806.21 & DZA    & 42.96 (0.08) & --  &  --  	&  5600 (50)	 	& 8.90 (0.02) & \\
                085430.49$-$250848.99 & DA   	& 31.88 (0.05) &   6720 (90) 	& 8.2 (0.1)	& 6650 (60)	 	& 8.25 (0.02) & \\
                090212.89$-$394553.32 & DAH  	& 27.46 (0.03) &  --  	&	-- & 8770 (100)	 	& 8.37 (0.02) & \\
                090633.51$-$262656.02 & DA 	& 41.34 (0.06) &  -- & -- & 4990 (40)	 	& 7.95 (0.03) & \\
                090734.25$-$360907.93 & DA   	& 25.32 (0.08) &  5500 (130) 	& 	8.2 (0.3)	& 5220 (60)	 	& 7.95 (0.04) & \\
                091228.06$-$264201.50 & DA  	& 27.48 (0.05) & 12\,730 (40) & 9.47 (0.03) & 13\,440 (280)	 	& 9.19 (0.02) & \\
                091600.94$-$421520.68 & DZH:  	& 44.35 (0.04) & --  &  --  	& 5130 (30)	 	& 8.05 (0.02) & \\
                091620.71$-$631117.21 & DA   	&  42.82 (0.02) &  10\,270 (40) & 8.50 (0.03)	& 10\,110 (100)	 	& 8.51 (0.02) & \\
                091708.67$-$454613.68 & DAZ  	& 35.31 (0.03) & 	-- & --	& 6330 (40)	 	& 8.02 (0.02) & \\
                091808.59$-$443724.25 & DAH  	& 35.27 (0.05) &  --  	&	-- & 5330 (40)	 	& 8.02 (0.03) & \\
                092449.05$-$491529.60 & DC: 	& 44.31 (0.04) &  --  	&	-- &  5420 (30)	 	& 8.08 (0.02) & \\
                093011.42$-$295943.38 & DA  	& 30.53 (0.07) & -- & -- & 5100 (60)	 	& 7.93 (0.05) & \\
                093659.79$-$372130.80 & DQ  & 38.10 (0.02) &   -- 	& 	--	& 9230 (90)	 	& 8.09 (0.02) & (p) \\
                093659.94$-$372126.91 & DA   	&  38.15 (0.02) &  8130 (60) 	& 	8.0 (0.1) & 7910 (60)	 	& 8.05 (0.02) & (l) \\
                093736.24$-$385223.21 & DA   	&  28.99 (0.05) &   5930 (40)	& 	8.43 (0.06) & 5660 (50)	 	& 8.00 (0.03) & \\
                094052.75$-$423225.46 & DC  	& 26.71 (0.07) &  --  	&	-- &  5860 (60)	 	& 8.14 (0.03) & \\
                094240.23$-$463717.68 & DAH  	& 48.83 (0.03) &  --  	&	-- & 5970 (30)	 	& 8.01 (0.02) & \\
                095522.89$-$711808.37 & DA   	&  32.73 (0.02) &  14\,420 (260) 	& 	7.87 (0.05) 	& 14\,280 (210)	 	& 7.80 (0.02) & (l) \\
                101039.30$-$471729.83 & DA   	&  26.94 (0.06) &   5980 (40) 	& 	8.24 (0.08) 	& 5850 (40)	 	& 8.12 (0.02) & \\
                101341.21$-$523400.86 & DA	& 25.25 (0.05) &  7230 (40) & 8.49 (0.06) & 6920 (60)	 	& 8.13 (0.02) & \\
                101812.80$-$343846.05 & DA   	&  30.49 (0.09) &   -- & --	& 5090 (50)	 	& 8.04 (0.04) & \\
                101947.34$-$340221.88 & DAH & 36.30 (0.05) &  --  	&	-- & 6480 (50)	 	& 8.37 (0.02) & \\
                103427.04$-$672239.24 & DA   	&  42.40 (0.02) &   19\,430 (150) & 8.44 (0.02) 	& 18\,780 (350)	 	& 8.39 (0.02) & \\
                103706.75$-$441236.96 & DAH 	& 25.57 (0.07) &  --  	&	-- & 5680 (50)	 	& 7.92 (0.03) & \\
                104646.00$-$414638.85 & DAH 	&  35.41 (0.04) & --  	&	-- & 6750 (40)	 	& 8.04 (0.02) & \\
                105735.13$-$073123.18 & DC	& 81.51 (0.02) &  --  	&	-- &  7100 (50)	 	& 8.25 (0.02) & (q) \\
                105747.61$-$041330.16 & DZ & 27.51 (0.06) & --  &  --  	& 6950 (60)	 	& 8.09 (0.03) & (r) \\
                105915.98$-$281955.96 & DAZ  	& 25.34 (0.06) & -- & -- & 6650 (60)	 	& 8.05 (0.03) & \\
                111717.11$-$441134.49 & DC 	& 37.47 (0.04) &  --  	&	-- &  5590 (30)	 	& 7.53 (0.02) & \\
                113216.54$-$360204.95 & DZH   & 27.44 (0.12) & --  &  --  	 & 4590 (70)	 	& 7.86 (0.06) & \\
                114122.38$-$350406.93 & DZA  	& 34.18 (0.09) &  --  &  --  	 & 4600 (40)	 	& 7.84 (0.04) & \\
                114734.45$-$745759.24 & DC:  & 50.08 (0.06) &  --  	&	-- &  3820 (80)	 	& 7.74 (0.05) & \\
                114901.67$-$405114.98 & DC 	& 25.7 (0.1) &  --  	&	-- &  4290 (60)	 	& 7.75 (0.05) & \\
                115020.14$-$255335.40 & DC 	& 34.05 (0.05) &  --  	&	-- &  6690 (60)	 	& 8.17 (0.02) & \\
                115403.49$-$310145.29 & DC  	&  25.39 (0.07) & --  	&	-- &  6110 (60)	 	& 8.11 (0.03) & \\
                121456.38$-$023402.84 & DZH  	& 26.28 (0.12) &  --  &  --     & 5220 (60)	 	& 8.17 (0.04) & (s) \\
                121616.94$-$375848.13 & DC  	&  26.3 (0.1) & --  	&	-- & 	4460	(70) &	7.88 (0.07) & \\
                121724.77$-$632945.73 & DZ 	& 26.65 (0.04) & --  &  --  	&      8000 (70) & 	8.09 (0.02) & \\    	
                $*$ 122257.77$-$742707.7	&      DA  	& 24.96 (0.07) &    6020 (50)	& 	8.6 (0.1)	&  5580 (60) & 	7.95 (0.04) & \\       
                123156.66$-$503247.99 & DA  	& 30.48 (0.03) & 19\,110 (20) & 8.0 (0.2) &	18\,010 (350) & 	7.94 (0.02) & \\           
                123445.37$-$444001.75 & DC  	&  35.12 (0.04) & --  	&	-- & 	6670 (70) & 	8.19 (0.03) & \\   
                124112.37$-$243428.54  &  DZ & 26.38 (0.08) & --  &  --  	& 	6550 (70) & 	8.25 (0.03) & \\
                124155.92$-$133501.27  &  DC  	&  27.82 (0.05) & --  	&	-- & 	8250	 (80) & 	8.00 (0.03) & \\     
                124504.52$-$491336.69  &  DQ 	&  34.41 (0.03) &  -- 	& 	--	&  8500 (70) & 8.06 (0.02) & \\ 	
                130744.29$-$792511.64   	 	&      DC  	&  25.4 (0.1) & --  	&	-- & 	4670 (80) & 	7.98 (0.07) & \\             
                131727.39$-$543808.28	     &      DA    & 40.57 (0.04) &  5710 (40) & 7.90 (0.08) & 5760 (30) & 	7.95 (0.02) 	& \\
                131830.01+735318.25      &      DC:  	& 27.4 (0.1) &  --  	&	-- & 	5000 (40) &	7.35 (0.04) & \\            
                131958.95$-$563928.42	     &      DC 	&  27.93 (0.05) & --  	&	-- & 	7010	 (50) & 	8.11 (0.02) & \\            
                132550.44$-$601508.04      &      DB  	&  27.82 (0.03) &   11\,080 (130) &	-- &     	11\,510 (120) & 	7.98 (0.03) & \\
                132756.43$-$281716.98      &      DQ  	&  27.48 (0.06) &   --	& 	--	&     	6440 (140) & 	7.60 (0.06) & \\  
                133216.49$-$440838.71      &      DC  	&  29.25 (0.09) & --  	&	-- & 	5710 (80) & 	8.17 (0.04) & \\            
                133314.60$-$675117.19      &      DZ     & 37.98 (0.05) & --  &  --  	&     	5510 (90) &	8.11 (0.05) & \\	
                134349.01$-$344749.39      &      DA  	&  27.69 (0.09) &  -- 	& 	-- &    5140 (80) & 	7.81 (0.05) & \\	
                134441.03$-$650942.13      &      DA  & 25.90 (0.09) & --  &  --  &    	4790 (130) & 	7.79 (0.09) & \\
                140115.27$-$391432.21   	 	&      DAH  	& 36.00 (0.09) &  --  	&	-- & 5510 (60) & 	8.43 (0.03) & \\
                140608.61$-$695726.60      &      DA     & 27.92 (0.04) &  6910 (40) & 7.99 (0.05) & 6770 (50) & 	7.95 (0.02) & \\
                141041.67$-$751030.18	     &      DZA     & 30.01 (0.08) &  --  &  --  	&     	4950 (40) & 	7.90 (0.04) & \\
                141159.17$-$592044.99	     &      DA 	& 69.44 (0.03) & 6780 (40) & 8.07 (0.05) &	6650 (40) & 	8.11 (0.02)	& \\
                141220.36$-$184241.64   	 	&      \textit{DAH}  	&  30.06 (0.09) & --  	&	-- & 5720 (90) & 	8.08 (0.05) & (t) \\
                141622.47$-$653126.81   	 	&      DA  	&  25.92 (0.05) &  9130 (80) 	& 	8.58 (0.08)	&     8610 (90) & 	8.47 (0.02) & \\
                142254.17$-$460549.72   	 	&      DC     & 26.45 (0.08) &  --  	&	-- & 	6480 (60) &	8.22 (0.03) & \\            
                142428.39$-$510233.63  	 	&      DQ 	&  31.59 (0.05) &  -- 	& 	--	&     6550 (60) & 	8.09 (0.03) & \\	
                143015.38$-$240326.12    &      \textit{DA}  	&  30.7 (0.1) &  -- 	& --	&     4870 (60) & 	7.90 (0.05)	& (i) \\  
                143019.96$-$252040.40	 	 	&      DA  	&  31.64 (0.06) &   6930 (40)	& 	8.33 (0.06)	& 	6740 (70) & 	8.32 (0.03) & \\
                143826.23$-$560110.20     &      DC  & 25.61 (0.05) &  --  	&	-- &    8210 (80) & 	8.24 (0.02) & \\             
                144710.68$-$694040.21     &      DC 	& 33.76 (0.07) &  --  	&	-- & 	4470 (30) & 	7.24 (0.02) & \\            
                150324.74$-$244129.02	     &      DA  	& 38.51 (0.05) & 6100 (30) & 8.7  (0.8) & 5670 (30) & 	7.60 (0.02) & \\
                151431.85$-$462555.28      &      DQZ  &  44.27 (0.03) &   --	& 	--	&    7540 (60) &	8.03 (0.02) & \\	
                151907.38$-$485423.83      &      DQZ & 28.26 (0.04) &  --  	& 	--	&     8870 (80) & 	8.07 (0.02) & \\	
                \hline
        \end{tabular}
        \\
\end{table*}

\setcounter{table}{2}

\begin{table*}
\centering
\scriptsize
        \caption{Spectral types and parameters of the white dwarf sample (continued)}
        \begin{tabular}{lllllllll}
                \hline
                WD\,J name  & SpT & Parallax (mas) & $T_{\rm eff}$ [K] & $\log(g)$ & $T_{\rm eff}$ [K] & $\log(g)$ & Note\\
                &  & & 3D Spectro &  	3D Spectro &  Gaia &  Gaia & \\
                \hline
                152915.63$-$642811.20      &      DA  	&  30.82 (0.07) &   5550 (30)	& 	8.00 (0.04)	&   5200 (60) & 	7.77 (0.04) & \\
                152926.39$-$141614.44   	 	&      DA  	&  26.7 (0.1) &   5310 (100)	& 	8.2 (0.2)	&   	5270 (90) & 	8.25 (0.06) & \\
                153044.96$-$620304.10   	 	&      DAZ  	&  26.56 (0.07) &  -- 	& 	--	& 5880 (60) & 	8.17 (0.03) & \\ 
                154053.08$-$485837.95  	 	&      DZA  	&  27.4 (0.1) &  -- 	& 	--	&     4830 (50) &	7.98 (0.04) & \\
                155131.68$-$385049.90   	 	&      DC  	&  28.1 (0.1) &  --	& --	&  	5290 (40) & 	8.07 (0.03) & \\
                160027.92$-$131949.93      &      DC    & 27.2 (0.1) &  --  	&	-- & 	5010 (100) & 	7.97 (0.08) & \\
                160137.01$-$383209.35   	 	&      DA  	&  30.70 (0.09) &  -- & -- 	& 	4910 (40) & 	7.69 (0.03) & \\            
                160454.29$-$720347.59	     &      DC  	&  27.06 (0.06) & --  	&	-- & 	4090 (40) 	& 6.75 (0.04) & \\
                162224.44$-$551132.01  	 	&      DA  	&  27.39 (0.07) &  5640 (200) & 8.0 (0.5) &   5400 (80) & 	7.96 (0.05) & \\    
                162558.78$-$344145.71  	 	&      DAH  	&  28.6 (0.1) & --  	&	-- & 5000 (60) & 	7.81 (0.04) & \\
                163029.74$-$373936.84  	 	&      DC  	&  30.1 (0.1) & --  	&	-- & -- & -- & \\             
                163058.32$-$281815.48  	 	&      DC  & 25.5 (0.2) &  --  	&	-- & 	3950 (140) & 	7.72 (0.09) & \\             
                163337.05$-$371314.28	     &      DC  	& 47.40 (0.07) &  --  	&	-- &    5430 (40) & 	8.24 (0.02) & \\             
                163626.53$-$873706.08  	 	&      DQ   &   26.42 (0.07) &  --	& 	--	&     5660 (70) & 	8.21 (0.04) & \\
                164725.24$-$544237.58	     &      DA  	& 45.20 (0.02) & 8800 (30) & 8.34 (0.02) & 8530 (70) & 	8.33 (0.02) & \\  
                165335.21$-$100116.33 &  DAe & 30.65 (0.04) & 7360 (40) & 7.84 (0.06) & 7350 (90) & 7.91 (0.03) & \\
                165538.10$-$232555.73  	 	&      DA  	&  26.15 (0.06) &  7120 (40) 	& 	8.09 (0.05) 	&   6990 (50) & 	8.10 (0.02) & \\
                165823.76$-$805857.14  	&      DC  	&  44.62 (0.05) & --  	&	-- &    4690 (30) & 	7.85 (0.03) & \\          	
                170054.19$-$690832.65      &      DA  	& 27.86 (0.05) & 8160 (40) & 8.59 (0.03) & 7950 (70) & 	8.47 (0.02) & \\	
                170427.96$-$005026.31 & DA & 37.04 (0.05) & 6650 (700) & 8.39 (0.08) & 6540 (50) & 8.30 (0.02) & \\
                170430.68$-$481953.11  	 	&      DC  	& 38.8 (0.1) &  --  	&	-- & 	5180 (40) & 	8.18 (0.03) & \\          	
                170641.36$-$264334.71      &      DAH  	& 76.65 (0.03) &  --  	&	-- & 	6130 (30)	 & 8.34 (0.01)  & (u) \\
                171436.16$-$161243.30      &      DAH  	& 26.98 (0.04) &  --  	&	-- & 	11\,140 (140)	 & 8.74 (0.02)  & \\    
                171652.09$-$590636.29      &      DAH 	& 33.51 (0.03) & -- & -- &     8600 (90) & 	8.37 (0.02) & \\        	
                172239.79$-$355441.65   	&      DA  	&  27.18 (0.08) &  7120 (50) 	& 	8.32 (0.08)	 &      7100 (130) & 	8.36 (0.04) & \\
                173351.73$-$250759.90   	 	&      DA  	&  26.8 (0.1) &  5520 (40) 	& 	8.00 (0.08)	& 	5560 (60) & 	8.17 (0.04) & \\
                173800.77$-$311237.21      &      DC 	& 25.3 (0.1) &  --  	&	-- & 	4660 (70) & 	7.97 (0.06) & \\             
                173837.46$-$342729.28      &      DA  	&  25.5 (0.1) &  -- 	& 	-- &      4830 (120) & 	7.83 (0.09) & \\
                174220.63$-$203935.92   	 	&      DC  	&  34.42 (0.07) &  --	& --	&  5590 (50) & 	8.17 (0.03) & \\             
                174246.61$-$650514.67      &      DC  	& 33.43 (0.04) &   --	& --	&  	8580 (90) & 	8.46 (0.02) & \\           	
                174349.28$-$390825.95      & DA  	&  46.83 (0.02) &   11\,700 (20) 	& 	7.89 (0.01)	&    11\,610 (210) & 	8.09 (0.03) & \\
                174611.08$-$625141.41   	 	&      DA  	& 29.04 (0.04) &   7530 (40) 	& 	8.00 (0.06)	& 	7400 (60) & 	7.99 (0.02) & \\
                174736.82$-$543631.16  	&      DC 	&  73.99 (0.05) & --  	&	-- & 	4360 (30) 	& 7.82 (0.02) 	& (v) \\            
                175325.53$-$840510.03  	 	&      DC  	&  26.27 (0.09) & --  	&	-- &    5110 (70) 	& 8.10 (0.05) & \\	
                175554.31$-$245648.94 & DA & 26.62 (0.03) & 12\,830 (10) & 8.395 (0.006) & 13\,000 (200) & 8.29 (0.02) & \\
                175931.34$-$620108.87      &      DA  	& 26.01 (0.04) & 17\,000 (70) & 9.14 (0.02) &     16\,220 (270) & 	9.06 (0.01) & \\
                180314.84$-$805750.43   	 	&      DC     &  29.7 (0.1) & --  	&	-- &    4800 (70) & 	8.25 (0.05) & \\	
                180315.18$-$371725.54   	 	&      DA  	&  37.84 (0.07) &  5500 (50) 	& 	8.1 (0.1)	&    5410 (50) & 	8.14 (0.03) & \\
                180345.86$-$752318.35   	 	&      DAH  	& 31.95 (0.05) &  --  	&	-- &	5600 (40) & 	8.03 (0.03) & \\   
                180853.83$-$704231.62   	 	&      DC  	& 28.1 (0.1) &  --  	&	-- & 	4720 (60) & 	8.02 (0.05) 	& \\          
                180901.95$-$410140.69      &      DC  	& 32.01 (0.06) &  --  	&	-- & 	5730 (100) & 	7.9 (0.6) & \\             
                181311.31$-$860811.23   	 	&      DA  	&  25.90 (0.08) &   --	& --	& 	4950 (70) & 	7.95 (0.06) & \\             
                181548.96+553232.22      &      DC:  	&  26.37 (0.05) & --  	&	-- & 	4630 (50) & 	7.19 (0.04) & \\	
                182159.54$-$595148.52      &      \textit{DA}  	& 33.16 (0.06) & --    & --   & 	4750 (30)	& 7.27 (0.03) & (c) \\          
                182228.37$-$653738.06   	 	&      DA  	&  27.88 (0.09) &   --	& 	--	&      5050 (40) & 	7.96 (0.04) & \\
                183351.29$-$694203.57      &      DA  	&  30.39 (0.02) &  8120 (50) & 7.87 (0.06) &	8010 (60) & 	7.39 (0.02) & \\	
                183852.85$-$441631.32   	 	&      DA  	&  29.57 (0.09) &  5770 (110) 	& 	8.5 (0.2)	&    5560 (100) & 	8.17 (0.06) & \\
                183856.35$-$535726.05   	 	&      DA  	& 28.0 (0.1) &   5260 (30) 	& 	8.00 (0.04)	&      5150 (60) & 	8.04 (0.04) & \\
                184650.69$-$452139.33   	 	&      DC  	&  35.6 (0.1) &  --	& --	&  	4860 (40) & 	7.92 (0.04) & \\             
                184947.86$-$095744.38 & DA & 30.61 (0.03) & 12\,130 (20) & 8.24 (0.01) & 12\,130 (160) & 8.05 (0.02) & \\
                185005.58$-$285117.29  	 	&      DA  	& 28.31 (0.08) &   5700 (180) 	& 	8.5 (0.4)	& 	5330 (90) & 	8.02 (0.07) & \\ 
                185709.09$-$265059.22	     &      DA  & 25.31 (0.06) & 7110 (100) & 8.2 (0.2) &	7020 (60) & 	7.97 (0.03) & \\     
                185934.75$-$162656.29 & DA & 25.86 (0.05) & 8510 (150) & 8.00 (0.05) & 8000 (90) & 8.0 (0.6) & \\
                190255.35$-$044012.64      &      DC  	& 28.6 (0.1) &  --  	&	-- & 	4670 (90) & 	8.03 (0.08) & \\             
                190525.34$-$495625.77      &      DZ     & 33.82 (0.02) & --  &  --  	&      10\,920 (120) & 	8.11 (0.02) & \\
                191100.25$-$382031.89      &      DC:  	&  35.7 (0.1) & --  	&	-- &    4080 (120) & 	7.68 (0.08) & \\  	
                191144.26$-$272954.76      &      DB  	& 28.87 (0.03) &    11\,680 (150)	& 	--	&   11\,480 (140) & 	8.02 (0.03) & \\
                191858.23$-$434920.40  	 	&      DC  	& 29.1 (0.1) &   --	& -- &	5360 (130) & 	8.51 (0.07) & \\             
                191936.23+452743.55      &      DC:  	&  35.64 (0.04) & --  	&	-- &    4780 (20) & 	7.31 (0.02) & \\             
                193538.63$-$325225.56   	 	&      DZAH  	& 29.3 (0.1) & --  &  --  	&     5310 (50) & 	7.97 (0.04) & \\
                194522.76$-$490420.23   	 	&      DC  	& 29.1 (0.1) &  --  	&	-- &    4320 (100) & 	7.81 (0.08)	& \\          	
                194549.13$-$153135.63 & DA & 32.35 (0.03) & 12\,590 (40) & 8.422 (0.008) & 12\,380 (170) & 8.39 (0.02) & \\
                195211.78$-$732235.48	     &      DC     & 31.2 (0.3) &  --  	&	-- & 	-- & 	-- & \\ 
                195616.36$-$525819.16  	&      DA     & 31.30 (0.08) &  7670 (620)  	&	8.65 (0.06) & 	-- & 	-- & Not in catalogue \\ 
                195639.81$-$511544.83  	 	&      DC  	&  31.6 (0.1) & --  	&	-- &    4640 (70) & 	7.93 (0.06) & \\             
                200348.80$-$474800.18  	 	&      DA  	&  32.73 (0.06) &   6060 (40)	& 	8.07 (0.07)	&    5920 (50) & 	7.97 (0.03) & \\	
                200707.98$-$673442.18   	 	&      DAH &  26.00 (0.05) & --  	&	-- &	7770 (70) & 	8.33 (0.02) & \\             
                201722.68$-$401043.73  	 	&      DZA     & 25.3 (0.1) & --  &  --  	&     4970 (80) & 	7.94 (0.0) & \\	
                201756.19$-$124639.44	     &      DC  	& 35.6 (0.1) &  --  	&	-- & 	4820 (50) & 	8.24 (0.04) & \\             
                202011.65$-$382445.66  	 	&      DA  	&   35.53 (0.05) &  7400 (40)	& 	8.44 (0.06)	&    7290 (70) & 	8.43 (0.02) & \\
                202016.78$-$652523.10  	 	&      DAZ  	& 25.99 (0.07) & 	-- & -- &     6340 (70) & 	8.30 (0.03) & \\ 
                202025.46$-$302714.65.              &      DC  	& 57.27 (0.02) & 	-- & -- &     9930 (110) & 	8.04 (0.02) & \\
                202030.93$-$420256.74	     &      DQ  &   25.02 (0.06) & -- 	& 	--	&     6970 (70) & 	8.02 (0.03) & \\	
                202748.03$-$563031.58  	 	&      DZ     & 28.0 (0.1) & --  &  --  	&     4140 (120) & 	7.82 (0.09) & \\
                202749.54$-$430115.21	     &      DC:  	&  47.02 (0.07) & --  	&	-- &    4880 (40) & 	8.39 (0.03) & \\
                202837.91$-$060842.77       & DA & 28.09 (0.03) & 11\,860 (100) & 8.49 (0.02) & 11\,340 (290) & 8.40 (0.04) & \\
                202956.94$-$643420.13  	 	&      DQ  	 & 26.79 (0.04) &  --  	& 	--	&     7290 (70) & 	8.03 (0.02) & \\
                204911.00$-$544617.50	     &      DA  	& 25.48 (0.04) & 7670 (30) & 8.02 (0.03) & 7550 (60) & 	7.91 (0.02) & \\
                \hline
        \end{tabular}
        \\
\end{table*}

\setcounter{table}{2}

\begin{table*}
\centering
\scriptsize
        \caption{Spectral types and parameters of the white dwarf sample (continued)}
        \begin{tabular}{llllllll}
                \hline
                WD\,J name  & SpT & Parallax (mas) & $T_{\rm eff}$ [K] & $\log(g)$ & $T_{\rm eff}$ [K] & $\log(g)$  & Note\\
                &  & & 3D Spectro &  	3D Spectro  &  Gaia &  Gaia & \\
                \hline
                205050.50$-$612235.61   	 	&      DA  	&  29.14 (0.05) &   7050 (80)	& 	8.28 (0.09)	&    6960 (70) & 	8.43 (0.03) & \\ 
                205213.41$-$250415.13      &      DC 	& 55.61 (0.04) &  --  	&	-- &    4910 (20) & 	7.85 (0.02)  &  \\             
                211240.64$-$292217.96  	&      \textit{DZQ} 	& 30.49 (0.04) &	-- & -- & 9770 (110) & 	8.11 (0.03)  & (w) \\        	
                212121.30$-$255716.33      &      DA  	& 40.78 (0.05) & 19\,450 (20) & 8.11 (0.05) & 19\,210 (370) & 	8.07 (0.02)  & \\
                212602.02$-$422453.76      &      DC:  	& 39.1 (0.3) &  --  	&	-- & 	5480 (30) & 	7.52 (0.03)   &  \\	
                213721.24$-$380838.22	     &      DC  	&  30.89 (0.06) & --  	&	-- &    6860 (70) & 	8.31 (0.03)   & \\
                214023.96$-$363757.44  	 	&      warm DQ  	&  25.09 (0.05) & --  	&	-- &    13\,190 (230) & 	8.84 (0.02)   & (x) \\
                214324.09$-$065947.99	     &      DA  	& 55.10 (0.03) & 9390 (80) & 8.5 (0.06) &  	8910 (80) 	& 8.42 (0.02)	&  \\
                214756.59$-$403527.79      &      DZQH  	& 35.8 (0.5) &  --  	&	-- & 	-- & -- & (y) \\
                $*$ 214810.74$-$562613.14  	 	&      DAH  	&  24.98 (0.08) & --  	&	-- &	5930 (60) & 	8.08 (0.03)    & \\	
                220437.98$-$312713.76	&      DA  	&  40.69 (0.07) &  -- 	& 	--	&   	4810 (30) & 	7.92 (0.03)    & \\	
                220552.11$-$665934.73	&      DAH 	&  31.82 (0.05) & --  	&	-- &	5260 (40) & 	7.84 (0.03)    & \\             
                220655.28$-$600135.32	&      DA  	&  26.82 (0.08) &   --	& 	--	&    5040 (40) & 	7.90 (0.04)    & \\
                223418.67$-$553403.40	&      DC  	&  26.5 (0.1) & --  	&	-- & 	4690 (70) & 	7.84 (0.05)    & \\             
                223601.50$-$554852.02 	&      DZ    & 31.34 (0.07) & --  &  --  	&     5130 (40) & 	8.00 (0.03)    & \\
                223607.66$-$014059.65 &      DAH  	&  25.63 (0.04) & --  	&	-- &	10\,020 (160) &	8.37 (0.03)    & \\             
                223634.58$-$432911.11	     &      DA  	& 33.00 (0.04) & 6730 (30) & 8.02 (0.04) & 6240 (40) & 	7.92 (0.02)    & \\	
                223700.03$-$542241.81	     &      DA & 33.93 (0.02) & 8320 (10) & 8.184 (0.008) &	8220 (70) & 	8.01 (0.02)    & \\	
                225335.70$-$143828.19  	 	&      DA  	&  27.4 (0.1) &   5500 (30)	& 	8.20 (0.05)	&     5320 (100) & 	8.10 (0.07)    & \\
                230232.34$-$330907.96       &      DC 	& 28.2 (0.1) &  --  	&	-- &    4710 (90) & 	7.90 (0.07)    & \\             
                230345.52$-$371051.56	     &      DZ     & 30.9 (0.1) & --  &  --  	&     4270 (90) &	7.88 (0.07)    & \\     
                234300.85$-$644737.90      &      DC  	&  26.89 (0.06) & --  	&	-- & 	5800 (50) & 7.98 (0.03)    & \\             
                234935.57$-$521528.02	     &      DC     & 32.36 (0.05) &  --  	&	-- & 	6250 (60) & 	8.42 (0.02)    & \\             
                235419.41$-$814104.96   	 	&      DZH  	& 37.10 (0.06) & -- & -- &  	4480 (40) & 	7.77 (0.04)    & \\
                235422.99$-$514930.65	     &      DC:  	& 32.90 (0.08) &  --  	&	-- & 	4470 (50) & 	7.81 (0.03)    & \\
                \hline
        \end{tabular}
        \\
\end{table*}

\subsection{DA white dwarfs}
\label{sec:DA}
The spectra for all observed DA white dwarfs are shown in Fig.~\ref{fig:DA1}. All DA white dwarfs with \textit{Gaia} T$_{\rm eff} >$ 5200\,K, and with more than one spectral line visible, were fitted spectroscopically using our fitting code described in Section~\ref{sec:models}, with best-fit atmospheric parameters corrected for 3D convection \citep{tremblay13c} identified in Table~\ref{tab:final_all}. We show fits to Balmer lines for the DA white dwarfs in Fig~\ref{fig:DA1_fit}. We do not fit the spectrum of \textbf{WD\,J0312$-$6444} as it is a known unresolved DA+DA binary \citep{Kilic2020}.

\begin{figure}
    \centering
	\includegraphics[width=\columnwidth]{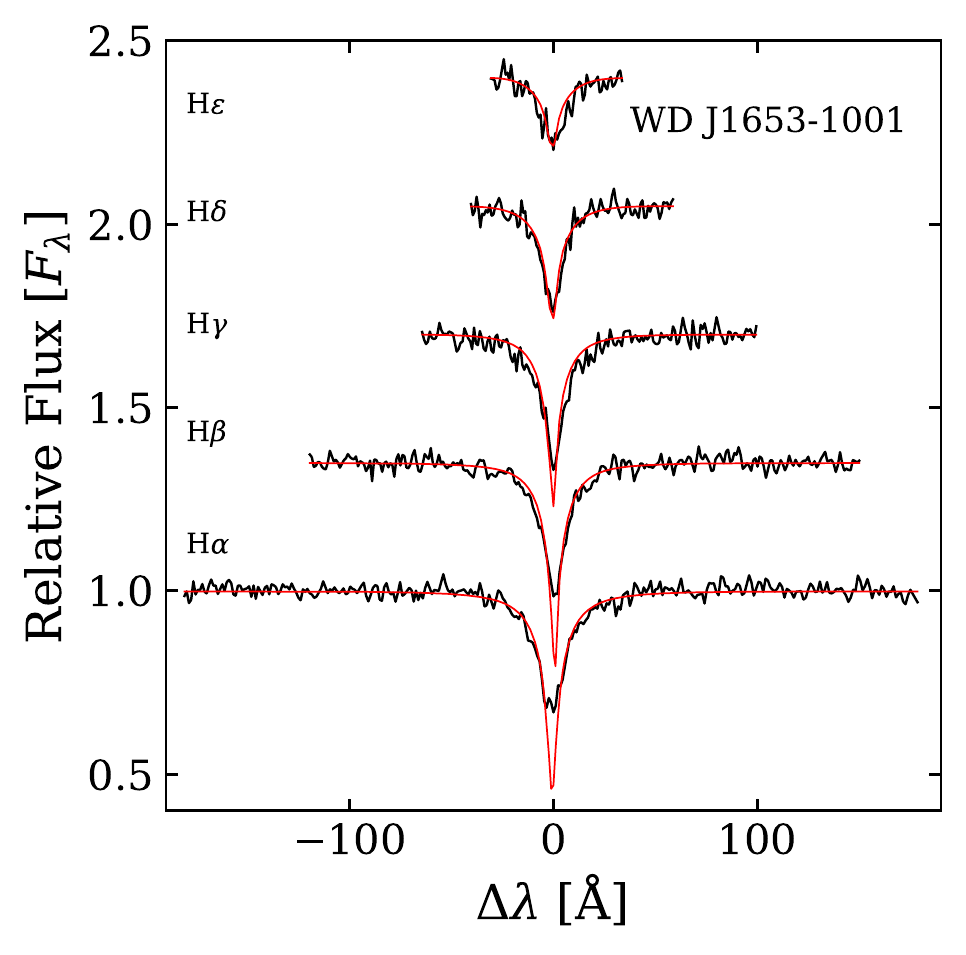}
	\caption{Spectroscopic fits to the normalised Balmer lines for the DAe white dwarf \textbf{WD\,J1653$-$1001}.}
    \label{fig:DAe_fits}
\end{figure}

\textbf{WD\,J1653$-$1001} is a DA white dwarf for which we make a tentative detection of emission in the core of the H~$\alpha$ and H~$\beta$ lines (see Fig.~\ref{fig:DAe_fits}). This emission appears to be similar to that seen in the DAe white dwarf \textbf{WD\,J0412+7549} observed in Paper I. Therefore we make the tentative classification of \textbf{WD\,J1653$-$1001} as a DAe. A discussion of these systems will be presented in Elms et al. (in prep.).

\subsection{Magnetic white dwarfs}

Fig.~\ref{fig:DAH1} shows 28 magnetic white dwarfs with hydrogen atmospheres that have spectral type DAH. It is not simple to determine the mass of a highly magnetic white dwarf by photometric fitting in the optical because of Zeeman splitting and displacement of spectral lines. Therefore the error bars of the $\log(g)$ values quoted in Table~\ref{tab:final_all} for cool magnetic white dwarfs may be slightly underestimated (Paper II).

\textbf{WD\,J0103$-$0522} was analysed in Paper I, where a quadratic wavelength shift of the $\pi$-component was observed, due to a complex field geometry, and has the largest \textit{Gaia} photometric surface gravity of any white dwarf in the sample. Even from the higher resolution X-Shooter observations, the line cores have round shapes and do not show evidence of multiple sub-components.

\textbf{WD\,J0317$-$8532B} is a 1.27\,$\pm$\,0.02~\,\Msun\ DAH which has a very high field strength of $\approx$ 340\,MG \citep{Barstow1995}, and is part of a wide double-degenerate binary system with a DA companion, \textbf{WD\,J0317$-$8532A}. This system has been studied extensively pre-\textit{Gaia}, as \textbf{WD\,J0317$-$8532B} is potentially a double-degenerate merger product due to its large mass (\citealt{Ferrario1997,Kulebi2010}). We have calculated the \textit{Gaia} best-fit parameters of the two components of this binary system (see Table~\ref{tab:final_all}), and have used these to determine the total ages of both stars (\citealt{Hurley2000,Cummings2018,Bedard2020}). The total age of the DAH \textbf{WD\,J0317$-$8532B} is 315\,$\pm$\,80\,Myr, and the total age of the companion is 450\,$\pm$\,40\,Myr, where errors are statistical and likely underestimated, especially for the hot magnetic component. These total ages are in agreement within 2$\sigma$ with single-star evolution for both objects. A merger could cause a cooling delay, such that the magnetic star would appear younger than its companion, and we cannot rule this out for \textbf{WD\,J0317$-$8532B} if there is a moderate cooling delay of the order of 200\,Myr.

\textbf{WD\,J1706$-$2643} was observed by \citet{Bagnulo2021} who detected a field strength of 8 MG. The field strengths of the remaining DAH white dwarfs have been estimated by visual comparison with theoretical $\lambda$-B curves \citep{Friedrich1996} and are displayed in Table~\ref{tab:Mag_st}. Uncertainties in field strength are estimated based on the width of the Zeeman split lines. 

\begin{table}
	\centering
        \caption{Magnetic field strengths for newly identified magnetic white dwarfs in the 40\,pc sample}
        \label{tab:Mag_st}
        \begin{tabular}{lll}
                \hline
                WD\,J name & SpT  & $\langle B \rangle$ (MG)  \\
                \hline
                  001349.89$-$714954.26 & DAH & 0.4 (0.2) \\
                  001830.36$-$350144.71 & DAH & 6.8 (0.4) \\
                  $*$ 014240.09$-$171410.85 & DAH & 15.1 (0.2) \\
                  025245.61$-$752244.56 & DAH & 22 (3) \\
                  035531.89$-$561128.32 & DAH & 2.3 (0.2)  \\
                  042021.33$-$293426.26 & DAH & 0.4 (0.2)  \\
                  050552.46$-$172243.48 & DAH & 3.9 (0.2)  \\
                  $*$ 054858.25$-$750745.20 & DZH & 1.1 (0.2) \\
                  075328.47$-$511436.98 & DAH & 19 (2) \\
                  075447.40$-$241527.71 & DAH & 10.5 (0.2) \\
                  090212.89$-$394553.32 & DAH & 21 (1) \\
                  091808.59$-$443724.25 & DAH & 0.4 (0.2) \\
                  094240.23$-$463717.68 & DAH & 3.4 (0.2) \\
                  101947.34$-$340221.88 & DAH & 110 (10) \\
                  103706.75$-$441236.96 & DAH & 0.3 (0.1) \\
                  104646.00$-$414638.85 & DAH & 3.6 (0.2) \\
                  113216.54$-$360204.95 & DZH & 0.25 (0.02) \\
                  121456.38$-$023402.84 & DZH & 2.1 (0.2) \\
                  140115.27$-$391432.21 & DAH & 7.7 (0.5) \\
                  141220.36$-$184241.64 & DAH & 21 (3) \\
                  162558.78$-$344145.71 & DAH & 4.0 (0.2) \\
                  171436.16$-$161243.30 & DAH & 55 (7) \\
                  171652.09$-$590636.29 & DAH & 0.7 (0.2) \\
                  180345.86$-$752318.35 & DAH & 0.2 (0.2) \\
                  193538.63$-$325225.56 & DZAH & 0.10 (0.01) \\
                  200707.98$-$673442.18 & DAH & 6.4 (0.2) \\
                  $*$ 214810.74$-$562613.14 & DAH & 12.4 (0.4) \\
                  220552.11$-$665934.73 & DAH & 2.2 (0.3) \\
                  223607.66$-$014059.65 & DAH & $>$\,250 \\
                  235419.41$-$814104.96 & DZH & 0.6 (0.2) \\
                \hline
        \end{tabular}
        \\
        Notes: Objects with an asterisk before their name have a parallax value outside of 40\,pc but may still be within that volume at 1$\sigma_\varpi$.
\end{table}

\textbf{WD\,J2236$-$0140} is magnetic, but its field strength cannot be well-constrained from the limited number of spectral features. There is a broad feature at $\approx$ 4400--4600~\AA. There is also a narrower, stationary component at 4140~\AA. The field strength is estimated to be 250 $< B <$ 750\,MG from these components, although H~$\alpha$ spectroscopy is needed to confirm this.

Fig.~\ref{fig:DZH1} shows seven magnetic metal-polluted white dwarfs. \textbf{WD\,J2354$-$8141} and \textbf{WD\,J1132$-$3602} show splitting of the Ca\,\textsc{ii} H line into two groups of two, and the Ca\,\textsc{ii} K line into six because of the large spin-orbit effect for the 4p state of Ca\,\textsc{ii} \citep{Kawka2011}. \textbf{WD\,J0916$-$4215} is potentially a highly magnetic DZH white dwarf with complex splitting of its spectral features. The field strengths of new DZH white dwarfs have been estimated and are displayed in Table~\ref{tab:Mag_st}. \textbf{WD\,J1935$-$3252} is weakly magnetic (100\,kG) with spectral type DZAH. 

The lower limit of detectable magnetic field strength depends on the object; the best case for a magnetic field detection is for an object with very narrow Ca lines and a high signal-to-noise. In this case, we find that field strengths of less than $\approx$ 50\,kG cannot be detected using X-Shooter spectroscopy.

For all magnetic white dwarfs, we estimate field strengths in Table~\ref{tab:Mag_st} from Zeeman splitting but do not derive spectroscopic atmospheric parameters, which is notoriously difficult \citep{kulebi2009}. Spectropolarimetry is required to determine the magnetic status of the remaining newly observed white dwarfs which do not display Zeeman splitting, a recent effort has been made towards this by \citet{Bagnulo2022} for young white dwarfs in 40\,pc.

\textbf{WD\,J0812$-$3529} has been classified as a DC in this work from a Goodman spectrum. \citet{Bagnulo2020} classify it as a DAH with a field strength of 30\,MG, determined from their high-quality spectropolarimetric observations. 

\subsection{DB white dwarfs}
The spectra for the two DB white dwarfs we observe are shown in Fig.~\ref{fig:DB1}. We derive the \Teff\ of these white dwarfs using 3D model atmospheres \citep{Cukanovaite2021}, and parameters are displayed in Table~\ref{table:dbparams}. These are in reasonable agreement with \textit{Gaia} values. These white dwarfs are at the cool end of the DB range, where spectroscopic fits are difficult \citep{koester2015, rolland2018}. We therefore fix $\log(g)$ to that determined from \textit{Gaia} photometry.

\begin{table}
	\centering
        \caption{Atmospheric parameters and chemical abundances of DB white dwarfs, with fixed $\log(g)$ determined from photometric fitting.}
        \label{table:dbparams}
        \begin{tabular}{llll}
                \hline
                WD\,J name & \Teff\ [K] & $\log(g)$ & $\log({\rm H/He})$ \\
                 &  (Spectro) & (\textit{Gaia})  & \\
                \hline
                1325$-$6015 & 11550 (120) & 7.98 (0.02) & $-$5.03 (0.08) \\
                1911$-$2729 & 11680 (150) & 8.02 (0.02) & $-$5.5 (0.3) \\
                \hline
        \end{tabular}
        \\
        Note: All quoted uncertainties represent the intrinsic fitting errors. We recommend adding systematics of 1\,per\,cent in \Teff\ to account for data calibration errors.
\end{table}

\subsection{DC white dwarfs}

The spectra of 69 DC white dwarfs are shown in Fig.~\ref{fig:DC1}. Nineteen of these were observed with the Goodman or FAST spectrographs, which both only provide spectra in the optical blue range 3000--6000~\AA\, such that H~$\alpha$ coverage is missing from the data. This is often the only diagnostic line for DA white dwarfs with low temperatures. Therefore, further spectroscopy may reveal that a subset of these DC systems are in fact DA white dwarfs. The coolest DA in the sample that was observed with Goodman is \textbf{WD\,J1317$-$5438}, which has a \Teff\ of $\approx$ 5800\,K. For white dwarfs below $\approx$ 5600\,K, the resolution and typical signal-to-noise ratio achieved with Goodman are not high enough to detect the H~$\beta$ line. Therefore the eleven optical blue-only DC with temperatures above 5600 K are likely to be genuine DC as we would see the H~$\beta$ line if they were DA. The remaining eight DC with lower temperatures could have unobserved H~$\alpha$ lines, and require further observations. These are classified as tentative DC (DC: spectral type in Table~\ref{tab:final_all}).

Three new white dwarf candidates from the north, \textbf{WD\,J1815+5532}, \textbf{WD\,J1919+4527}, and \textbf{WD\,J1318+7353}, are all confirmed as white dwarfs spectroscopically. They are classified as tentative DC (DC:) as their OSIRIS spectra are noisy, and potential spectral features cannot be excluded.

On the \textit{Gaia} HR diagram (see Fig.~\ref{fig:HR_Full}), \textbf{WD\,J1952$-$7322} is shown to have the faintest absolute \textit{Gaia} G-band magnitude for any DC white dwarf within 40\,pc. The spectrum of \textbf{WD\,J1952$-$7322} displays hints of mild optical collision-induced absorption (CIA), which would be consistent with a mixed H and He atmospheric composition and IR-faint categorisation \citep{Bergeron2022}. Only \textit{Gaia} photometry is available for this white dwarf, so its parameters cannot be constrained given the degeneracy between $\log({\rm H/He})$ and \Teff\ with such broad band-passes. \textbf{WD\,J1630$-$2818} shows signs of mild optical CIA in its spectrum. For both of these white dwarfs, we therefore do not infer \Teff\ and $\log(g)$ from \textit{Gaia} photometry.

\textbf{WD\,J1147$-$7457} is a potential ultra-cool ($<$ 4000\,K) DC white dwarf and a candidate halo white dwarf, as it has a tangential velocity of $\approx$ 160 km/s. 

\textbf{WD\,J1604$-$7203} is a low-probability ($P_{\rm WD}$ = 0.28) white dwarf candidate in the \citet{Gentile2021} catalogue. It has a \textit{Gaia} photometric $\log(g)$ of 6.75, and a \Teff\ of 4090\,K, when fitted as a single star. This object is likely a double degenerate system (see Section~\ref{sec:binary} for discussion).

There are Ca\,\textsc{ii} H+K emission features in the spectrum of \textbf{WD\,J0519$-$7014} which are not associated with the white dwarf and are due to less than ideal sky subtraction as the result of contamination from the Large Magellanic Cloud. This white dwarf is still classified as a DC, as these emission features are not from the star itself.

\subsection{DAZ white dwarfs}
\label{sec:DAZ}
Fig.~\ref{fig:DAZ1} shows the spectra of ten DAZ white dwarfs. \textbf{WD\,J0358+2157} (reported in Paper I) and \textbf{WD\,J0426$-$4153} are both highly metal-polluted DAZ white dwarfs that will have a dedicated analysis in a future study (Cutolo et al., in prep.), and therefore no spectral fits are presented here.

We fit the other eight DAZ stars using the combined photometry and spectroscopy method of \citet{Koester2010}. The fitting of \Teff\ and $\log(g)$ relies on photometry from \textit{Gaia}, GALEX \citep{GALEX2005}, PanSTARRS \citep{PanSTARRS2016}, SkyMapper \citep{SkyMAPPER2005}, 2MASS \citep{2MASS2006} and WISE \citep{WISE2010}. Not all photometry was available for every object. The best-fit parameters, including $\log({\rm Ca/H})$ abundances, of the remaining 8 DAZ white dwarfs are displayed in Table~\ref{table:dazparams}.

\begin{table}
	\centering
        \caption{Atmospheric parameters and chemical abundances of newly observed DAZ white dwarfs, where \Teff\ and $\log(g)$ have been determined from a combination of spectroscopic and photometric fitting.}
        \label{table:dazparams}
        \begin{tabular}{llll}
                \hline
                WD\,J name & \Teff\ [K] & $\log(g)$ & $\log({\rm Ca/H})$\\
                \hline
                0143$-$6718 & 6230 (10) & 7.91 (0.01) & $-$11.05 \\
                0343$-$5125 & 6710 (10)  & 7.99 (0.01)  & $-$9.60 \\
                0445$-$4232 & 6650 (10) &	7.92 (0.01)  & $-$10.70 \\
                0626$-$1850 & 7280 (10) & 7.96 (0.01)  & $-$10.50 \\
                0917$-$4546 & 6260 (10) &	7.97 (0.01) & $-$10.30 \\
                1059$-$2819 & 6530 (10) & 7.99 (0.01) & $-$9.30 \\
                1530$-$6203 & 5860 (10) &	8.15 (0.02) & $-$11.00 \\
                2020$-$6525 & 6120 (10) &	8.20 (0.02) & $-$10.65 \\
                \hline
        \end{tabular}
        \\
        Note: All quoted uncertainties represent the intrinsic fitting errors. We recommend adding systematics of 1\,per\,cent in \Teff\ to account for data calibration errors.
\end{table}

\subsection{DZ and DZA white dwarfs}
\label{sec:DZ}

We show 24 DZ, DZA, DZH and DZAH white dwarf spectra in Figs.~\ref{fig:DZ1}--\ref{fig:DZA1}. We fit the combined spectroscopy and photometry for 19 of these objects. \textbf{WD\,J0548$-$7507} and \textbf{WD\,J2354$-$8141} are DZH white dwarfs and are not fitted due to the complexity of the splitting of their lines. We also do not fit the potentially high-field DZH \textbf{WD\,J0916$-$4215}. The X-Shooter spectra of \textbf{WD\,J2147$-$4035} and \textbf{WD\,J1214$-$0234} have already been fitted by \citet{Elms2022} and \citet{Hollands2021}, respectively. In this section, we discuss all DZ and DZA white dwarfs for which we fit their combined spectroscopy and photometry using the model atmosphere code of \citet{Koester2010}. 

The fitting of \Teff\ and $\log(g)$ relies on photometry from \textit{Gaia}, GALEX, PanSTARRS, SkyMapper, 2MASS and WISE. Not all photometry was available for every object. We detect Ca in all DZ and DZA spectra in our sample.

\begin{figure*}
    \centering
	\includegraphics[width=\textwidth]{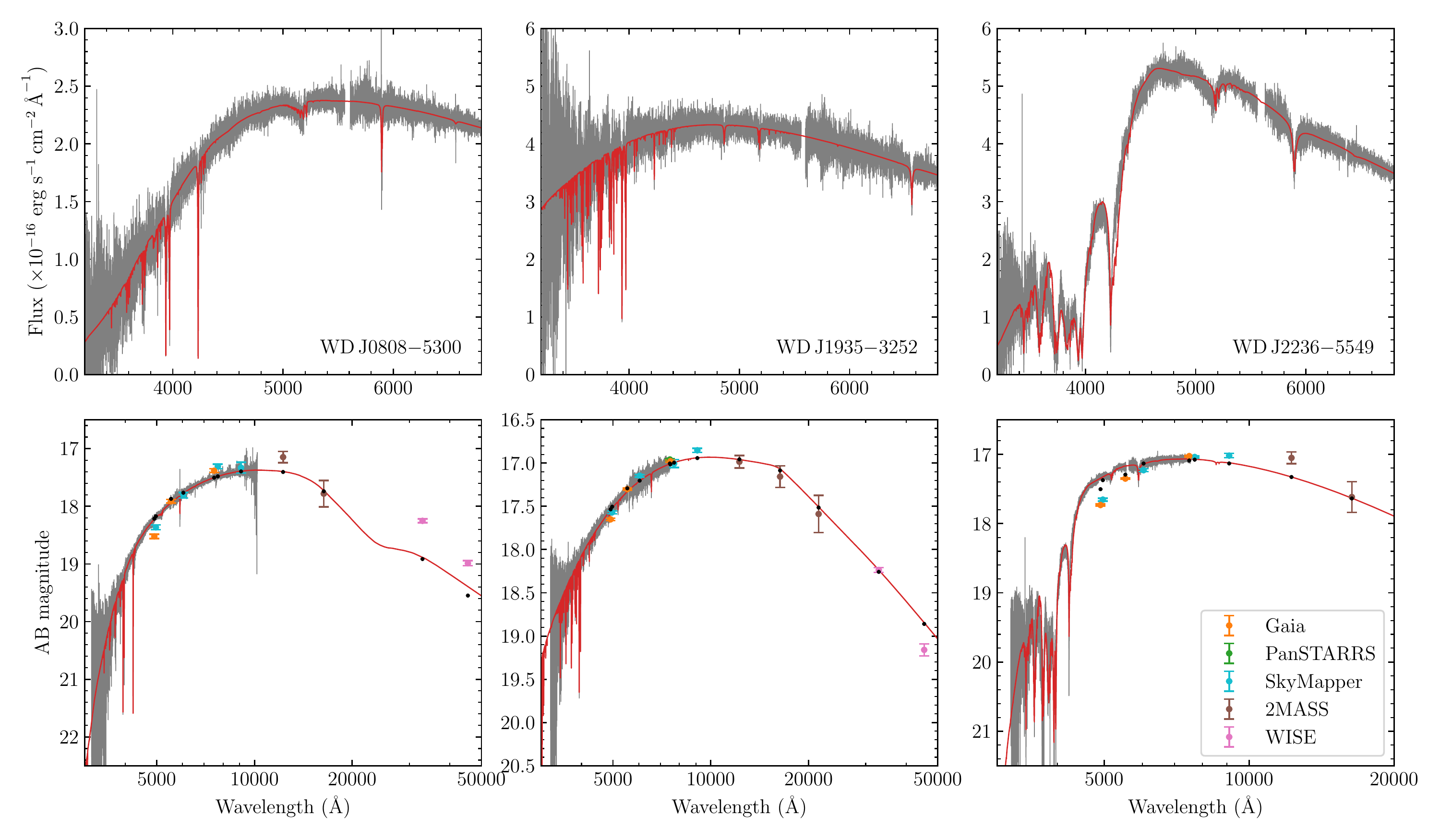}
	\caption{Simultaneous fits of spectroscopy and photometry for three metal-rich DZ and DZA white dwarfs: \textbf{WD\,J0808$-$5300} (left panels), \textbf{WD\,J1935$-$3252} (middle panels) and \textbf{WD\,J2236$-$5549} (right panels). The top row of panels compare our best fit models to normalised spectroscopic observations. The spectroscopic observations are re-calibrated onto the models but are still in physical flux units. The bottom panels compare our best fit models to catalogue photometry over a wider wavelength range than the available spectroscopy provides.}
    \label{fig:DZ_fits}
\end{figure*}

\textbf{WD\,J1057$-$0413}, \textbf{WD\,J1217$-$6329}, \textbf{WD\,J1905$-$4956}, and \textbf{WD\,J2236$-$5548} are DZ white dwarfs with He-dominated atmospheres where no H is detected. Ca was detected in the atmosphere of \textbf{WD\,J1057$-$0413} by \citet{Coutu2019}, and we additionally detect Mg and Fe in this white dwarf. \textbf{WD\,J2236$-$5548} is a cool DZ which shows strong metal lines and has a He-dominated atmosphere, we have constrained abundances for five metals: Ca, Na, Mg, Fe, and Cr (See Fig.~\ref{fig:DZ_fits} for fit). 

\textbf{WD\,J0044$-$1148}, \textbf{WD\,J0554$-$1035}, \textbf{WD\,J1241$-$2434}, and \textbf{WD\,J1333$-$6751} are all DZ white dwarfs with He-dominated atmospheres and trace H that is inferred indirectly from their spectra. There is no visible H~$\alpha$ line in these spectra, however we observe narrow and sharp metal lines. The electron density in the atmosphere, and therefore the opacity of the atmosphere, is significantly increased by the presence of H which causes the metal lines to appear narrower. \textbf{WD\,J0044$-$1148} has a companion separated by a few arcseconds (see Table~\ref{tab:binaries_know}). \textbf{WD\,J0554$-$1035} was identified as a DZ with Ca in Paper I; we also measure the $\log({\rm H/He})$ abundance that was not previously constrained. There is a blend of Fe lines in the spectra of \textbf{WD\,J1241$-$2434} and \textbf{WD\,J1333$-$6751}.

\textbf{WD\,J0818$-$1512}, \textbf{WD\,J1132$-$3602}, \textbf{WD\,J2027$-$5630}, and \textbf{WD\,J2303$-$3710} have very narrow Ca lines, indicating a H-dominated atmosphere. Therefore their abundances presented in Table~\ref{table:dzparams} are in relation to hydrogen, despite their spectral classification of DZ. There is Zeeman splitting in the spectrum of \textbf{WD\,J1132$-$3602} which indicates a magnetic field of about 280\,kG, which has been accounted for in the modelling. \textbf{WD\,J2027$-$5630} is a potential ultra-cool DZ, with a combined spectroscopic and photometric \Teff\ of around 3700\,K.

\textbf{WD\,J0808$-$5300}, \textbf{WD\,J0850$-$5848}, \textbf{WD\,J1141$-$3504}, \textbf{WD\,J1410$-$7510}, \textbf{WD\,J1540$-$4858}, \textbf{WD\,J1935$-$3252}, and \textbf{WD\,J2017$-$4010} are DZA white dwarfs with sharp metal lines and a very narrow H~$\alpha$ line, indicating nearly pure-H atmospheres (Fig.~\ref{fig:DZA1}).

\textbf{WD\,J0850$-$5848} has a high photometric $\log(g)$ of $\approx$ 8.9 when using mixed H/He models, and a combined spectroscopic and photometric $\log(g)$ of $\approx$ 8.7. We infer a white dwarf mass of 1.045\,$\pm$\,0.005~\,\Msun, and a progenitor mass of 5.4\,$\pm$\,0.1~\,\Msun~\citep{Cummings2018}. The spectrum of \textbf{WD\,J0850$-$5848} does not indicate the presence of CIA, so we infer that this is indeed a massive white dwarf, and is among the most massive metal-polluted white dwarfs ever observed.

\textbf{WD\,J1410$-$7510} and \textbf{WD\,J1540$-$4858} both display sharp Fe lines. The DZAH \textbf{WD\,J1935$-$3252} displays strong metal lines from four elements: Ca, Mg, Fe and Al, and has a weak magnetic field of 100\,kG (see Fig.~\ref{fig:DZ_fits} for fit).

\textbf{WD\,J0808$-$5300} displays atmospheric CIA of H$_2-$H$_2$ and H$_2-$H, seen in infrared photometry from 2MASS and WISE. This white dwarf is polluted by Ca, Na, Mg, Fe, Al and Cr. We detect an absorption feature caused by MgH molecules at around 5200\,\AA, a feature that has been detected in white dwarfs with mixed H/He atmospheres \citep{Blouin2019, Kaiser2021}. To our knowledge, we have made the first detection of MgH in a H-dominated atmosphere white dwarf. The hybrid fit to this white dwarf is shown in Fig.~\ref{fig:DZ_fits}.

The abundances of Li, Na, Mg, K, Ca, Cr and Fe for the DZH white dwarf \textbf{WD\,J1214$-$0234} are calculated in \citet{Hollands2021} using the X-Shooter spectrum shown in Fig.~\ref{fig:DZH1}.

\setlength{\tabcolsep}{4.5pt}

\begin{table*}
	\centering
		\scriptsize
        \caption{Atmospheric best-fit parameters and chemical abundances of DZ and DZA white dwarfs, where \Teff\ and $\log(g)$ have been determined from a combination of spectroscopic and photometric fitting. Weakly magnetic DZH and DZAH are also fitted. Upper table: Best-fit parameters for white dwarfs with He-dominated atmospheres. Lower table: Best-fit parameters for white dwarfs with H-dominated atmospheres.}
        \label{table:dzparams}
        \begin{tabular}{lllllllllll}
                \hline
                WD\,J name & SpT & \Teff\ [K] & $\log(g)$ & $\log({\rm H/He})$ & $\log({\rm Ca/He})$ & $\log({\rm Na/He})$ & $\log({\rm Mg/He})$ & $\log({\rm Fe/He})$ & $\log({\rm Cr/He})$\\
                \hline
                0044$-$1148 & DZ & 5310 (30) & 7.99 (0.02) & $-$1.23 (0.03) & $-$11.53 (0.04) & --  & -- & -- & --  \\
                0554$-$1035 & DZ & 6230 (20) & 8.04 (0.01) & $-$4.52 (0.05) & $-$11.78 (0.03) & --  & -- & -- & --  \\
                1057$-$0413 & DZ & 6500 (20) & 8.03 (0.01) & -- & $-$10.30 (0.01) & --  & $-$8.88 (0.02) & $-$9.60 (0.03) & --  \\
                1217$-$6329 & DZ & 7420 (80) & 7.96 (0.03) & -- & $-$10.43 (0.05) & --  & -- & -- & --   \\
                1241$-$2434 & DZ & 6310 (30) & 8.13 (0.01) & $-$2.78 (0.04) & $-$11.42 (0.01) & --  & -- & $-$10.29 (0.03) & --  \\
                1333$-$6751 & DZ & 5640 (60) & 8.17 (0.03) & $-$1.97 (0.02) & $-$11.41 (0.03) & --  & -- & $-$10.62 (0.04) & --  \\
                1905$-$4956 & DZ & 10\,600 (40) & 8.08 (0.01) & -- & $-$8.99 (0.03) & --  & -- & -- & -- \\
                2236$-$5548 & DZ & 5350 (10) & 8.17 (0.01) & -- & $-$9.17 (0.01) & $-$9.16 (0.01) & $-$7.41 (0.01) & $-$8.64 (0.01) & $-$9.9 (0.1) \\
                \hline
                \hline
                WD\,J name & SpT & \Teff\ [K] & $\log(g)$ & $\log({\rm Ca/H})$ & $\log({\rm Na/H})$ & $\log({\rm Mg/H})$ & $\log({\rm Fe/H})$ & $\log({\rm Al/H})$ & $\log({\rm Cr/H})$\\
                \hline
                0808$-$5300 & DZA & 4910 (10) & 8.34 (0.01) & $-$9.74 (0.02) & $-$9.60 (0.02)  & $-$8.16 (0.02) & $-$9.05 (0.03) & $-$9.54 (0.03) & $-$10.48 (0.03) \\
                0818$-$1512 & DZ & 4720 (10) & 7.68 (0.01) &  $-$11.50 (0.04) & -- & -- & -- & -- & -- \\
                0850$-$5848 & DZA & 5430 (20) & 8.73 (0.01) &   $-$10.65 (0.01) & -- & -- & -- & -- & -- \\
                1132$-$3602 & DZH & 4990 (10) & 8.12 (0.01) &  $-$10.84 (0.03) & -- & -- & -- & -- & -- \\
                1141$-$3504 & DZA & 4880 (20) & 8.07 (0.01) &  $-$11.11 (0.02) & -- & -- & -- & -- & -- \\
                1410$-$7510 & DZA & 5180 (10) & 8.011 (0.007) &  $-$10.64 (0.01) & -- & -- & $-$9.36 (0.02) & -- & -- \\
                1540$-$4858 & DZA & 5000 (30) & 8.10 (0.02) &  $-$10.57 (0.03) & -- & -- & $-$9.77 (0.03) & -- & -- \\
                1935$-$3252 & DZAH & 5430 (10) & 8.00 (0.01) &  $-$9.68 (0.02) & -- & $-$7.89 (0.03) & $-$8.61 (0.02) & $-$9.12 (0.04) & -- \\
                2017$-$4010 & DZA & 5250 (20) & 8.08 (0.01) &  $-$10.62 (0.03) & -- & -- & -- & -- & -- \\
                2027$-$5630 & DZ & 3750 (130) & 7.7 (0.1) &  $-$12.6 (0.1) & -- & -- & -- & -- & -- \\
                2303$-$3710 & DZ & 4790 (50) & 8.28 (0.03) &  $-$10.76 (0.06) & -- & -- & -- & -- & -- \\
                \hline
        \end{tabular}
        \\
        Note: All quoted uncertainties represent the intrinsic fitting errors. We recommend adding systematics of 1\,per\,cent in \Teff\ to account for data calibration errors.
\end{table*}

\subsection{DQ white dwarfs}
\label{sec:DQ}

We observed nine DQ white dwarfs (Fig.~\ref{fig:DQ1}). We fitted all objects with the \citet{Koester2010} model atmosphere code using an iterative procedure. Results from the fitting procedure are in Table~\ref{table:dqparams}. The fitting of \Teff\ and $\log(g)$ relies on photometry from \textit{Gaia}, GALEX, SkyMapper and 2MASS. Not all photometry was available for every object. 

Two of the DQ white dwarfs in the sample, \textbf{WD\,J0801$-$2828} and \textbf{WD\,J1636$-$8737}, display CH molecular absorption features in their spectra near 4300~\AA. We classify \textbf{WD\,J0801$-$2828} and \textbf{WD\,J0817$-$6808} as peculiar DQ (DQpec) white dwarfs. This classification describes cool DQ below 6000\,K with molecular absorption bands with central wavelengths that have been shifted 100$-$300~\AA\ from the positions of the C$_{2}$ Swan bands \citep{Hall2008}. The warm DQ \textbf{WD\,J2140$-$3637} is discussed further in Section~\ref{sec:WDJ2140}.

\subsection{DQZ and DZQ white dwarfs}
\label{sec:DQZ}

\textbf{WD\,J1514$-$4625} and \textbf{WD\,J1519$-$4854} are classified as DQZ, and \textbf{WD\,J2112$-$2922} is classified as DZQ. All three show both carbon absorption features and metal lines in their spectra (see Fig.~\ref{fig:DQ3}). In all three cases, we detect metals from the Ca\,\textsc{ii} H+K lines, and carbon from the C$_{2}$ Swan bands. The field of view of the Goodman spectrograph is 10 arcmin, and \textbf{WD\,J1514$-$4625} and \textbf{WD\,J1519$-$4854} were both observed by Goodman and are separated by over a degree on the sky, so they are not a duplicate observation. These stars are unlikely to be DQ + DZ binaries, as all three stars have photometric $\log(g)$ values close to or above the canonical value of 8.0 for single stars. \citet{Elms2022} make a tentative detection of carbon in the ultra-cool DZ \textbf{WD\,J2147$-$4035}; this star would notionally be a DZQpecH (Fig.~\ref{fig:DZH1}). These objects are discussed further in Section~\ref{sec:dqz_dzq_disc}.

\begin{table*}
	\centering
        \caption{Atmospheric parameters and chemical abundances of DQ, DQZ and DZQ white dwarfs. \Teff\ and $\log(g)$ have been determined from iterative spectroscopic and photometric fitting. The warm DQ \textbf{WD\,J2140$-$3637} is not included here, as we assume it has a C-dominated atmosphere when fitting, rather than a He-dominated atmosphere (see Section~\ref{sec:WDJ2140}).}
        \label{table:dqparams}
        \begin{tabular}{lllllll}
                \hline
                WD\,J name & SpT & \Teff\ [K] & $\log(g)$ & $\log({\rm C/He})$ & $\log({\rm H/He})$ & $\log({\rm Ca/He})$\\
                \hline
                0801$-$2828 & DQpec & 5970 (10) &	7.96 (0.01) & $-$5.90 (0.01) & $-$4.25  & --\\
                0817$-$6808 & DQpec & 4620 (20) & 8.02 (0.02) & $-$7.70 (0.01) & -- & -- \\
                0936$-$3721 & DQ  & 8890 (20) & 7.96 (0.01) & $-$4.94 (0.02) & --  & --\\
                1245$-$4913 & DQ  & 8120 (20) &	7.94 (0.01) & $-$5.30 (0.02) & -- & --\\
                1327$-$2817 & DQ  & 7510 (50) &	7.90 (0.02) & $-$5.74 (0.01) & -- & --\\
                1424$-$5102 & DQ  & 6340 (30) &	7.98 (0.01) & $-$7.45 (0.01) &-- & --\\
                1514$-$4625 & DQZ  & 7470 (20) & 7.99 (0.01) & $-$5.96 (0.02) & --& $-$11.7 \\
                1519$-$4854 & DQZ  & 8960 (20) & 8.06 (0.01) & $-$4.60 (0.02) & --& $-$11.6 \\
                1636$-$8737 & DQ  & 5370 (40) &	8.11 (0.02) & $-$7.60 (0.01) & $-$3.40  & --\\
                2020$-$4202 & DQ  & 6870 (30) &	7.99 (0.01) & $-$6.6 (0.2) &-- & -- \\
                2029$-$6434 & DQ  & 7120 (20) &	7.97 (0.01) & $-$6.30 (0.01) & -- & --\\
                2112$-$2922 & DZQ & 8960 (40) & 7.87 (0.01) & $-$4.80 (0.01) & -- & $-$11.6 \\
                \hline
        \end{tabular}
        \\
        Note: All quoted uncertainties represent the intrinsic fitting errors. We recommend adding systematics of 1\,per\,cent in \Teff\ to account for data calibration errors.
\end{table*}

\subsection{Main-sequence stars}
\label{sec:STAR}

Fig.~\ref{fig:STAR1} shows two white dwarf candidates with $P_{\rm WD}$ equal to 1 from \citet{Gentile2021} that turned out to be main-sequence stars following spectroscopic observations: \textbf{WD\,J0924$-$1818} and \textbf{WD\,J1732$-$1710}. The issues of contamination from \textit{Gaia} DR2 white dwarf samples \citep{Gentile2019} have mostly been solved in DR3 \citep{Gentile2021}, such that there are now minimal contaminant sources in our sample ($<$ 1\,per\,cent of this 40\,pc south sample has main-sequence contaminants). It is likely that these sources have spurious \textit{Gaia} parallaxes which places them on the white dwarf sequence of the HR diagram, hence their high $P_{\rm WD}$ values. Both stars have high excess flux error values in \textit{Gaia}, indicating either variability or issues with photometry.

\section{Discussion}
\label{sec:disc}

\subsection{Comparison with the overall 40 pc sample}

\begin{figure}
    \centering
	\includegraphics[width=\columnwidth]{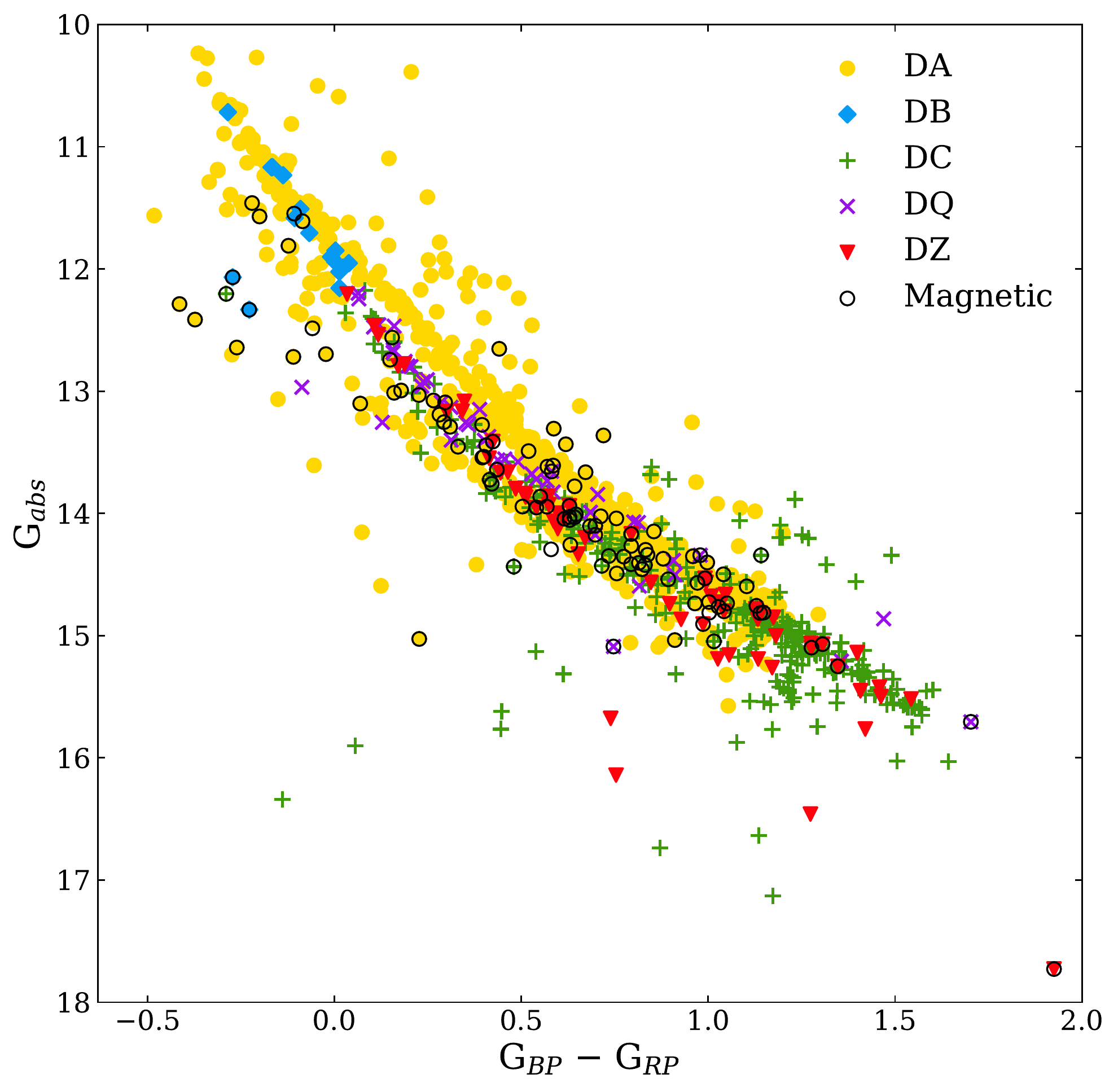}
	\caption{A \textit{Gaia} DR3 Hertzsprung-Russell (HR) diagram for the full spectroscopic 40\,pc sample of 1058 white dwarfs. Magnetic stellar remnants have black contours. Data are colour- and symbol-coded by their primary spectral type classification only, for simplicity.}
    \label{fig:HR_Full}
\end{figure}

The \textit{Gaia} DR3 HR diagram for the volume-limited 40\,pc spectroscopic white dwarf sample is shown in Fig.~\ref{fig:HR_Full}. The faintest and reddest white dwarf in the sample is \textbf{WD\,J2147$-$4035}, at the bottom right of Fig.~\ref{fig:HR_Full} \citep{Elms2022}.

The mean \textit{Gaia} photometric \Teff\ of our sub-sample of 246 white dwarfs presented in this work is 6930\,K, whereas for the full 40\,pc sample the mean \textit{Gaia} \Teff\ is 7530\,K. Both samples have a standard deviation of $\approx$ 3000\,K. We expect our sub-sample to have a lower mean \Teff\ than in 40\,pc overall because our new observations are biased towards fainter white dwarfs at lower \Teff\ that had not previously been observed spectroscopically.

The mean \textit{Gaia} photometric mass of both our sub-sample and the overall 40\,pc sample is 0.63~\,\Msun. The mean mass is biased by the cool white dwarfs with \Teff\ $<$ 5000\,K for which masses may have been incorrectly calculated from models (see Fig.~\ref{fig:teff_logg}). The mean mass for white dwarfs with \Teff\ $>$ 5000\,K is 0.66~\,\Msun\ (Paper II).

\begin{figure}
    \centering
	\includegraphics[width=\columnwidth]{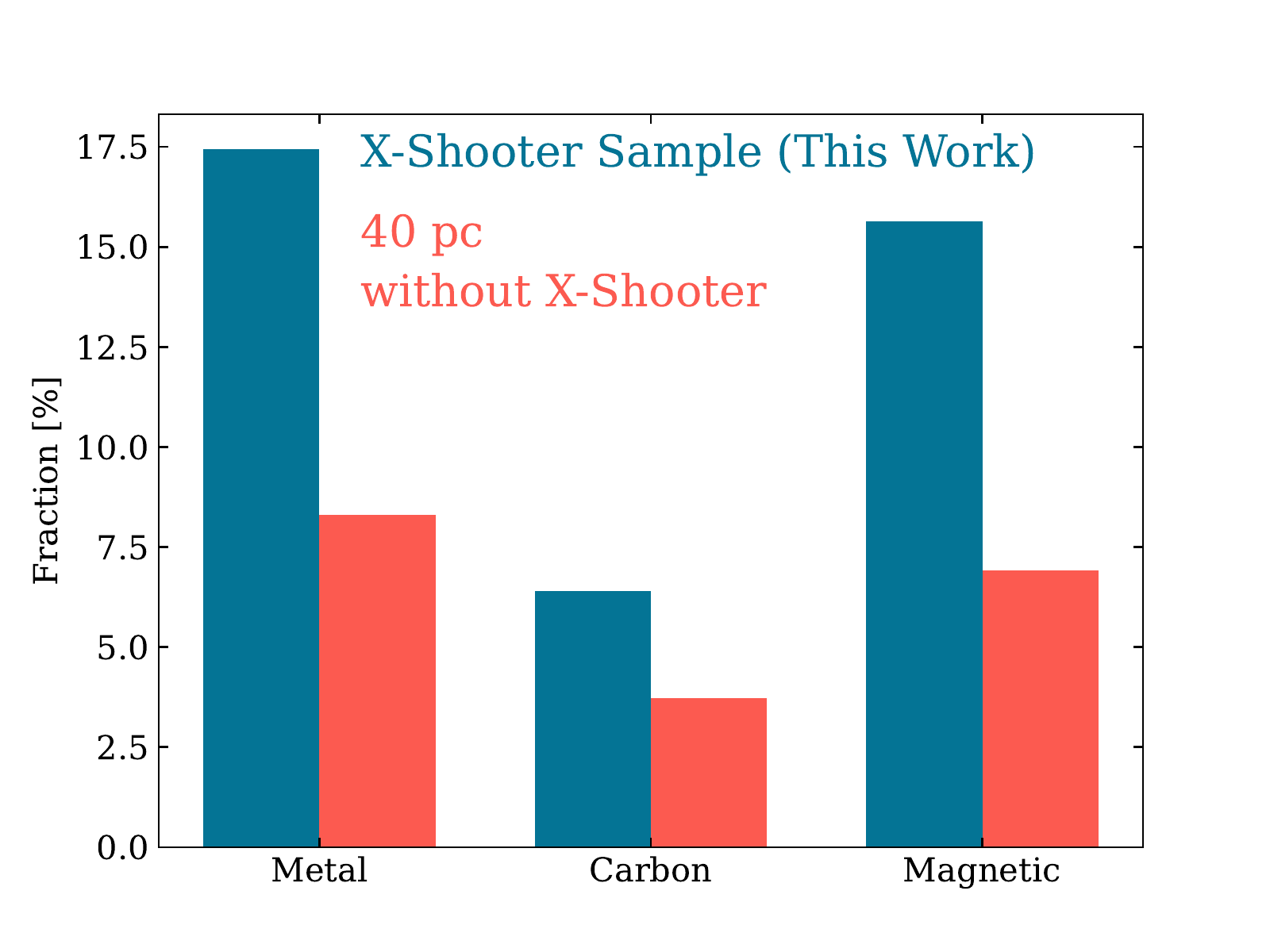}
	\caption{Incidence of different atmospheric compositions between a sample of 179 X-Shooter observations presented in this work, and the full 40\,pc sample not including X-Shooter observations. We consider white dwarfs with trace metals in their atmospheres, carbon in their atmospheres, and magnetic white dwarfs.}
    \label{fig:spt_bar}
\end{figure}

Within this work, we have a sample of 179 white dwarfs observed with X-Shooter. This X-Shooter sample provides a set of white dwarf spectra with a large wavelength coverage and high signal-to-noise ratio. Metal-polluted, carbon-rich, and magnetic white dwarfs are over-represented in this X-Shooter sub-sample compared to the remaining 40\,pc white dwarfs (not including those observed with X-Shooter), as shown in Fig.~\ref{fig:spt_bar}. An over-abundance of magnetic and of metal-polluted white dwarfs may be due to the resolution of X-Shooter, a medium-resolution spectrograph, compared to the observations for the existing 40\,pc sample, providing us with the opportunity to detect low levels of metal abundances and weaker Zeeman splitting. Since our X-Shooter sub-sample is biased towards lower \Teff, there might also be a greater incidence of metal-pollution, trace carbon and magnetism due to this bias. It is critical to obtain higher resolution and quality spectra of 40\,pc white dwarfs to update fractions of metal-polluted and magnetic white dwarfs and determine the underlying distributions for this volume-limited sample. 

Using Keck HIRES high-resolution spectra, \citet{Zuckerman2003} observed that 25\,per\,cent of DA white dwarfs with \Teff\ below 10\,000\,K were metal-polluted. In our 40\,pc south sub-sample, we observe a metal-pollution rate of around 15\,per\,cent for DA white dwarfs with \Teff\ below 10\,000\,K. It is possible that we do not see such a high fraction of polluted white dwarfs as reported in \citet{Zuckerman2003} due to the intrinsic fainter nature of our sub-sample. Our sub-sample also uses medium-resolution spectroscopy rather than high-resolution, so less metal lines will be detected.

\subsection{Metal-polluted DQ White Dwarfs}
\label{sec:dqz_dzq_disc}
Both \citet{Coutu2019} and \citet{Farihi2022} observe a significant deficit in the frequency of metal pollution in DQ stars, and observe only a 2\,per\,cent pollution rate in DQs. To explain this deficit, \citet{Hollands2022} and \citet{Blouin2022} model the effect of metal pollution on the presence of Swan bands in DQ white dwarf spectra, and show that for above a relatively low level of pollution, Swan bands will be suppressed such that a DQZ would present as a DZ. Therefore, the only metal-polluted DQ stars that can be observed spectroscopically should have relatively low levels of pollution \citep{Blouin2022}, which aligns with what we observe in the 40\,pc sample. Another explanation for this observed deficit is that DQ white dwarfs at all temperatures are the product of binary evolution, altering their circumstellar environments and reducing the occurrence of planetary debris \citep{Farihi2022}. 

Thirty\,per\,cent of the white dwarf population in 40\,pc have He-rich atmospheres, and DZ and DQ white dwarfs independently correspond to about 18\,per\,cent of those white dwarfs with He-rich atmospheres. If the presence of carbon and metals in white dwarfs are independent of each other, the percentage of He-rich white dwarfs in a volume-limited sample with both metal and carbon lines should be about 3\,per\,cent. Therefore in 40\,pc we expect to find 8\,$\pm$\,3 metal-polluted DQ white dwarfs.

The white dwarf \textbf{WD\,J0916+1011} is classified as a DQZ by \citet{Kleinman2013} and is at a distance of 38.6\,pc. \textbf{WD\,J2147$-$4035} is a white dwarf with spectral type DZQH \citep{Elms2022} and its spectrum is presented in Fig.~\ref{fig:DZH1}. The white dwarf \textbf{Procyon B} is not in the \textit{Gaia} DR3 catalogue, however it is at a distance of $\approx$ 3.5\,pc and was classified as a DQZ following the detection of Mg lines in its UV spectrum \citep{Provencal2002}.

Adding \textbf{Procyon B}, \textbf{WD\,J0916+1011} and \textbf{WD\,J2147$-$4035} to the two newly observed DQZ white dwarfs and the DZQ in this paper gives six out of 253 He-rich white dwarfs in the 40\,pc sample that display both metal lines and carbon lines. We therefore do not detect a notable deficit in the numbers of these white dwarfs, but we note that the numbers are too small to draw meaningful conclusions. \citet{Coutu2019} use a sample of SDSS spectra which have lower signal-to-noise than the X-Shooter and Goodman spectra in our sample, possibly explaining why they see less metal-pollution in DQs, or Swan bands in DZs, than we observe in 40\,pc, potentially missing those stars with very weak Swan bands and stronger metal features such as \textbf{WD\,J2112$-$2922}. 

\subsection{WDJ2140-3637: A warm DQ white dwarf}
\label{sec:WDJ2140}

\textbf{WD\,J2140$-$3637} is a warm DQ white dwarf that has been previously identified in \citet{Bergeron2021}. Warm DQ white dwarfs have spectra dominated by C\,\textsc{i} lines in the optical, and tend to have He-dominated atmospheres \citep{Koester2019} compared to the C/O-dominated magnetic hot DQ white dwarfs at \Teff\ $>$ 18\,000\,K \citep{Dufour2007}. \citet{Bergeron2021} showed that \textbf{WD\,J2140$-$3637} belongs to a massive warm DQ white dwarf sequence identified by \citet{Coutu2019} and they state that it has the largest carbon abundance of any warm DQ. 

\begin{figure}
    \centering
	\includegraphics[width=\columnwidth]{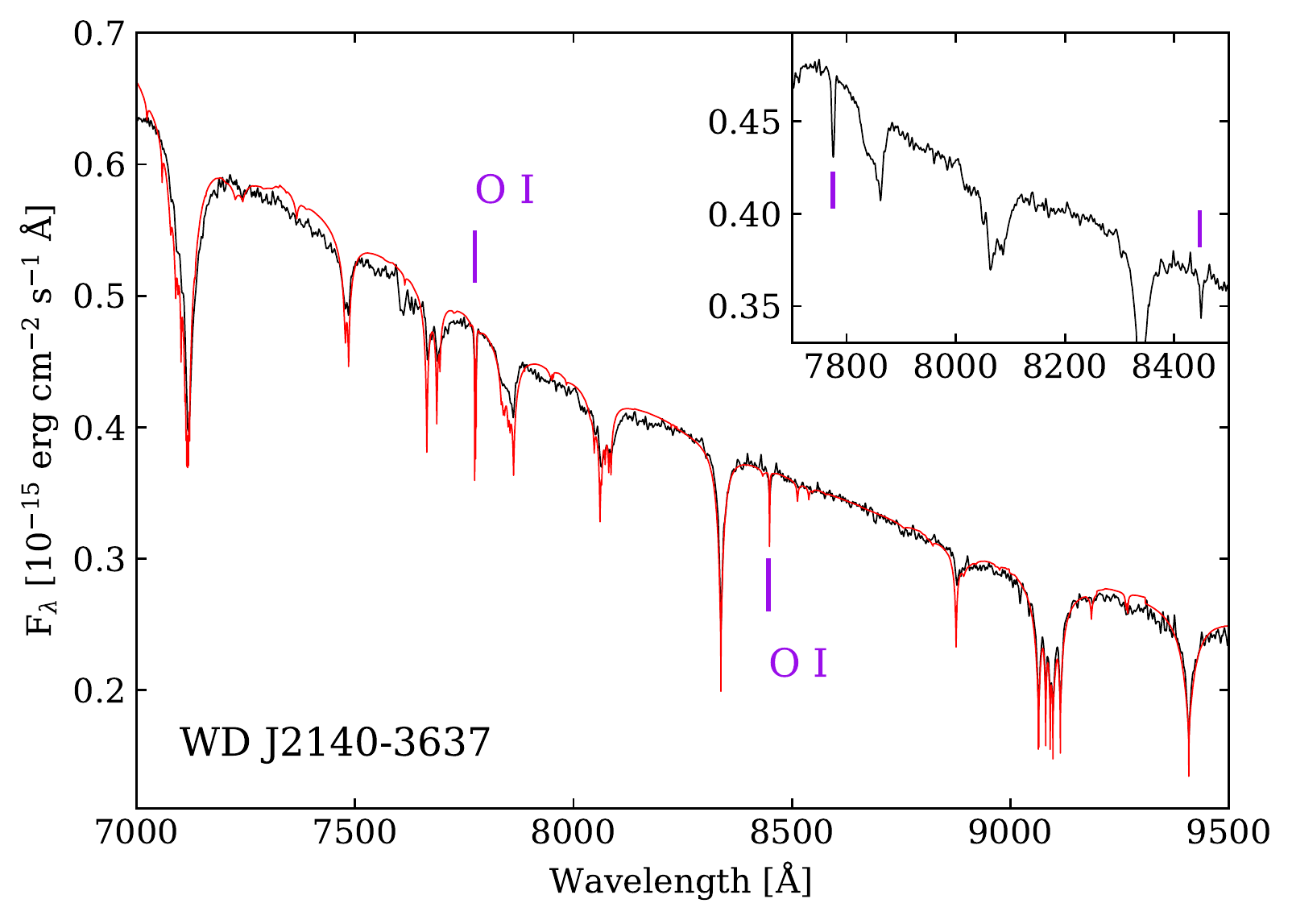}
	\caption{X-Shooter spectrum of \textbf{WD\,J2140$-$3637} plotted with the combined photometric and spectroscopic fit using \citet{Koester2010} models. The O\,\textsc{i} absorption features around 7775~\AA\ and 8446~\AA\ are highlighted with purple ticks. The spectrum is convolved by a Gaussian with a FWHM of 1~\AA\ and shifted by 45~km/s. An inset plot shows the region around the oxygen absorption features.}
    \label{fig:WDJ2140}
\end{figure}

We observe an O\,\textsc{i} triplet absorption feature at 7772, 7774, and 7775~\AA, and an O\,\textsc{i} feature around 8446~\AA, which are labelled in Fig.~\ref{fig:WDJ2140}. As with atmospheric carbon, the presence of oxygen in the atmosphere of \textbf{WD\,J2140$-$3637} is likely due to dredge-up by an extending convection zone in the upper helium layer of a CO-core white dwarf with small total masses of H and He. We have made the first detection of oxygen in the atmosphere of \textbf{WD\,J2140$-$3637}.

We fit this object using the same models as for the other DQ stars in this sample \citep{Koester2010}, and find \Teff\ = 11\,800\,$\pm$\,200\,K and $\log(g)$ = 8.77\,$\pm$\,0.01. Assuming carbon is the dominant atmospheric element, we estimate the following abundances: $\log({\rm H/C})< -$3.50, $\log({\rm He/C}) <$ 1.00, $\log({\rm N/C}) < -$2.50, $\log({\rm O/C}) = -$2.10\,$\pm$\,0.10. The limit for He due to an absence of spectral features means we cannot exclude that He is more abundant than C. Therefore this white dwarf is potentially the first warm non-magnetic DQ which has a carbon-dominated atmosphere. 

Warm DQ white dwarfs may be the cooled down counterparts of hot DQ stars, which are thought to originate from double CO-core white dwarf mergers  \citep{Dunlap2015,Williams2016,Cheng2019,Coutu2019}. The mass of \textbf{WD\,J2140$-$3637} determined from our fitting is 1.06\,$\pm$\,0.01~\,\Msun. 

\subsection{Comparison of DA Spectroscopic and Photometric Parameters} 
\label{sec:comparison}

\begin{figure*}
    \centering
	\includegraphics[width=\textwidth]{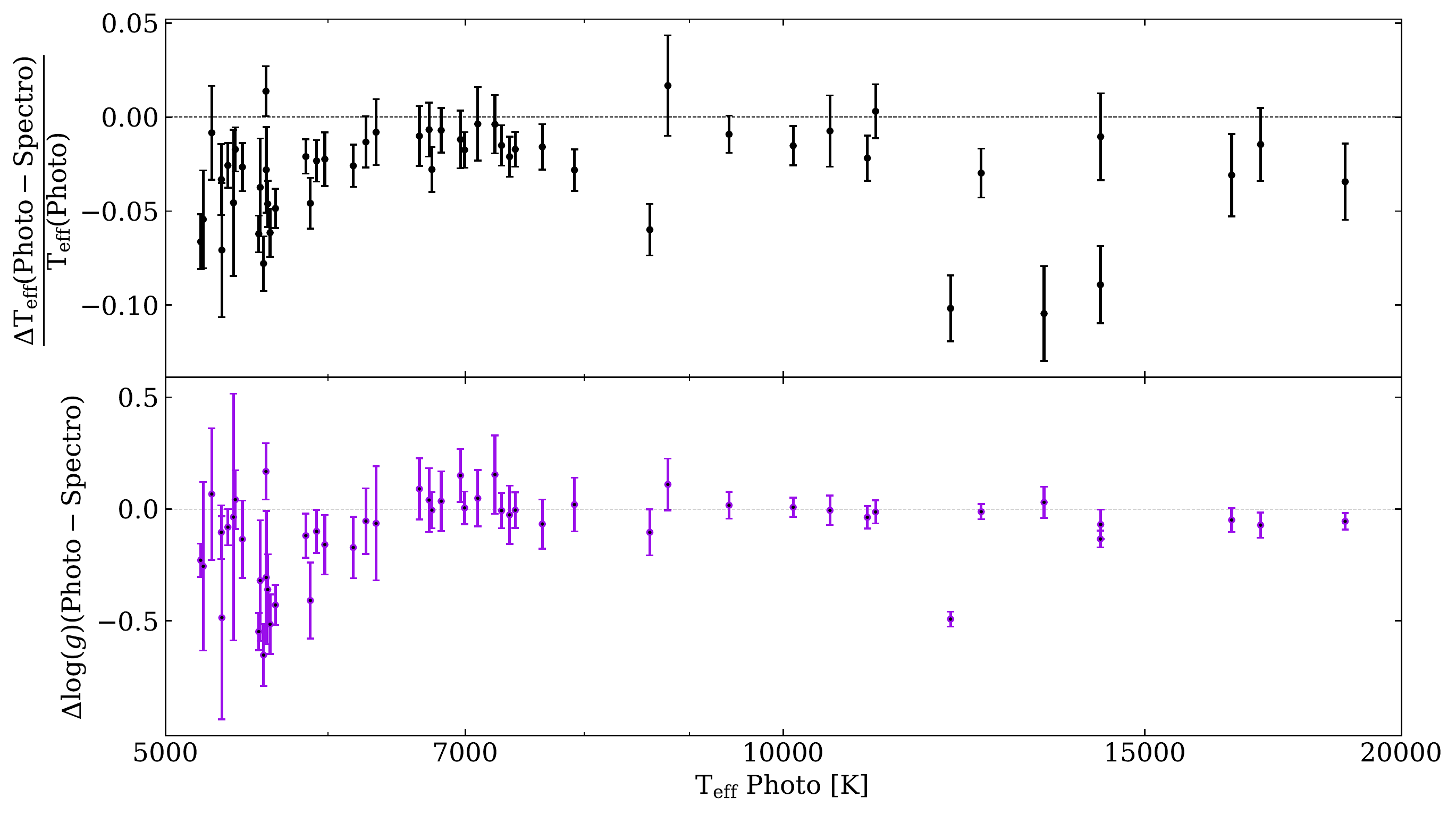}
	\caption{Differences between \textit{Gaia} photometric (\textit{Photo}) and spectroscopic (\textit{Spectro}) \Teff\ (top) and $\log(g)$ (bottom) for DA white dwarfs observed with X-Shooter, against \textit{Gaia} photometric \Teff\ \citep{Gentile2021}. The spectroscopic fitting method is that which was used to fit all DA white dwarfs in this paper (see Section~\ref{sec:spectro_params}).}
    \label{fig:gaia_vs_nicola}
\end{figure*}

For the homogeneous sub-sample of DA white dwarfs with X-Shooter spectroscopy, Fig.~\ref{fig:gaia_vs_nicola} displays the differences in \Teff\ of the spectroscopic fitting method adopted in this paper compared to \textit{Gaia} photometric parameters. There is no clear systematic differences for DA white dwarfs above 8000\,K due to low number statistics. We observe a clear systematic offset between X-Shooter spectroscopic solutions and \textit{Gaia} photometric parameters in the region 6000 $<$ \Teff\ $<$ 8000\,K, where \textit{Gaia} photometric temperatures are systematically lower by 1.5\,$\pm$\,0.8\,per\,cent (see Fig.~\ref{fig:gaia_vs_nicola}). The region \Teff\ $<$ 6000\,K is excluded because there is a known issue with photometric fits for these low-temperature white dwarfs (see Fig.~\ref{fig:teff_logg}).

In Paper I, using a different spectroscopic data set from WHT for a similar sample of cool DA white dwarfs within 40\,pc, a similar offset was found between spectroscopic and photometric temperatures. It was concluded that \textit{Gaia} colours are systematically too red, or the spectroscopic solutions too warm. Radius measurements using \textit{Gaia} photometry and astrometry depend on a comparison between observed and predicted absolute magnitude, the latter itself a function of \Teff. Therefore, an under-prediction of photometric \Teff\ would result in an over-prediction of radius, hence a systematic decrease in $\log(g)$ given the mass-radius relation. As a consequence, any systematic offset in $\log(g)$ values between both techniques is in part a consequence of the offset in \Teff.

In summary, from this work and the recent literature \citep[Paper I]{Genest-Beaulieu2019,Tremblay2019,Cukanovaite2021}, there is a clear offset between photometric and spectroscopic \Teff\ solutions for DA white dwarfs that is present when using different homogeneous spectroscopic data sets (e.g. WHT, X-Shooter, SDSS) and photometric data sets (e.g. \textit{Gaia} DR2 and DR3, Pan-STARRS, SDSS). This offset appears to be of a similar percentage for temperatures between 5500\,K and 30\,000\,K, where the 1.5\,per\,cent value found in this work is very similar to the offset found for warm non-convective (\Teff\ > 15\,000\,K) DA white dwarfs from SDSS in \citet{Tremblay2019}. Finally, a similar offset is seen for DB white dwarfs \citep{Cukanovaite2021}.

\subsection{Binary Systems and Binary Candidates}
\label{sec:binary}

Table~\ref{tab:binaries} lists all new candidate unresolved binary systems in our 40\,pc south sub-sample, where we selected objects with \textit{Gaia} $\log(g) <$ 7.72 when fitted as single stars. A white dwarf with a mass lower than $\approx$ 0.50~\,\Msun\ ($\log(g) \lesssim$ 7.80) could not have formed through single-star evolution within the age of the universe, therefore these low $\log(g)$ solutions indicate binarity. We do not include very cool white dwarfs that are significantly below $T_{\rm eff} =$ 4500\,K in our candidate list, as they have a low-mass problem such that low $\log(g)$ values for some of these stars may not indicate binarity (Paper II). We do not consider the DZ (\textbf{WD\,J0818$-$1512}) and DQ (\textbf{WD\,J1327$-$2817}) stars that have low photometric $\log(g)$ values from their pure-He or mixed H/He atmosphere fits \citep{Gentile2021} to be candidate binary systems, as their combined spectroscopic and photometric fits including metals/carbon in Tables~\ref{table:dzparams} and \ref{table:dqparams} increase their $\log(g)$ values significantly.

In Paper II, a system is also considered a candidate unresolved binary when the difference between the spectroscopic and photometric $\log(g)$ values is greater than 0.5 dex. For three DA white dwarfs with \Teff\ $<$ 6000\,K, the difference between spectroscopic and photometric $\log(g)$ values is greater than 0.5 dex. The photometric $\log(g)$ value for these stars is close to the canonical value of 8.0 in all cases, and the spectroscopic $\log(g)$ values are higher. We do not infer binarity in these systems and suggest instead that spectroscopic fitting of low \Teff\ DA white dwarfs may, in some cases, produce larger $\log(g)$ values than expected. We include some DA white dwarfs in our table that have low photometric $\log(g)$ but larger spectroscopic $\log(g)$, as these are still candidate binary systems independent of their spectroscopic best-fit parameters.

\textbf{WD\,J1604$-$7203} is a cool (\Teff\ $\approx$ 4000\,K) DC white dwarf that has the lowest photometric $\log(g)$ in the entire 40\,pc sample, of 6.75\,$\pm$\,0.04 dex. Despite having a photometric $T_{\rm eff} <$ 4500\,K, we include it in our binary candidate list (Table~\ref{tab:binaries}) due to its remarkably low photometric $\log(g)$. Even allowing for binary evolution and mass loss resulting in a low-mass white dwarf component, current He-core white dwarf evolution models \citep{Istrate2016} would not allow a low-mass white dwarf to cool down to such low surface temperature within the age of the universe. The best explanation for such a low photometric $\log(g)$ is that this is likely a multiple-degenerate system (double or triple), with its exact nature difficult to constrain given the known systematic photometric under-estimate of mass in very cool white dwarfs (Paper II), and the lack of spectral lines.

\textit{Gaia} DR3 provides the renormalised unit weight error (RUWE) parameter, which should be around 1.0 for single stars \citep{Belokurov2020}. If the RUWE is significantly greater than 1.0, this indicates a poor astrometric solution, possibly due to contamination that might have also affected the photometry. \textbf{WD\,J1318+7353} and \textbf{WD\,J2126$-$4224} have RUWE values of 3.5 and 9.1 respectively, indicating that they may be binary systems or otherwise variable.

Table~\ref{tab:binaries_know} lists all other white dwarfs we observe that are part of a binary system, and was built based on mixed spectral types and common proper-motion pairs. All common proper-motion companions with no confirmed spectral types lie on the main-sequence of the \textit{Gaia} HR diagram. The companions of \textbf{WD\,J1406$-$6957} and \textbf{WD\,J1945$-$4904} are candidate cool M-dwarfs with indicative spectral type M7 \citep{Reyle2018}. The small number of unresolved WD+MS binaries in 40\,pc are missing from \citet{Gentile2021}.

\citet{Zuckerman2014} investigated metal-polluted WD+MS star binary systems in order to elucidate the frequency of wide-orbit planets as a function of the semi-major axis of a binary. They found that over a certain range of semi-major axes, the presence of a secondary star suppressed the formation and/or long-term stability of an extended planetary system around the primary.   Specifically, for binary star sky plane separations between about 120 and 2500\,AU, white dwarfs are significantly less likely to be polluted with heavy elements than single white dwarfs or binaries with sky plane separations $>$2500\,AU.  

White dwarfs in Table~\ref{tab:binaries_know} are consistent with this pattern. Eighteen Table~\ref{tab:binaries_know} white dwarfs are not a DQ, or in a double degenerate, or have sky plane separations less than 120\,AU.  Of these 18, 13 have semi-major axes between 120 and 2500\,AU; only one is metal polluted.  For sky plane separations $>$2500\,AU, one in five of the white dwarfs are polluted.

One can combine the results from the \citet{Zuckerman2014} and the present paper. In an annulus between about 190 and 2800\,AU (a ratio of semi-major axes $\approx$15), there are 28 non-polluted and no polluted white dwarfs, whereas, based on statistics from the 40\,pc southern sub-sample presented in this work, 4 should be polluted.

\begin{table}
	\centering
        \caption{New unresolved double degenerate binary candidates in our 40\,pc subsample (this work).}
        \label{tab:binaries}
        \begin{tabular}{llll}
                \hline
                WD\,J name & SpT & \textit{Gaia} \Teff & \textit{Gaia} $\log(g)$ \\
                \hline
                0551$-$2609 & DC & 4750 (40) & 7.30 (0.03) \\
                1117$-$4411 & DC & 5590 (30) & 7.53 (0.02) \\
                1318+7353 & DC & 5000 (40) &	7.35 (0.04)  \\
                1447$-$6940 & DC & 4470 (30) & 	7.24 (0.02)  \\
                1503$-$2441 & DA & 5670 (30) & 	7.60 (0.02) \\
                1601$-$3832 & DA & 4910 (40) & 	7.69 (0.03) \\
                1604$-$7203 & DC & 4090 (40) & 6.75 (0.04)  \\
                1815+5532 & DC & 4630 (50) & 7.19 (0.04)  \\
                1821$-$5951 & DA & 4750 (30) & 7.27 (0.03)\\
                1833$-$6942 & DA & 8010 (60) & 7.39 (0.02) \\
                1919+4527 & DC & 4780 (20) & 7.31 (0.02)  \\
                2126$-$4224 & DC & 5480 (30) & 	7.52 (0.03)  \\
                \hline
        \end{tabular}
        \\
\end{table}

\begin{table}
	\centering
	\scriptsize
        \caption{Binary systems in our 40\,pc subsample (this work).}
        \label{tab:binaries_know}
        \begin{tabular}{llll}
                \hline
                \textit{Gaia} DR3 ID & WD\,J name & SpT & Sep \\
                 & (where  &  & (arcsec) \\
                  & applicable) & & \\
                \hline
                \hline
                2377344185944929152 & 0044$-$1148 & DZ & 4.3\\
                2377344185944929280 & & & \\
                \hline
                2486388560866377856 & 0212$-$0804 & DA & 3.7 \\
                2486388560866377728 & & dM (a) & \\
                \hline
                4672306015773211008 & 0312$-$6444 & DA+DA (b) & -- \\
                \hline
                4613612951211823616 & 0317$-$8532A & DA (c) & 6.9 \\
                4613612951211823104 & 0317$-$8532B & DAH (d) & \\
                \hline
                4678664766393827328 & 0416$-$5917 & DA (e) & 13.1 \\
                4678664766393829504 & & dK (f) & \\
                \hline
                2925551818747071488 & 0646$-$2246 & DC & 5.2 \\
                2925551853106808832 & & & \\
                \hline
                5624029566946316928 & 0907$-$3609 & DA & 10.8 \\
                5624029566946047616 & & & \\
                \hline
                5436014972680358272 & 0936$-$3721 & DA (g) & 4.2 \\
                5436014972680358784 & 0936$-$3721 & DQ (h) & \\
                \hline
                6133033635916500608 & 1234$-$4440 & DC & 38.1 \\
                6133033601555979648 & & G (f) & \\
                \hline
                6188345358621778816 & 1327$-$2817 & DQ & 5.2 \\
                6188345358621678592 & & dK (i) & \\
                \hline
                5845312191917620224 & 1333$-$6751 & DZ & 283 \\
                5845300239052540416 & & & \\
                \hline
                5846206030463663232 & 1406$-$6957 & DA & 25.2 \\
                5846206202262355712 & & & \\
                \hline
                6272326022391660928 & 1430$-$2403 & DA & 36.6 \\
                6272325816233230848 & & & \\
                \hline
                6271903947364173056 & 1430$-$2520 & DA & 8.5 \\
                6271903943069412608 & & & \\
                \hline
                4053455379420643584 & 1738$-$3427 & DA & 3.5 \\
                4053455379465036800 & & & \\
                \hline
                5909739660590724224 & 1746$-$6251 & DA & 430 \\
                5909762269301963264 & & G (f) & \\
                \hline
                6725656144031366144 & 1809$-$4101 & DC & 214\\
                6725655937872937472 & & & \\
                \hline
                4073522222505044224 & 1857$-$2650 & DA & 70.2 \\
                4073522012035886848 & & & \\
                \hline
                6671045050707117568 & 1945$-$4904 & DC & 49.5 \\
                6671044947630014464 & & & \\
                \hline
                6665685378201412992 & 1956$-$5258 & DA & 4.7 \\
                6665685343840128384 & & dM (j) & \\
                \hline
                6470278694244646912 & 2049$-$5446 & DA & 23.3 \\
                6470278694244647168 & & dK (k) & \\
                \hline
                6578917727331681536 & 2126$-$4224 & DC & 208 \\
                6578729710843028608 & & dM (j) & \\
                \hline
                6485572518732377856 & 2343$-$6447 & DC & 41.4 \\
                6485572557387287680 & & dK (f) & \\
                \hline
        \end{tabular}
        \\
        Note: References here are different to Table~3. (a) \citet{Gaidos2014}, (b) \citet{Kulebi2010}, (c) \citet{Kilic2020}, (d) \citet{Barstow1995}, (e) \citet{Bedard2017}, (f) \citet{Gray2006}, (g) \citet{Gianninas2011}, (h) \citet{Dufour2005}, (i) \citet{Bidelman1985}, (j) \citet{Smethells1974}, (k) \citet{Houk1978}. WD\,J031225.70$-$644410.89 is an unresolved single \textit{Gaia} source.
\end{table}

\section{Conclusions}
\label{conclusions}

The volume-limited 20\,pc sample has been, up until \textit{Gaia} DR2, the largest volume-limited sample of white dwarfs \citep{Hollands2018_Gaia}. In Paper I and Paper II, a sample of northern hemisphere white dwarfs within 40\,pc was presented, with a high level of spectroscopic completeness. In this work, we have described the spectral types of 246 white dwarfs within 1$\sigma_\varpi$ of 40\,pc, of which 209 were previously unobserved and five have updated spectral types from higher quality spectroscopic observations. We have identified many new magnetic white dwarfs, some of which display complex Zeeman splitting, and have estimated their field strengths. We have observed metal-polluted white dwarfs, including \textbf{WD\,J2236$-$5548} and \textbf{WD\,J0808$-$5300} which are polluted by five and six metals, respectively. We have re-observed the warm DQ white dwarf \textbf{WD\,J2140$-$3637} and detected oxygen in its atmosphere for the first time. We report three new white dwarfs which are metal-polluted and display carbon absorption lines (DQZ and DZQ spectral types). We have also presented new candidate unresolved binary systems from their photometric over-luminosity.

We have fitted DA white dwarfs spectroscopically as well as photometrically. We noted that there is a similar offset in \Teff\ for spectroscopic parameters using both southern X-Shooter (this work) and northern WHT (Paper I) data sets, when compared to \textit{Gaia} photometric fitting. 

The volume-limited 40\,pc sample of \textit{Gaia} white dwarfs now has a very high level of spectroscopic completeness and we have used this sample to perform a statistical analysis of the local population of white dwarfs \citep{Cukanovaite2022}. We have confirmed the classification of 1058 white dwarfs out of 1083 candidates from DR3. The 40\,pc sample provides an eight-fold increase in volume over the previous 20\,pc sample \citep{Hollands2018_Gaia}, which did not have the level of spectroscopic completeness that the 40\,pc sample now has. The completeness of the \textit{Gaia} DR3 white dwarf catalogue as well as the selection of \citet{Gentile2021} are expected to be very high for single white dwarfs. 

Creating significantly larger volume-limited samples than 40\,pc requires MOS surveys such as WEAVE, 4MOST and DESI \citep{4MOST,WEAVE,DESI2022}, which may take decades to cover the whole sky. Therefore, the 40\,pc sample will be the benchmark volume-limited white dwarf sample for many years to come. A full statistical analysis of the 40\,pc sample is being prepared and will be presented in a future paper (Paper IV).

\section*{Acknowledgements}
This work is based on observations collected at the European Southern Observatory under ESO programmes 0102.C-0351, 1103.D-0763, and 105.20ET.001. Based on observations made with the Gran Telescopio Canarias (GTC; programme GTC103-21A), installed in the Spanish Observatorio del Roque de los Muchachos of the Instituto de Astrof\'isica de Canarias, in the island of La Palma. Based on observations obtained at the Southern Astrophysical Research (SOAR) telescope, which is a joint project of the Minist\'{e}rio da Ci\^{e}ncia, Tecnologia e Inova\c{c}\~{o}es (MCTI/LNA) do Brasil, the US National Science Foundation’s NOIRLab, the University of North Carolina at Chapel Hill (UNC), and Michigan State University (MSU). This work has made use of data from the European Space Agency (ESA) mission \textit{Gaia} (\url{https://www.cosmos.esa.int/gaia}), processed by the \textit{Gaia} Data Processing and Analysis Consortium (DPAC, \url{https://www.cosmos.esa.int/web/gaia/dpac/consortium}). Funding for the DPAC has been provided by national institutions, in particular the institutions participating in the \textit{Gaia} Multilateral Agreement. Research at Lick Observatory is partially supported by a generous gift from Google. 

This project has received funding from the European Research Council (ERC) under the European Union’s Horizon 2020 research and innovation programme (Grant agreement No. 101020057). PET, BTG, IP and TRM were supported by grant ST/T000406/1 from the Science and Technology Facilities Council (STFC). MAH was supported by grant ST/V000853/1 from the STFC. The authors acknowledge financial support from Imperial College London through an Imperial College Research Fellowship grant awarded to CJM. SGP acknowledges the support of a STFC Ernest Rutherford Fellowship. MRS thanks for support from ANID, – Millennium Science Initiative Program – NCN19\_171 and FONDECYT (grant 1221059). RR has received funding from the postdoctoral fellowship program Beatriu de Pin\'os, funded by the Secretary of Universities and Research (Government of Catalonia) and by the Horizon 2020 programme of research and innovation of the European Union under the Maria Sk\l{}odowska-Curie grant agreement No 801370. ST and ARM acknowledge support from MINECO under the PID2020-117252GB-I00 grant. ARM acknowledges support from Grant RYC-2016-20254 funded by MCIN/AEI/10.13039/501100011033 and by ESF Investing in your future. DDM acknowledges support from the Italian Space Agency (ASI) and National Institute for Astrophysics (INAF) under agreements I/037/12/0 and 2017-14-H.0 and from INAF project funded with Presidential Decree N.43/2018. TC was supported by the Leverhulme Trust Grant (ID RPG-2020-366).

\section*{Data Availability Statement}
The raw X-Shooter data underlying this article are available in the ESO archive, at \url{http://archive.eso.org/cms.html}. Any reduced spectra from any spectrograph used in this article will be shared on reasonable request to the corresponding author.

\bibliographystyle{mnras}
\bibliography{mybib} 

\clearpage
\onecolumn
{\bf \large \noindent Affiliations}\\
    \begin{description}
    \item $^{1}$ Department of Physics, University of Warwick, Coventry, CV4 7AL, UK \\
    \item$^{2}$ European Southern Observatory, Karl Schwarzschild Stra{\ss}e 2, Garching, 85748, Germany \\
    \item$^{3}$ Department of Physics and Astronomy, University of Sheffield, Sheffield, S3 7RH, UK \\
    \item$^{4}$ Institut f\"ur Theoretische Physik und Astrophysik, University of Kiel, 24098 Kiel, Germany \\
    \item$^{5}$ Department of Earth, Planetary, and Space Sciences, University of California, Los Angeles, CA 90095, USA \\
    \item$^{6}$ Department of Physics and Astronomy, University College London, London WC1E 6BT, UK \\
    \item$^{7}$ Department of Astronomy \& Institute for Astrophysical Research, Boston University, 725 Commonwealth Ave., Boston, MA 02215, USA \\
    \item$^{8}$ Lunar and Planetary Laboratory, Sonett Space Science Building, University of Arizona, Tucson, Arizona 85721, USA \\
    \item$^{9}$ Astronomisches Rechen-Institut, Zentrum f\"ur Astronomie der Universit\"at Heidelberg, D-69120 Heidelberg, Germany \\
    \item$^{10}$ Department of Physics and Astronomy, University of California, Los Angeles, CA 90095-1562, USA \\
    \item$^{11}$ Astromanager LLC, Hilo, HI 96720, USA \\
    \item$^{12}$ Astrophysics Group, Department of Physics, Imperial College London, Prince Consort Rd, London, SW7 2AZ, UK \\
    \item$^{13}$ INAF - Capodimonte Astronomical Observatory Naples Via Moiariello 16, I-80131 Naples, Italy \\
    \item$^{14}$ Center for Astrophysics and Space Sciences, University of California, San Diego, CA 92093-0424, USA \\
    \item$^{15}$ Gemini Observatory, Hilo, HI 96720, USA \\
    \item$^{16}$ Departament de F\'{\i}sica, Universitat Polit\`{e}cnica de Catalunya, c/Esteve Terrades 5, 08860 Castelldefels, Spain \\
    \item$^{17}$ Institut d'Estudis Espacials de Catalunya, Ed. Nexus-201, c/Gran Capit\`a 2-4, 08034 Barcelona, Spain \\
    \item$^{18}$ Departamento de F{\'i}sica, Universidad T{\'e}cnica Federico Santa Mar{\'i}a, Avenida Espa{\~n}a 1680, Valpara{\'i}so, Chile \\
    \item$^{19}$ Millennium Nucleus for Planet Formation, NPF, Valpara{\'i}so, Chile \\
    \item$^{20}$ INAF-Osservatorio Astrofisico di Torino, Strada dell’Osservatorio 20, 10025, Pino Torinese, Italy \\
    \item$^{21}$ Institute for Gravitational Wave Astronomy, School of Physics and Astronomy, University of Birmingham, Birmingham, B15 2TT, UK \\
    \item$^{22}$ Earth and Planets Laboratory, Carnegie Institution for Science, 5241 Broad Branch Rd NW, Washington, DC 20015, USA \\
    \end{description}

\appendix

\section{Online Figures}

\begin{figure*}
	\includegraphics[viewport= 1 20 520 720, scale=0.85]{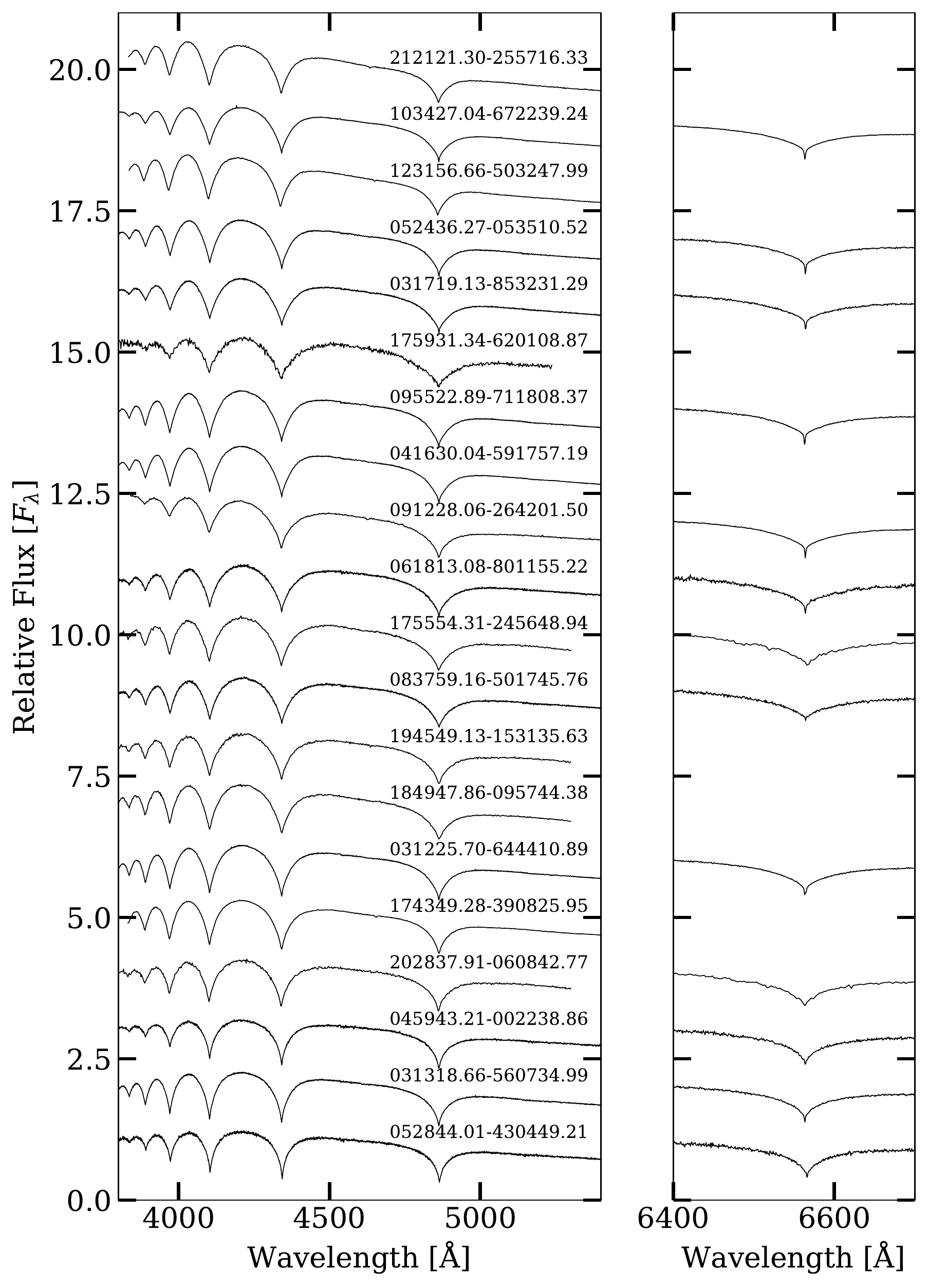}
	\vspace*{10mm}
	\caption{Spectroscopic observations of 100 DA white dwarfs ordered with decreasing photometric temperature (1/5). Temperature range: 19\,500\,K $>$ \Teff\ $>$ 10\,500\,K.}
        \label{fig:DA1}
\end{figure*}

\renewcommand{\thefigure}{A\arabic{figure}}
\setcounter{figure}{0}

 \begin{figure*}
	\includegraphics[viewport= 1 20 520 720, scale=0.85]{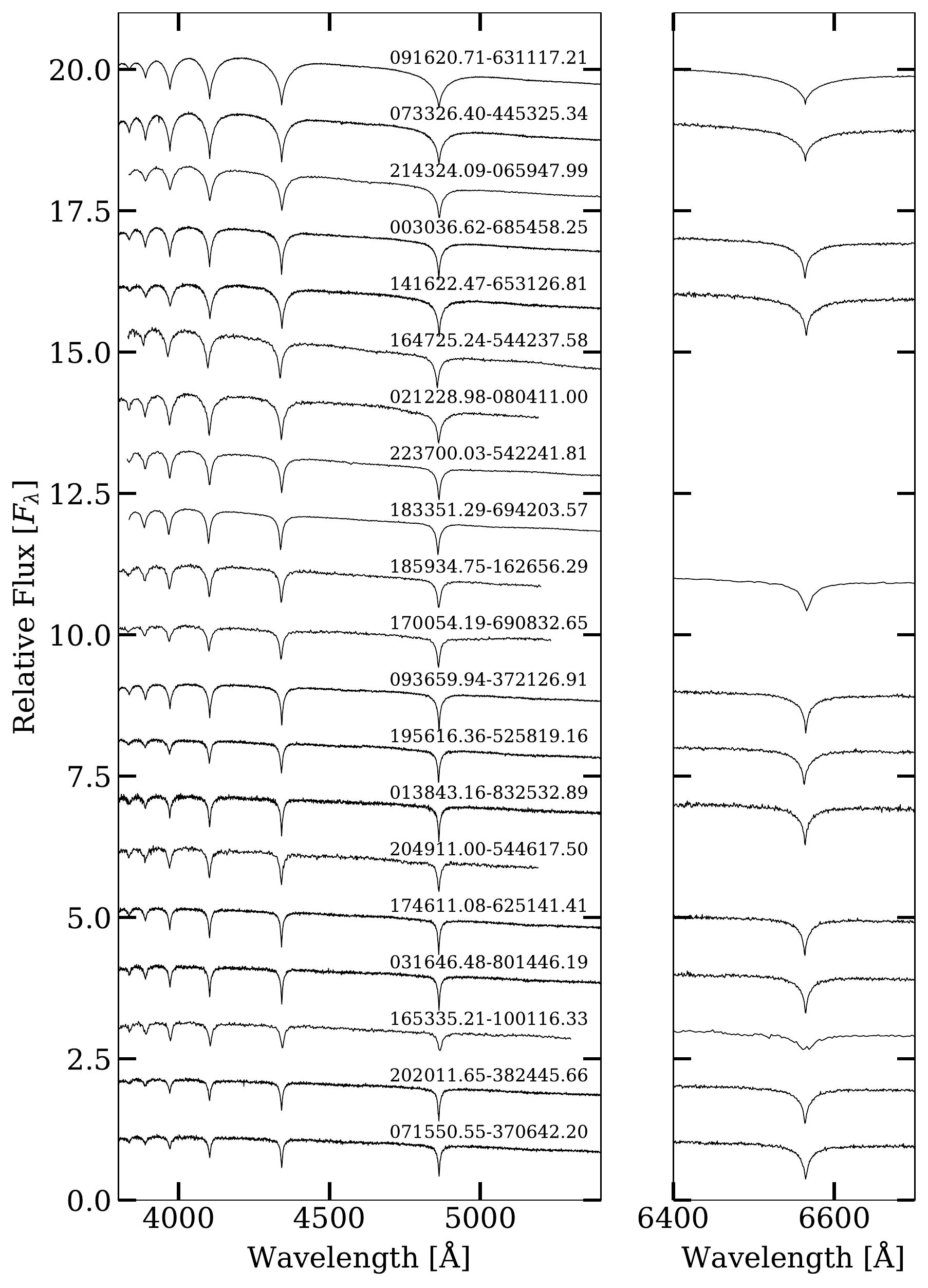}
	\vspace*{10mm}
	\caption{Spectroscopic observations of 100 DA white dwarfs ordered with decreasing photometric temperature (2/5). Temperature range: 10\,500\,K $>$ \Teff\ $>$ 7200\,K.}
        \label{fig:DA2}
\end{figure*}

\renewcommand{\thefigure}{A\arabic{figure}}
\setcounter{figure}{0}

 \begin{figure*}
	\includegraphics[viewport= 1 20 520 720, scale=0.85]{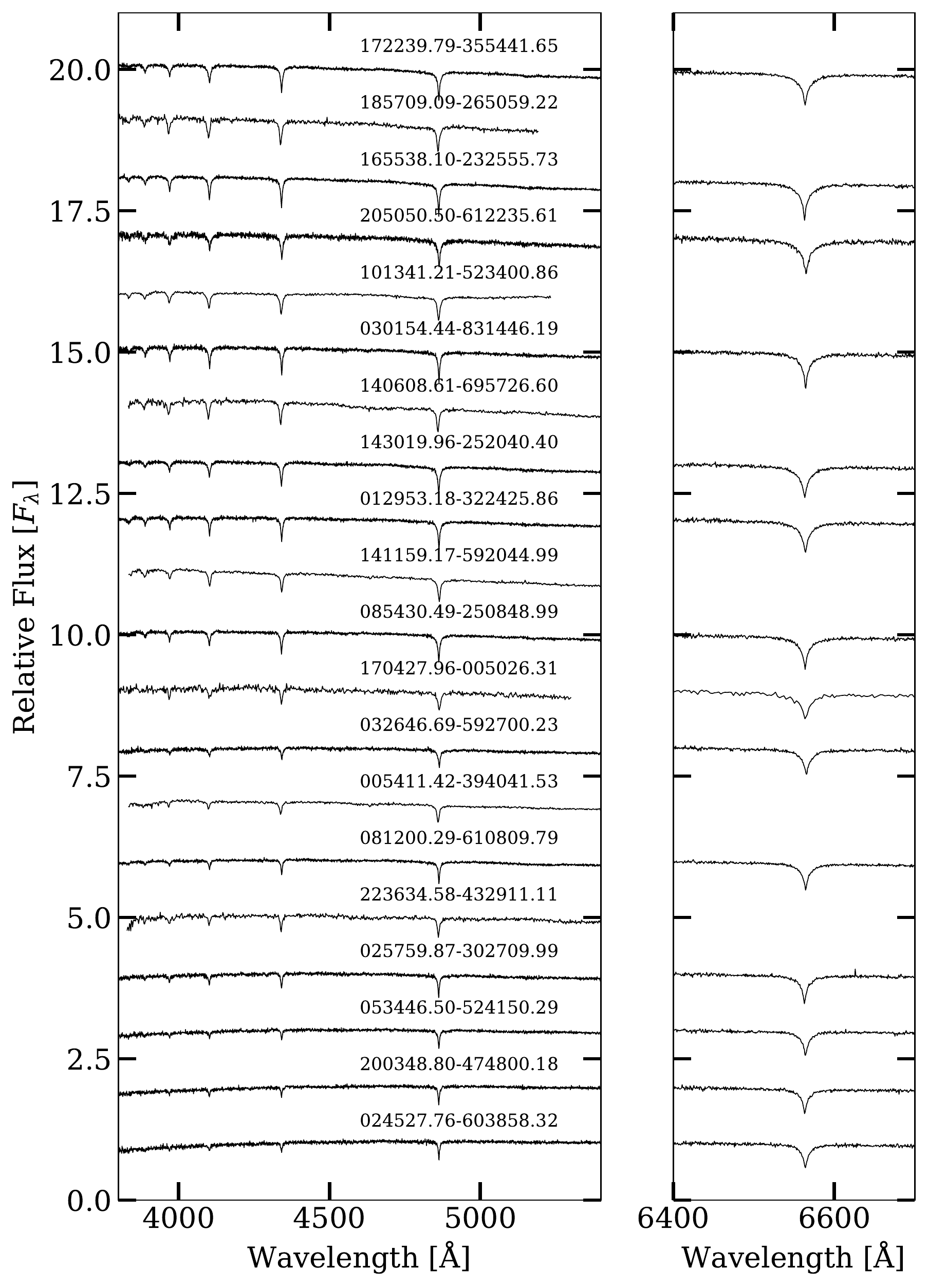}
	\vspace*{10mm}
	\caption{Spectroscopic observations of 100 DA white dwarfs ordered with decreasing photometric temperature (3/5). Temperature range: 7200\,K $>$ \Teff\ $>$ 5900\,K.}
        \label{fig:DA3}
\end{figure*}

\renewcommand{\thefigure}{A\arabic{figure}}
\setcounter{figure}{0}

 \begin{figure*}
	\includegraphics[viewport= 1 20 520 720, scale=0.85]{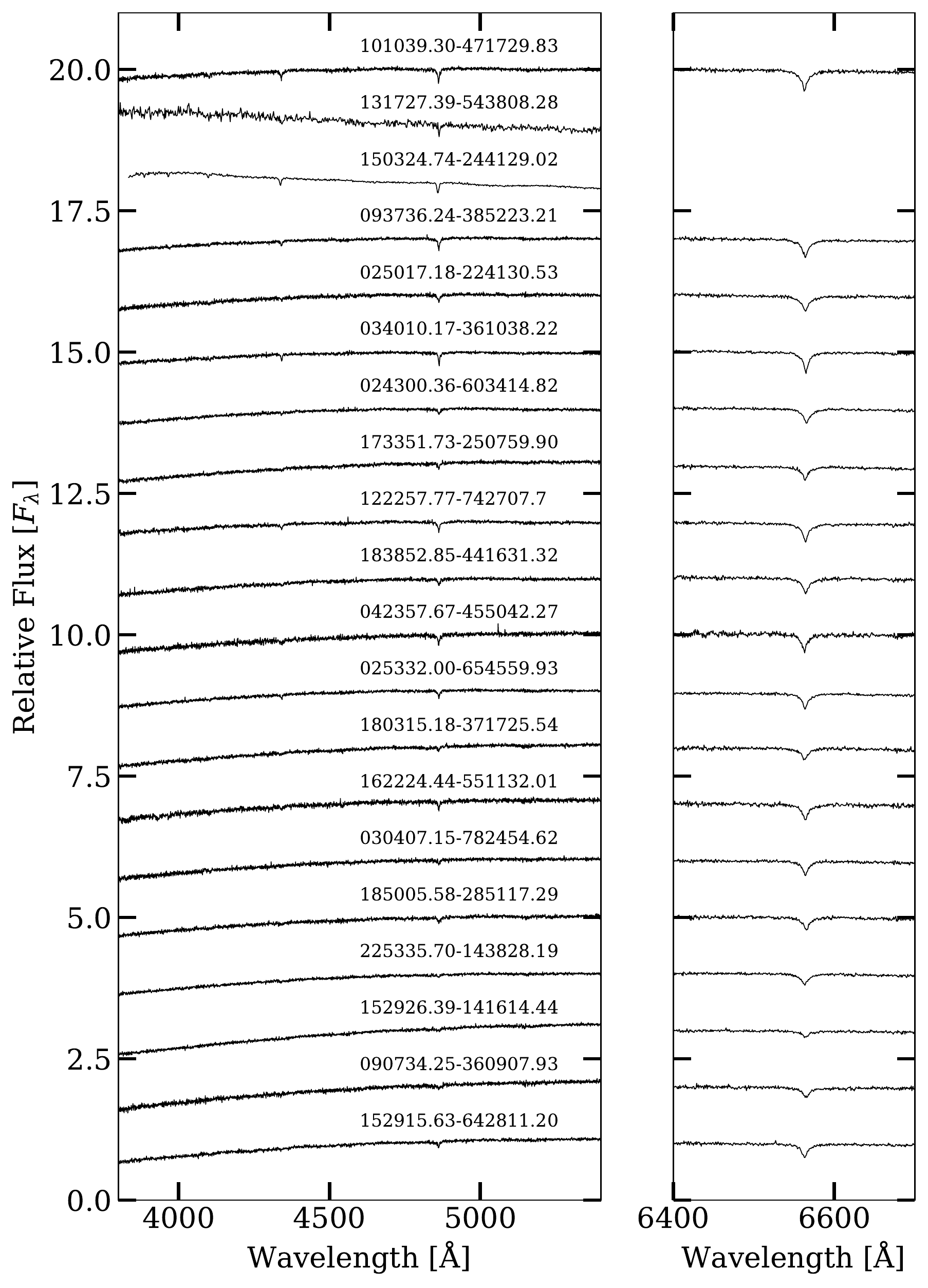}
	\vspace*{10mm}
	\caption{Spectroscopic observations of 100 DA white dwarfs ordered with decreasing photometric temperature (4/5). Temperature range: 5900\,K $>$ \Teff\ $>$ 5200\,K.}
        \label{fig:DA4}
\end{figure*}

\renewcommand{\thefigure}{A\arabic{figure}}
\setcounter{figure}{0}

 \begin{figure*}
	\includegraphics[viewport= 1 20 520 720, scale=0.85]{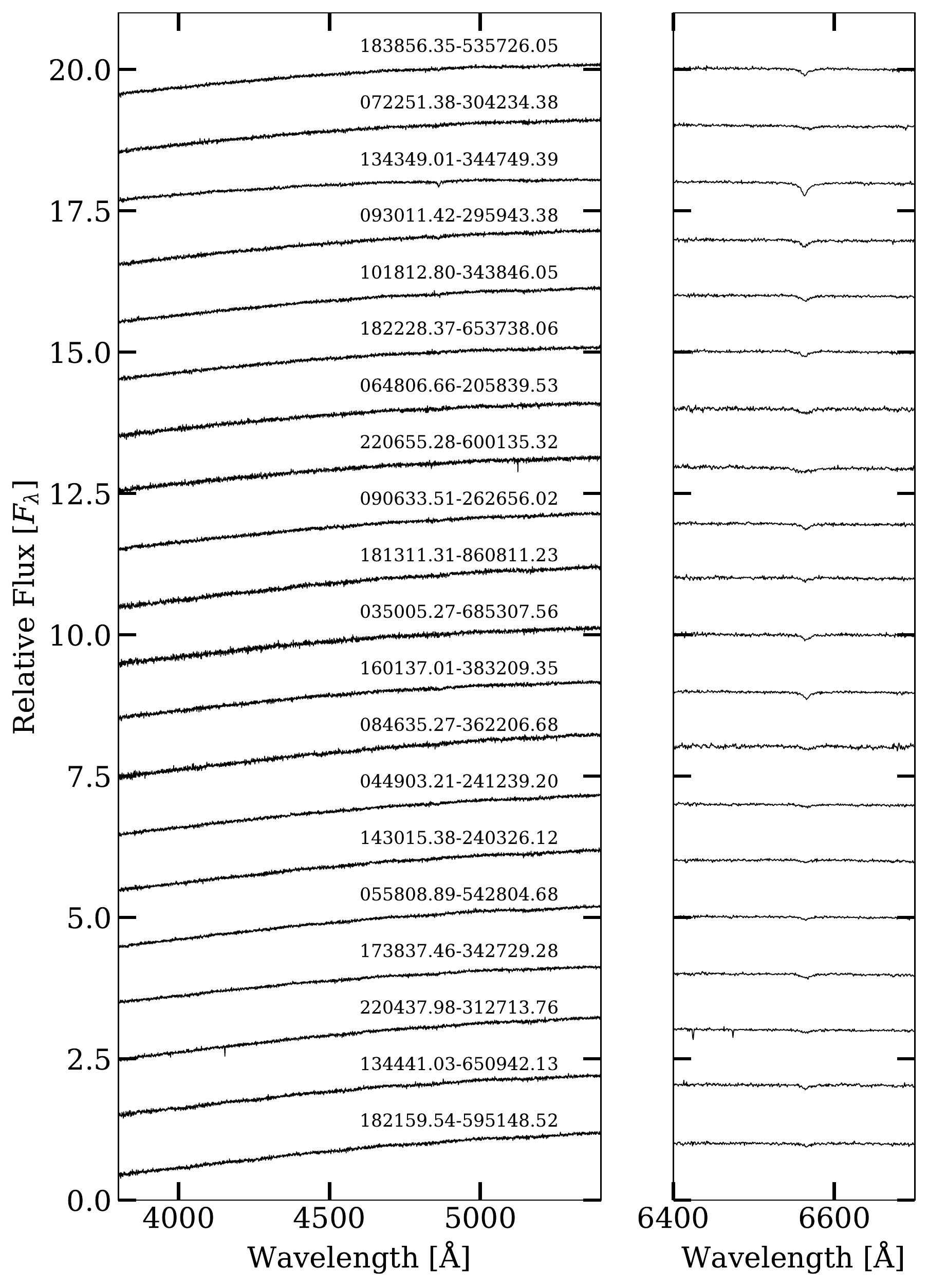}
	\vspace*{10mm}
	\caption{Spectroscopic observations of 100 DA white dwarfs ordered with decreasing photometric temperature (5/5). Temperature range: 5200\,K $>$ \Teff\ $>$ 4800\,K.}
        \label{fig:DA5}
\end{figure*}

 \begin{figure*}
	\includegraphics[viewport= 1 20 520 720, scale=0.85]{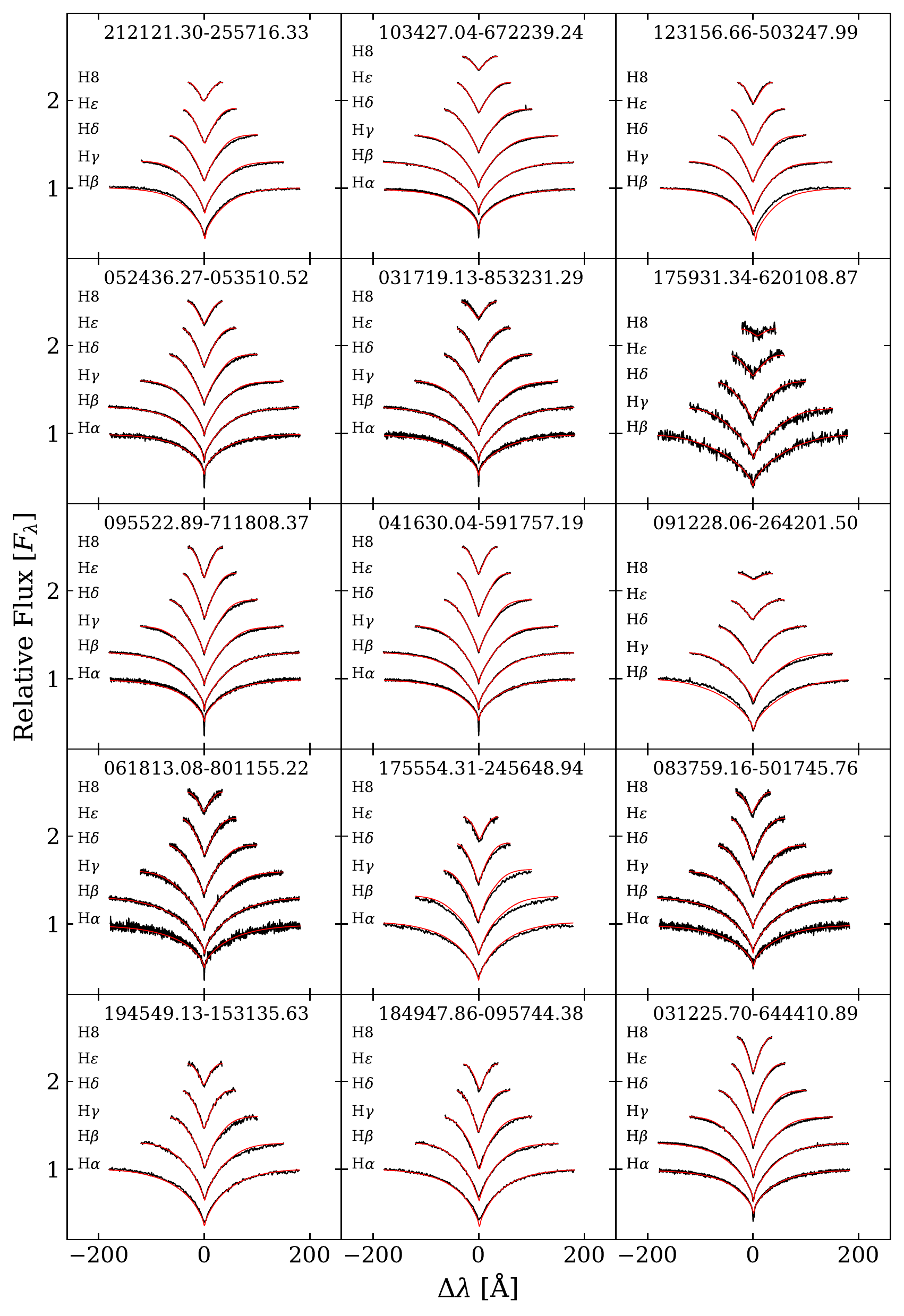}
	\vspace*{10mm}
	\caption{Spectroscopic fits to the normalised Balmer lines for 81 DA white dwarfs ordered with decreasing photometric temperature (1/6). Temperature range: 19\,500\,K $>$ \Teff\ $>$ 12\,000\,K.}
        \label{fig:DA1_fit}
\end{figure*}

\renewcommand{\thefigure}{A\arabic{figure}}
\setcounter{figure}{1}

 \begin{figure*}
	\includegraphics[viewport= 1 20 520 720, scale=0.85]{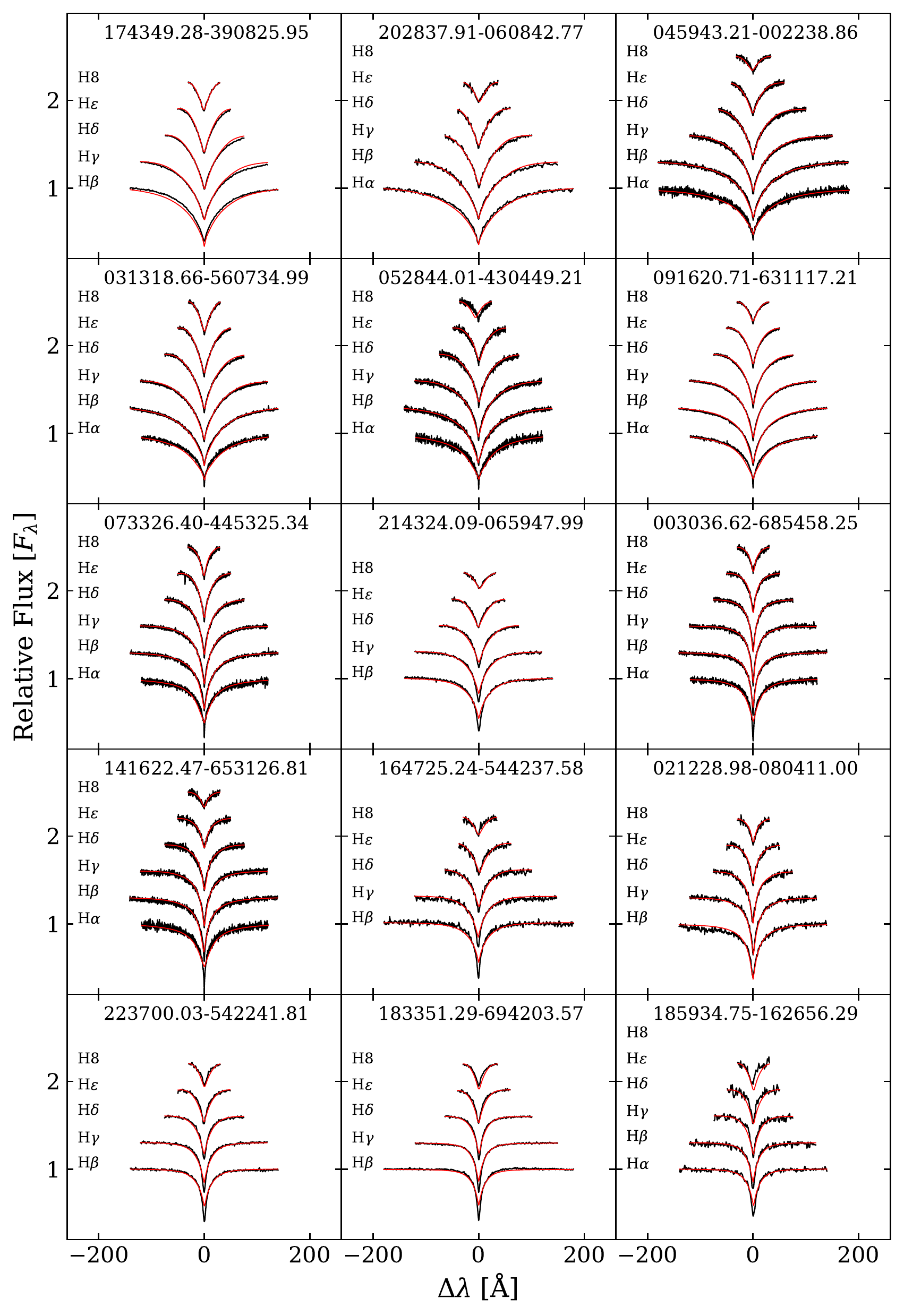}
	\vspace*{10mm}
	\caption{Spectroscopic fits to the normalised Balmer lines for 81 DA white dwarfs ordered with decreasing photometric temperature (2/6). Temperature range: 12\,000\,K $>$ \Teff\ $>$ 8000\,K.}
        \label{fig:DA2_fit}
\end{figure*}

\renewcommand{\thefigure}{A\arabic{figure}}
\setcounter{figure}{1}

 \begin{figure*}
	\includegraphics[viewport= 1 20 520 720, scale=0.85]{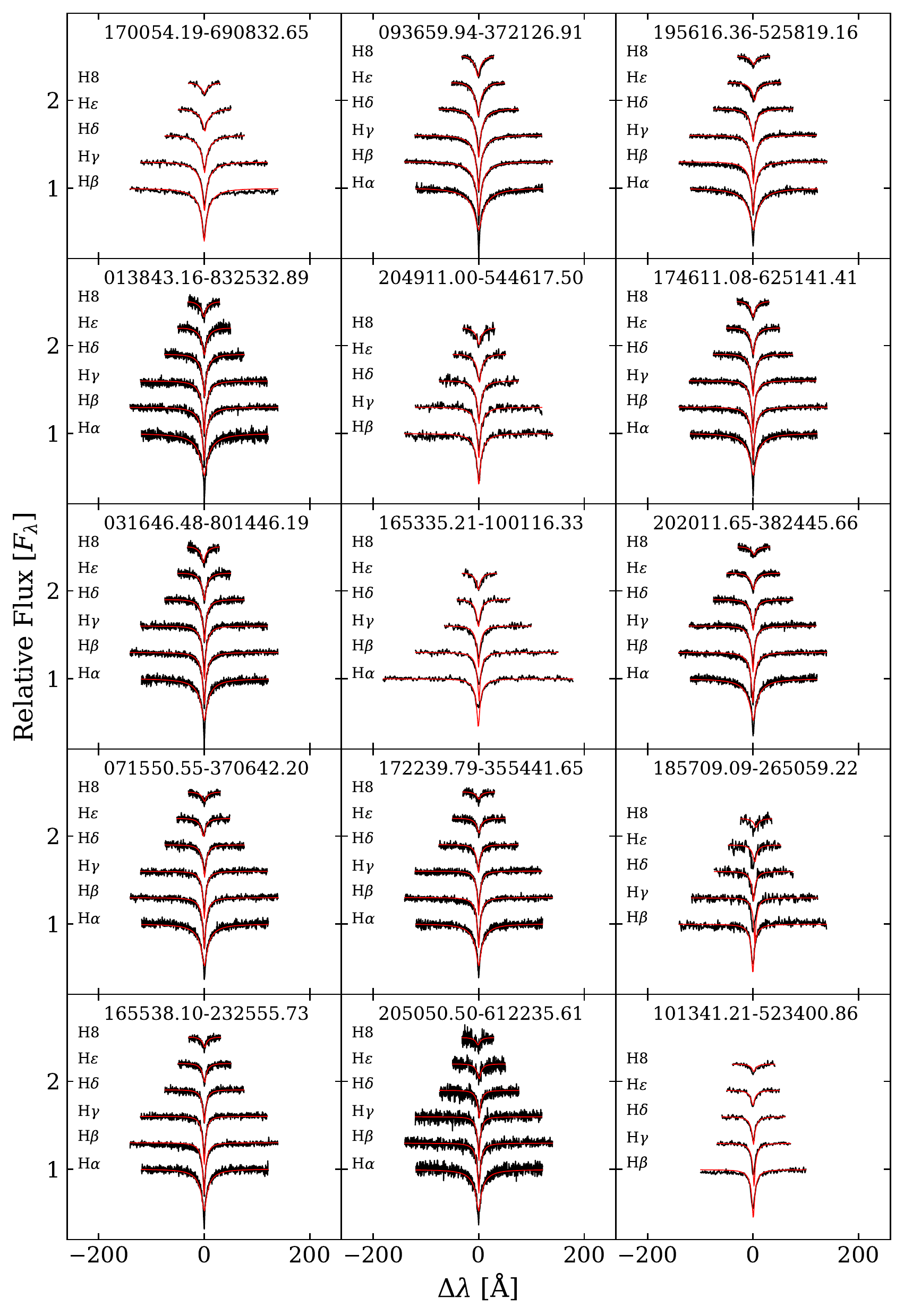}
	\vspace*{10mm}
	\caption{Spectroscopic fits to the normalised Balmer lines for 81 DA white dwarfs ordered with decreasing photometric temperature (3/6). Temperature range: 8000\,K $>$ \Teff\ $>$ 6900\,K.}
        \label{fig:DA3_fit}
\end{figure*}

\renewcommand{\thefigure}{A\arabic{figure}}
\setcounter{figure}{1}

 \begin{figure*}
	\includegraphics[viewport= 1 20 520 720, scale=0.85]{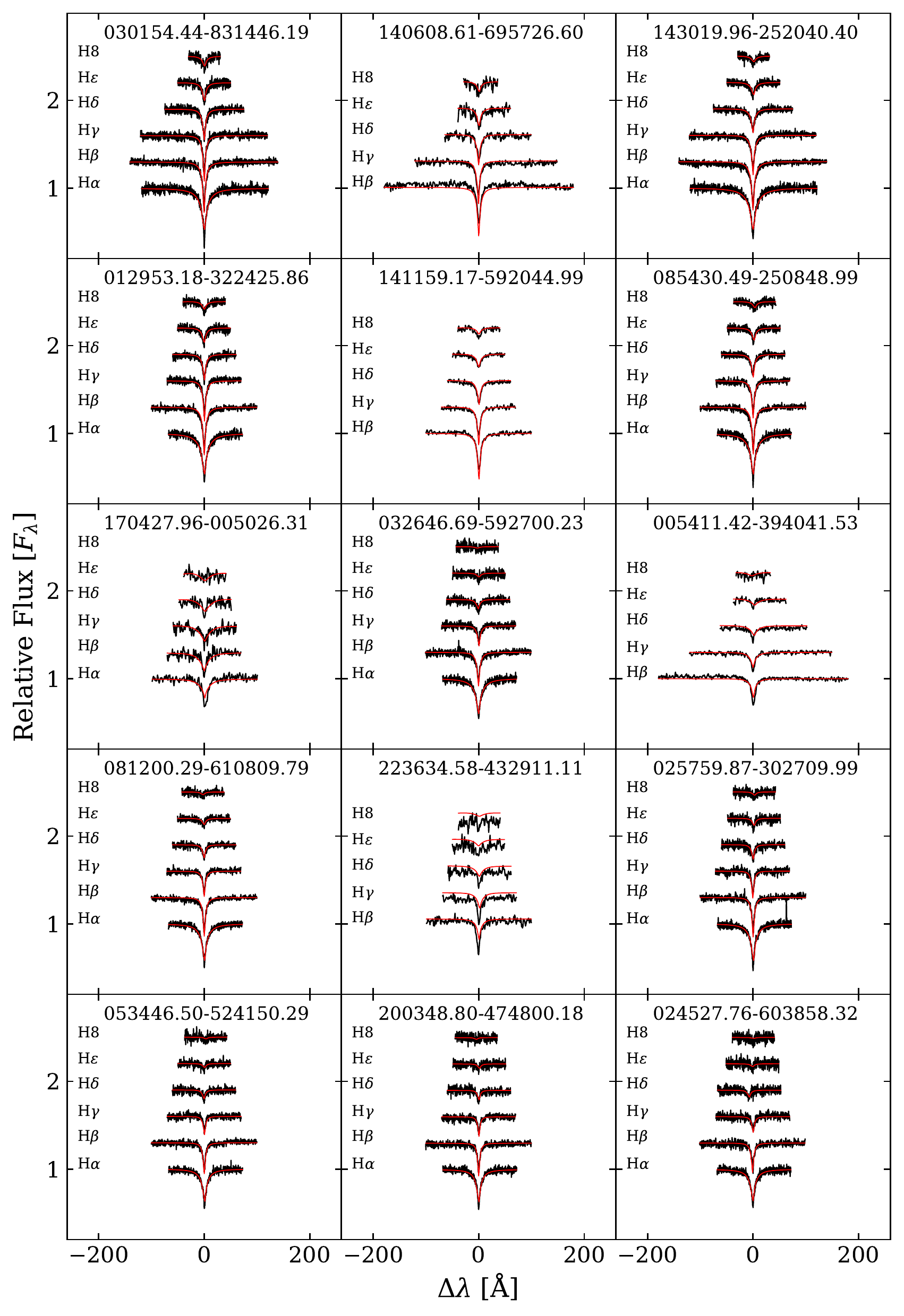}
	\vspace*{10mm}
	\caption{Spectroscopic fits to the normalised Balmer lines for 81 DA white dwarfs ordered with decreasing photometric temperature (4/6). Temperature range: 6900\,K $>$ \Teff\ $>$ 5900\,K.}
        \label{fig:DA4_fit}
\end{figure*}

\renewcommand{\thefigure}{A\arabic{figure}}
\setcounter{figure}{1}

 \begin{figure*}
	\includegraphics[viewport= 1 20 520 720, scale=0.85]{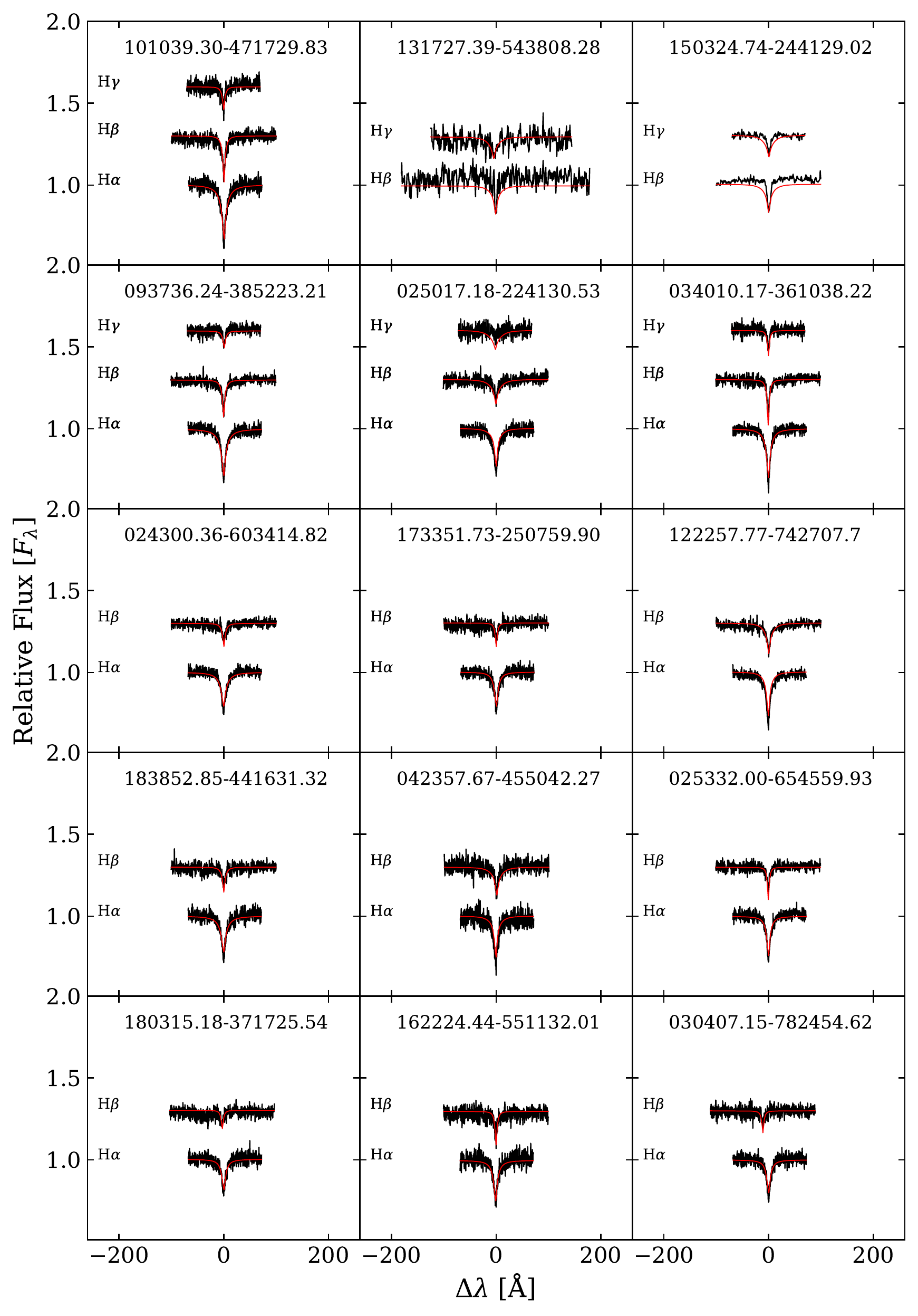}
	\vspace*{10mm}
	\caption{Spectroscopic fits to the normalised Balmer lines for 81 DA white dwarfs ordered with decreasing photometric temperature (5/6). Temperature range: 5900\,K $>$ \Teff\ $>$ 5400\,K.}
        \label{fig:DA5_fit}
\end{figure*}

\renewcommand{\thefigure}{A\arabic{figure}}
\setcounter{figure}{1}

 \begin{figure*}
	\includegraphics[viewport= 1 20 520 320, scale=0.85]{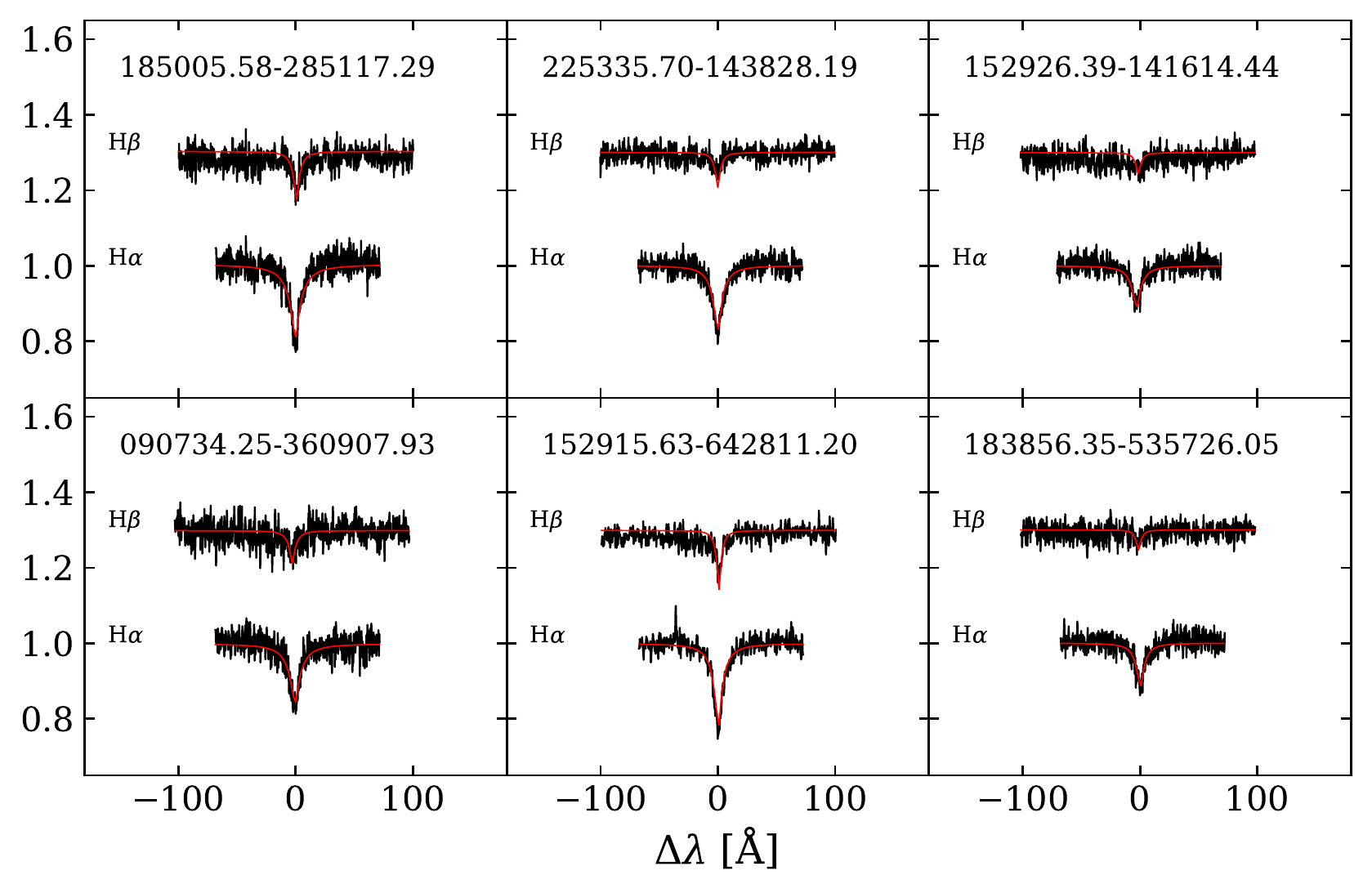}
	\vspace*{10mm}
	\caption{Spectroscopic fits to the normalised Balmer lines for 81 DA white dwarfs ordered with decreasing photometric temperature (6/6). Temperature range: 5400\,K $>$ \Teff\ $>$ 5200\,K.}
        \label{fig:DA6_fit}
\end{figure*}

 \begin{figure*}
	\includegraphics[viewport= 1 20 520 720, scale=0.85]{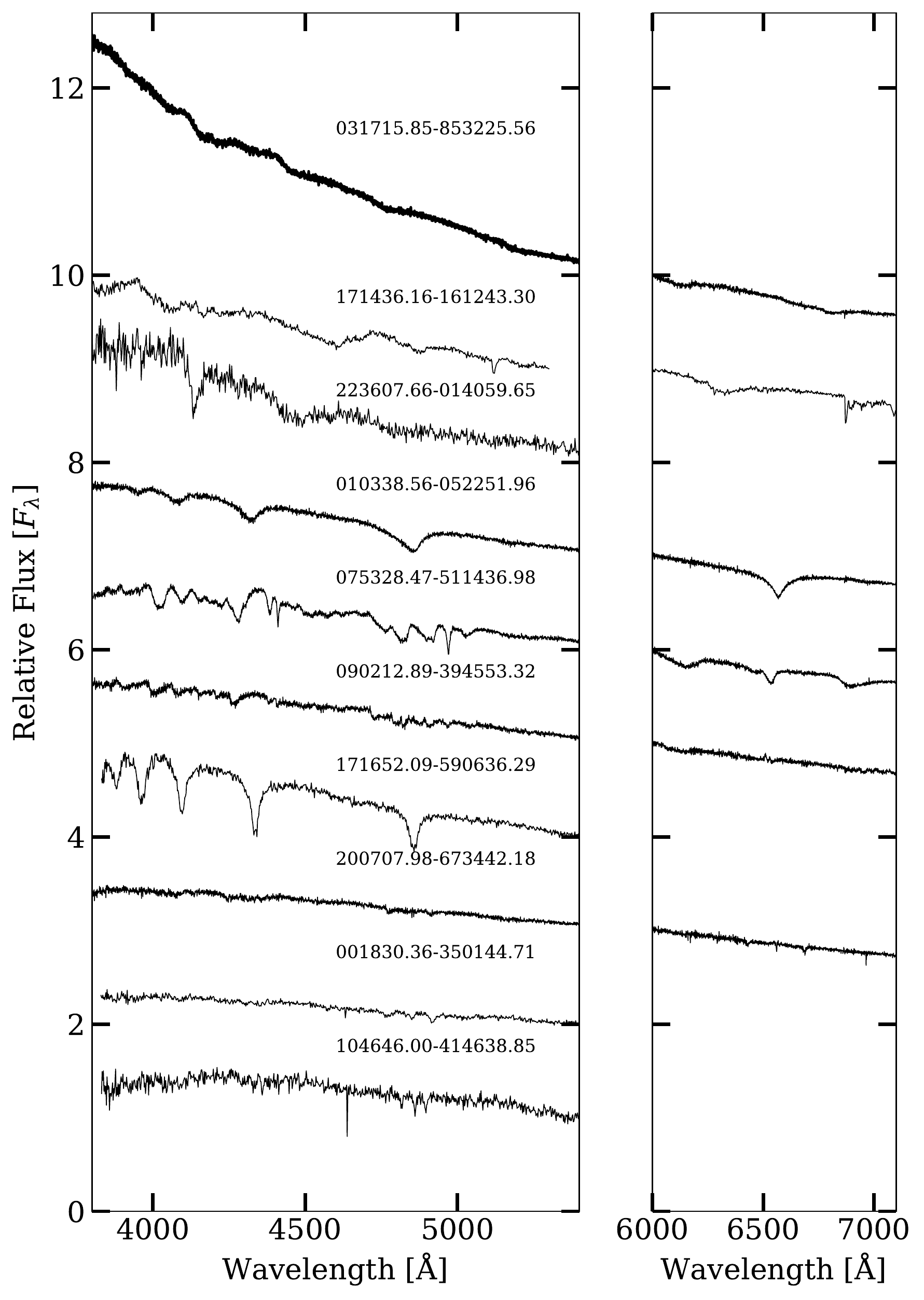}
	\vspace*{10mm}
	\caption{Spectroscopic observations of 28 DAH magnetic white dwarfs ordered with decreasing photometric temperature (1/3). Temperature range: 26\,500\,K $>$ \Teff\ $>$ 6700\,K.}
        \label{fig:DAH1}
\end{figure*}

\renewcommand{\thefigure}{A\arabic{figure}}
\setcounter{figure}{2}

 \begin{figure*}
	\includegraphics[viewport= 1 20 520 720, scale=0.85]{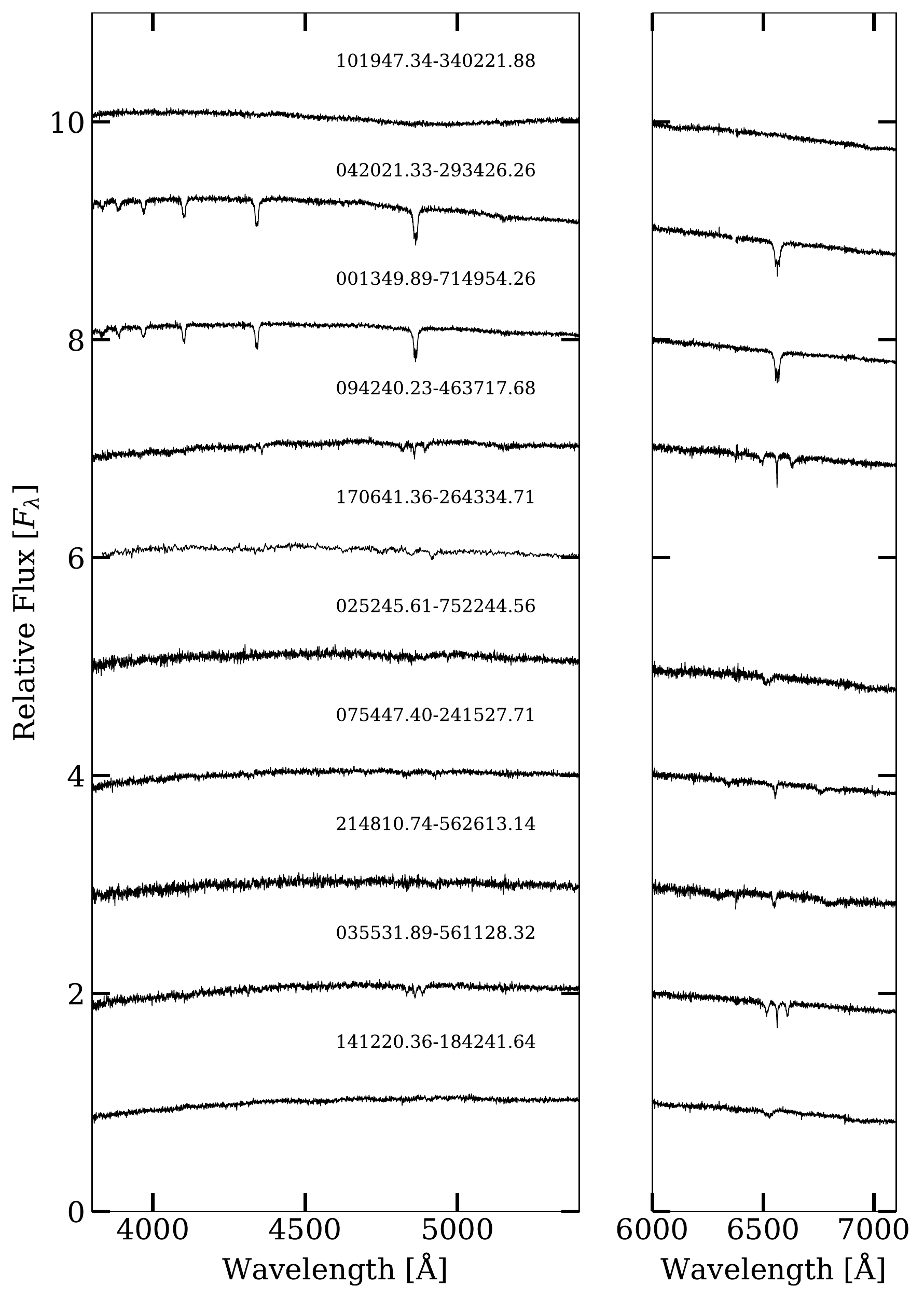}
	\vspace*{10mm}
	\caption{Spectroscopic observations of 28 DAH magnetic white dwarfs ordered with decreasing photometric temperature (2/3). Temperature range: 6700\,K $>$ \Teff\ $>$ 5700\,K.}
        \label{fig:DAH2}
\end{figure*}

\renewcommand{\thefigure}{A\arabic{figure}}
\setcounter{figure}{2}

 \begin{figure*}
	\includegraphics[viewport= 1 20 520 720, scale=0.85]{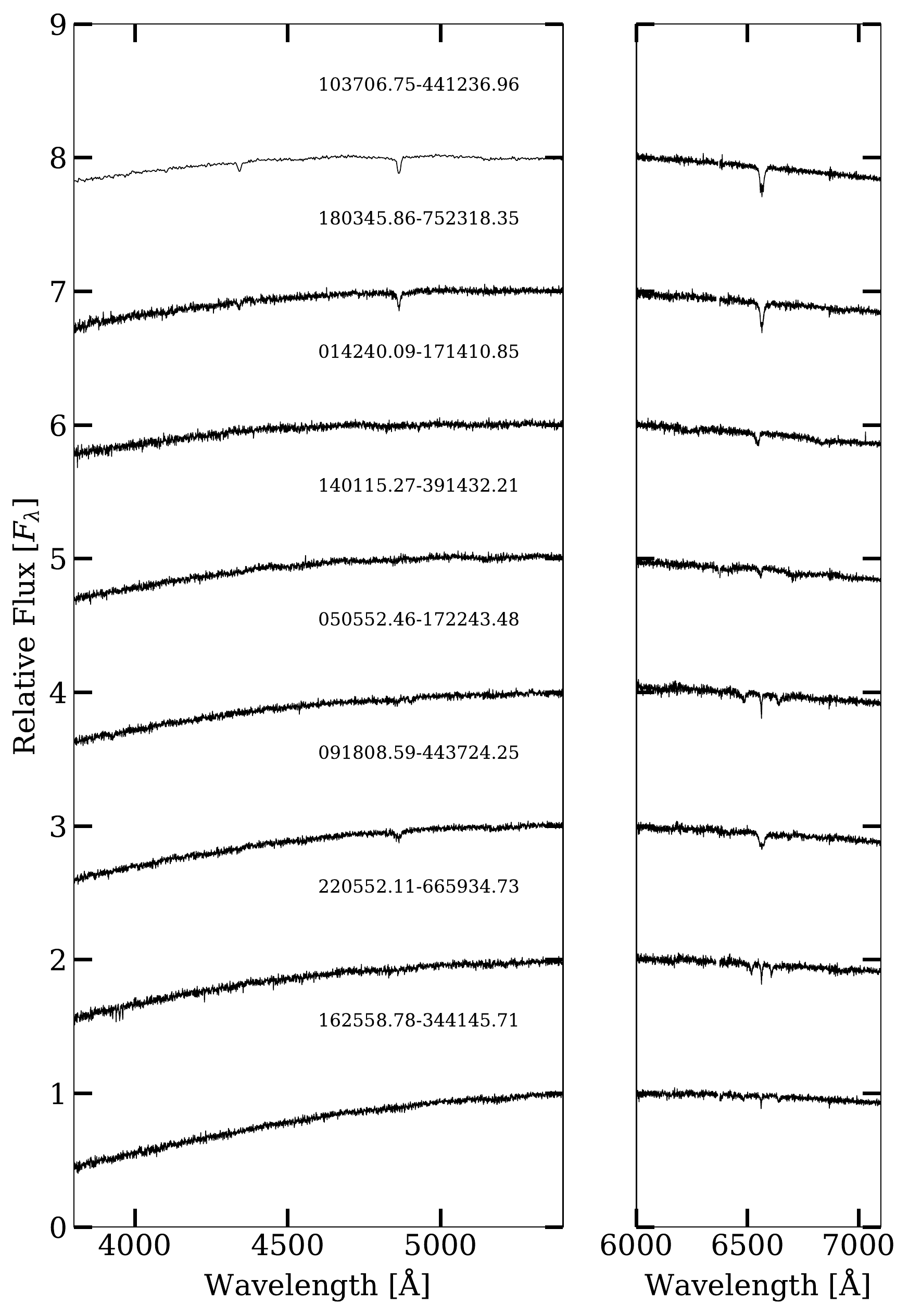}
	\vspace*{10mm}
	\caption{Spectroscopic observations of 28 DAH magnetic white dwarfs ordered with decreasing photometric temperature (3/3). Temperature range: 5700\,K $>$ \Teff\ $>$ 5000\,K.}
        \label{fig:DAH3}
\end{figure*}

 \begin{figure*}
	\includegraphics[viewport= 1 20 520 720, scale=0.85]{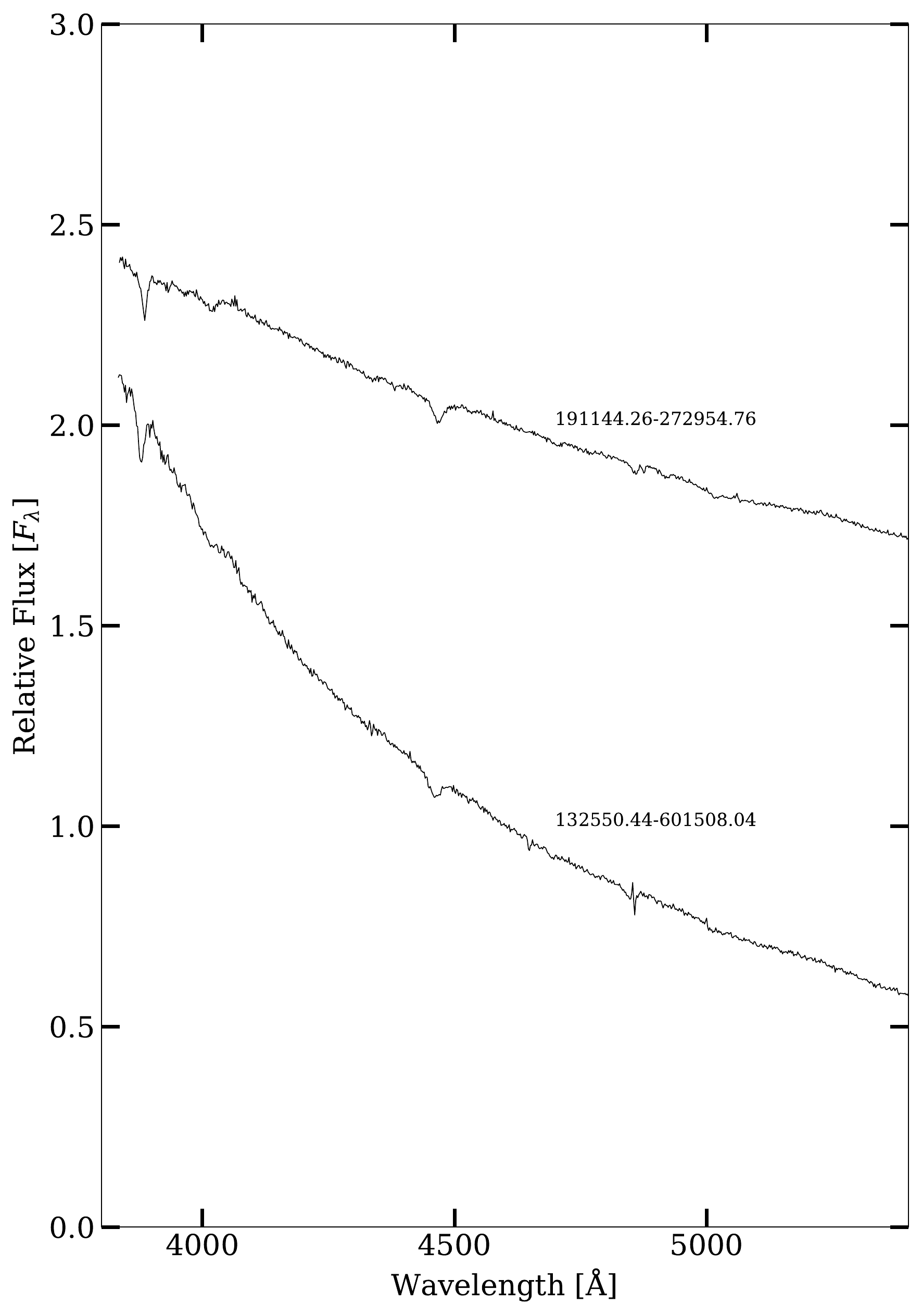}
	\vspace*{10mm}
	\caption{Spectroscopic observations of 2 DB white dwarfs.}
        \label{fig:DB1}
\end{figure*}

\begin{figure*}
	\includegraphics[viewport= 1 20 520 720, scale=0.85]{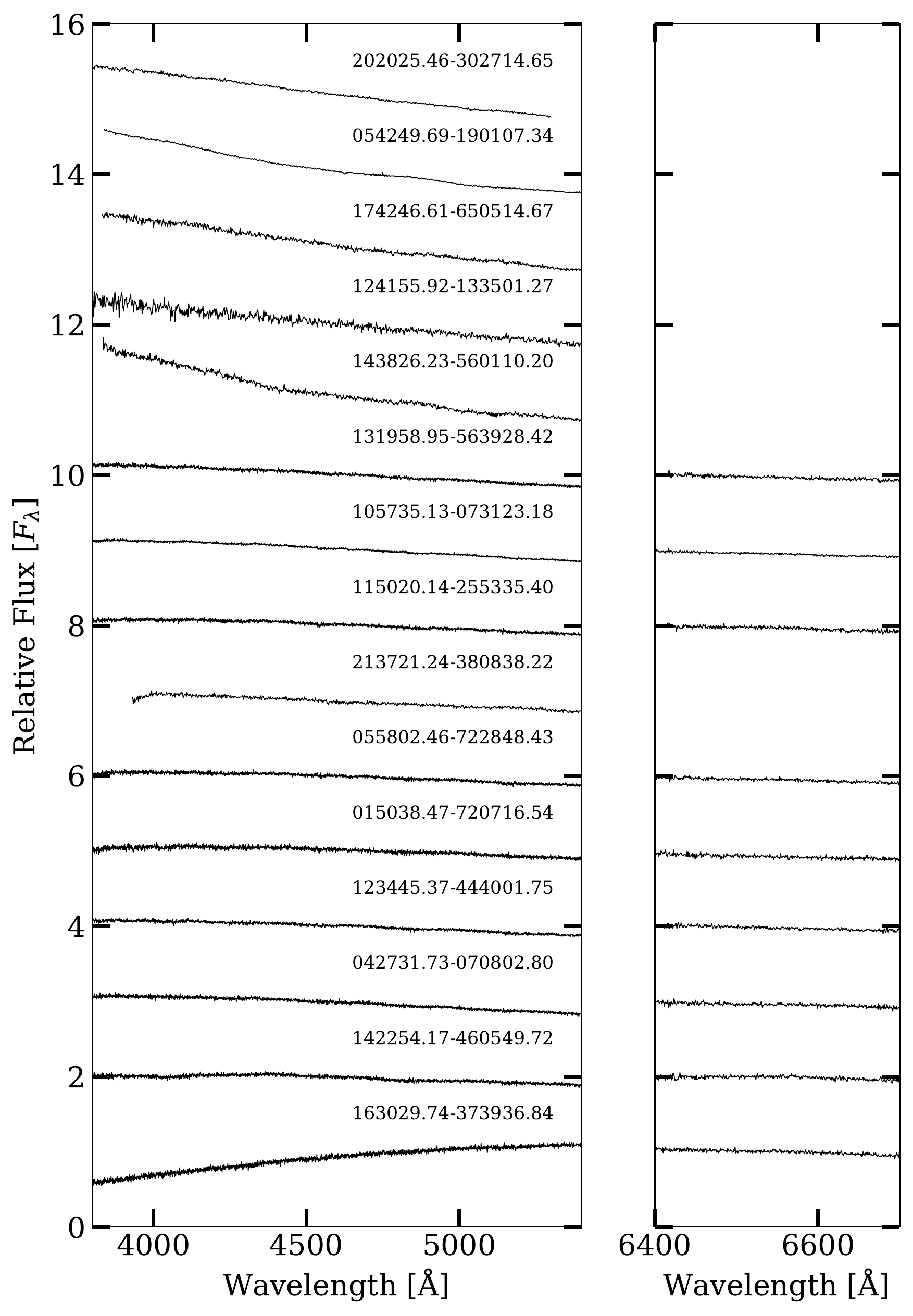}
	\vspace*{10mm}
	\caption{Spectroscopic observations of 69 DC white dwarfs ordered with decreasing photometric temperature (1/4). Temperature range: 10\,500\,K $>$ \Teff\ $>$ 6600\,K.}
        \label{fig:DC1}
\end{figure*}

\renewcommand{\thefigure}{A\arabic{figure}}
\setcounter{figure}{4}

\begin{figure*}
	\includegraphics[viewport= 1 20 520 720, scale=0.85]{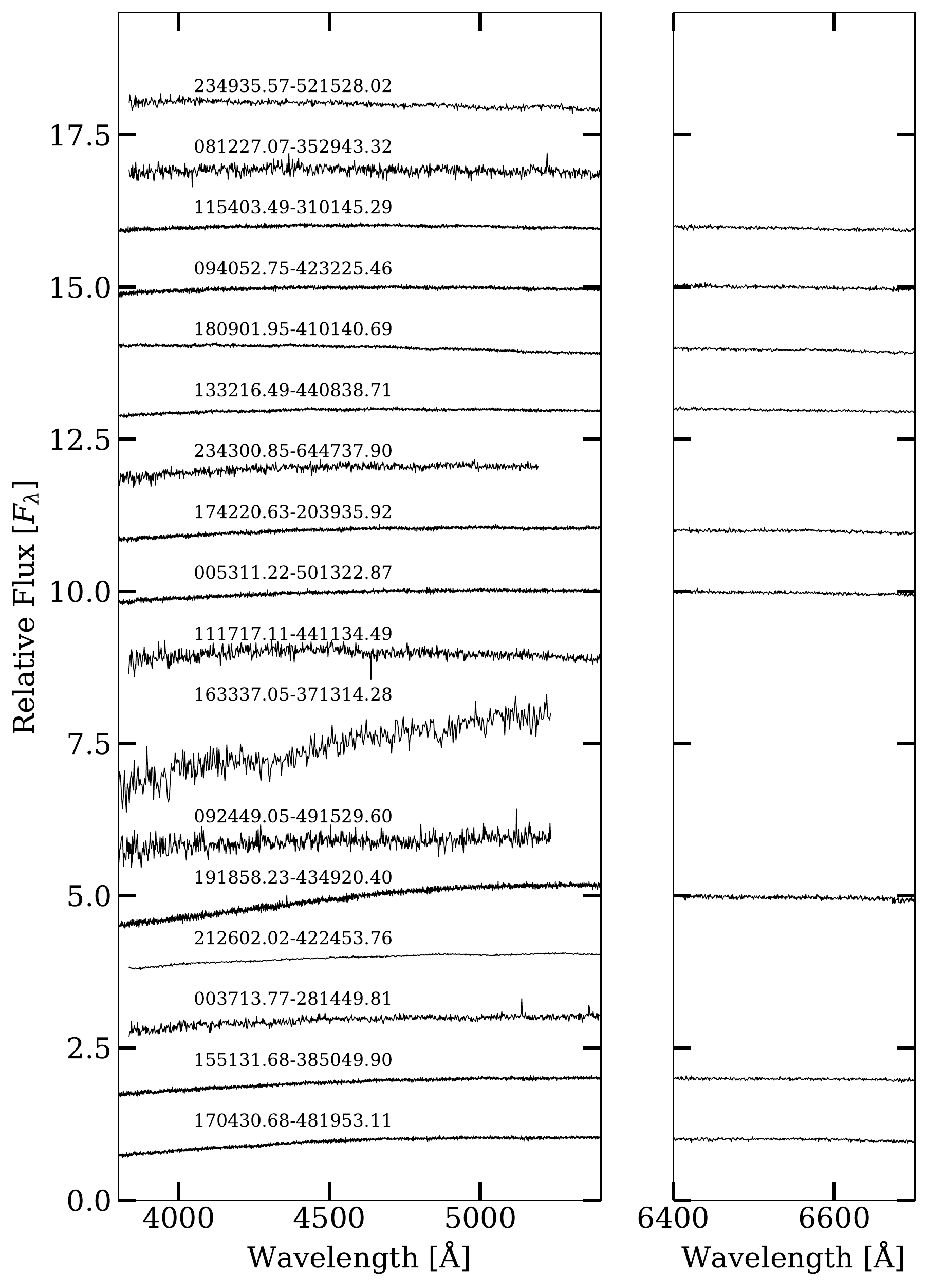}
	\vspace*{10mm}
	\caption{Spectroscopic observations of 69 DC white dwarfs ordered with decreasing photometric temperature (2/4). Temperature range: 6600\,K $>$ \Teff\ $>$ 5200\,K.}
        \label{fig:DC2}
\end{figure*}

\renewcommand{\thefigure}{A\arabic{figure}}
\setcounter{figure}{4}

\begin{figure*}
	\includegraphics[viewport= 1 20 520 720, scale=0.85]{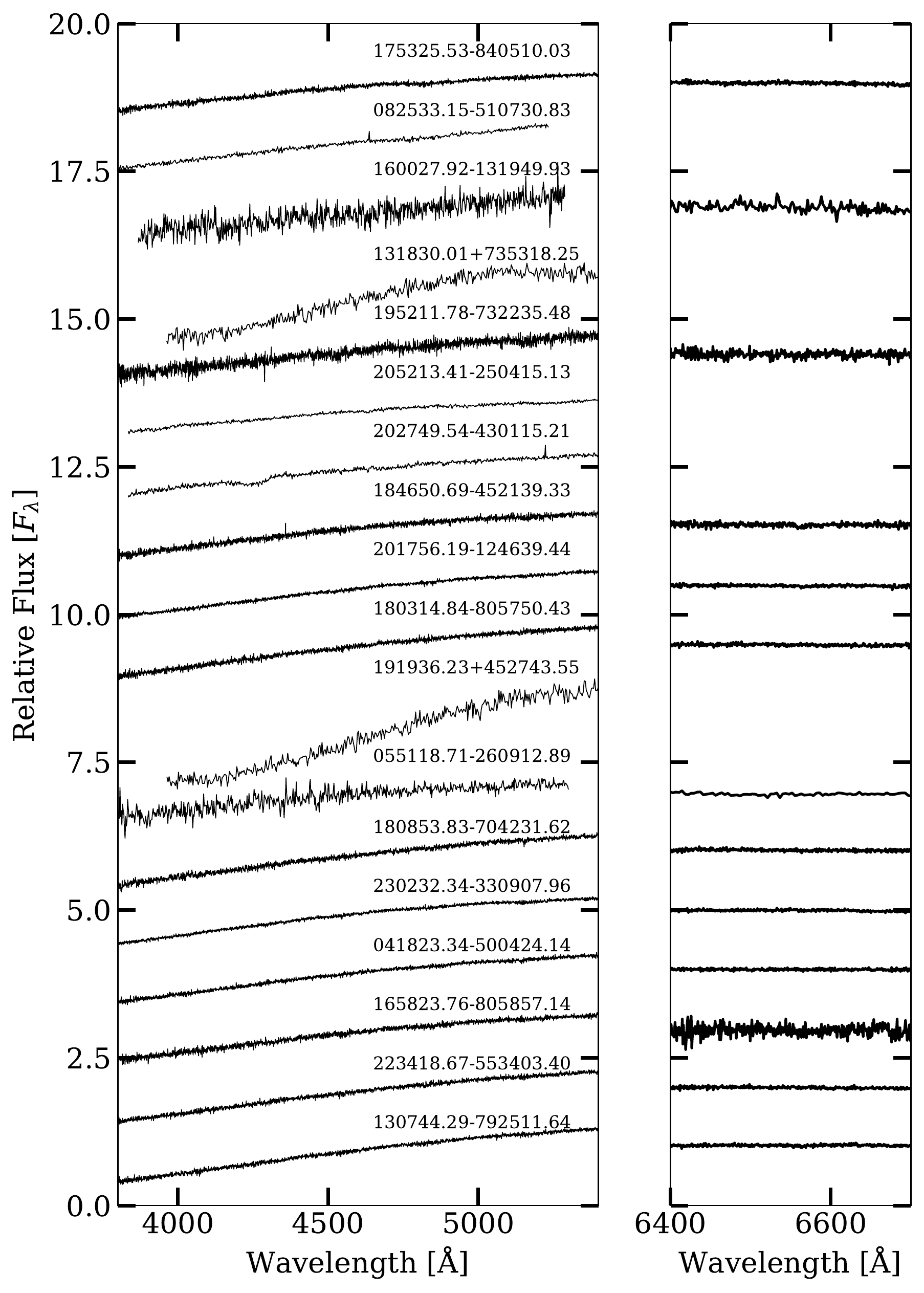}
	\vspace*{10mm}
	\caption{Spectroscopic observations of 69 DC white dwarfs ordered with decreasing photometric temperature (3/4). Temperature range: 5200\,K $>$ \Teff\ $>$ 4700\,K.}
        \label{fig:DC3}
\end{figure*}

\renewcommand{\thefigure}{A\arabic{figure}}
\setcounter{figure}{4}

\begin{figure*}
	\includegraphics[viewport= 1 20 520 720, scale=0.85]{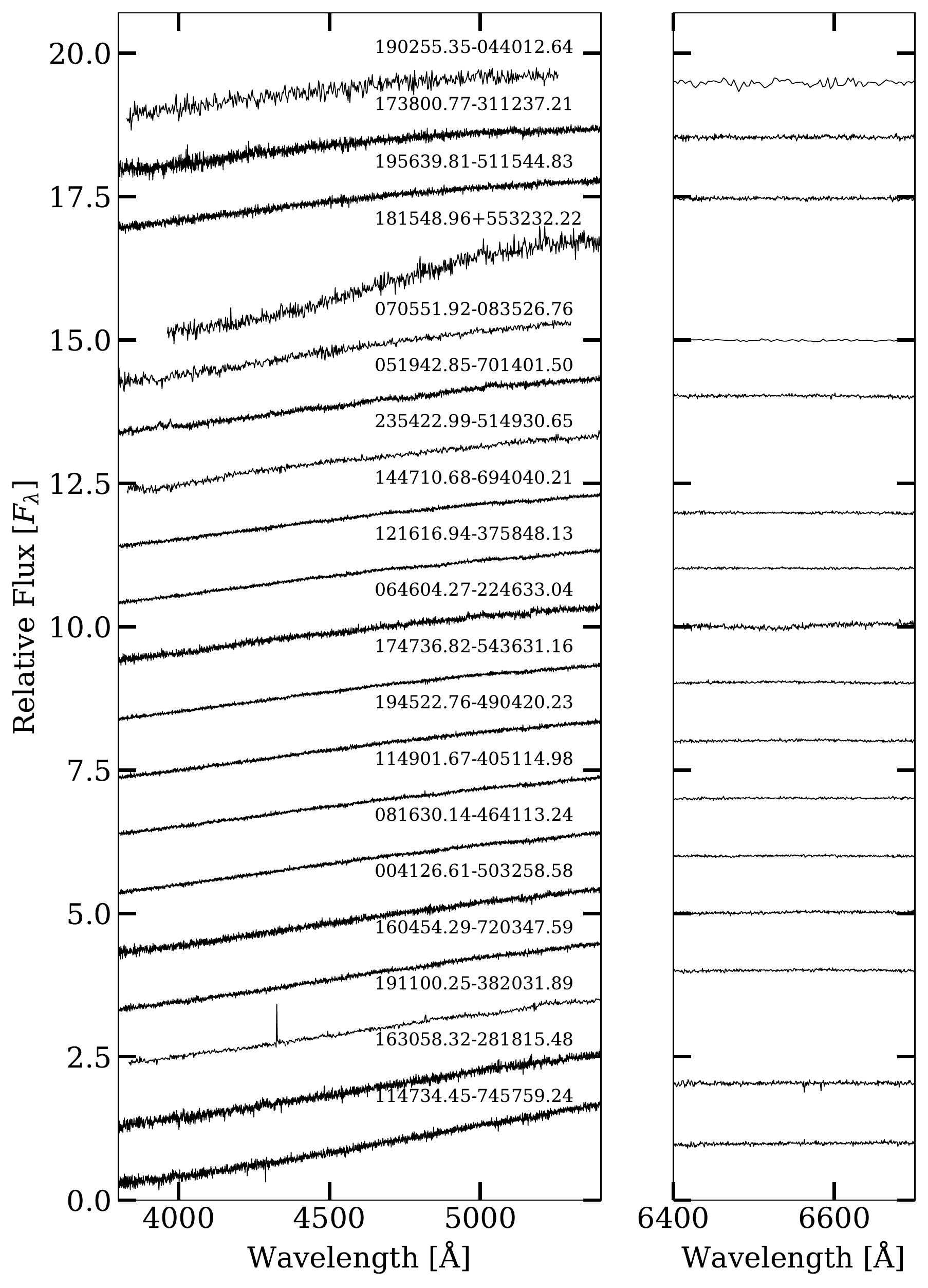}
	\vspace*{10mm}
	\caption{Spectroscopic observations of 69 DC white dwarfs ordered with decreasing photometric temperature (4/4). Temperature range: 4700\,K $>$ \Teff\ $>$ 3800\,K.}
        \label{fig:DC4}
\end{figure*}

 \begin{figure*}
	\includegraphics[viewport= 1 20 520 720, scale=0.85]{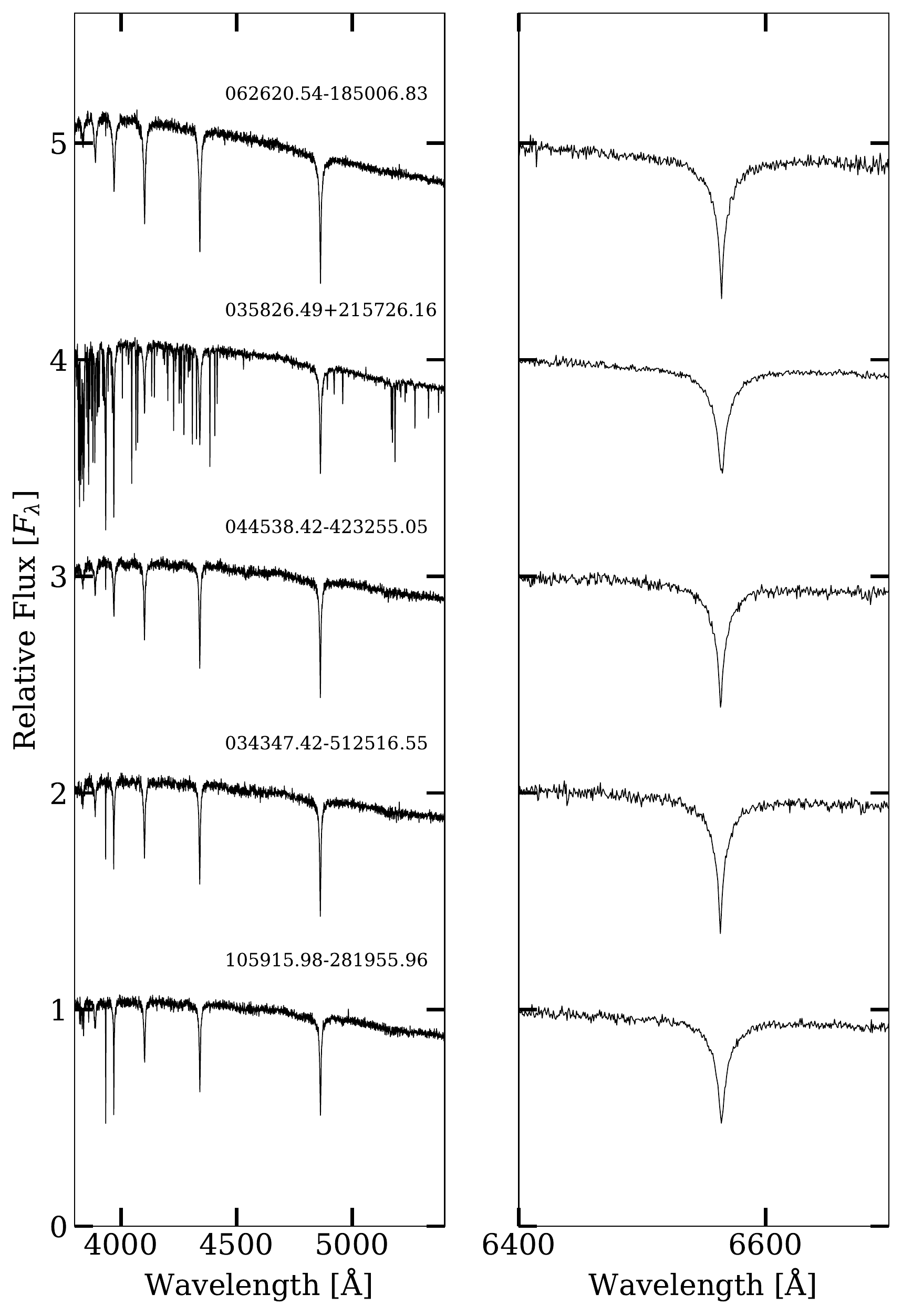}
	\vspace*{10mm}
	\caption{Spectroscopic observations of 10 DAZ white dwarfs ordered with decreasing photometric temperature (1/2). Temperature range: 7300\,K $>$ \Teff\ $>$ 6600\,K.}
        \label{fig:DAZ1}
\end{figure*}

\renewcommand{\thefigure}{A\arabic{figure}}
\setcounter{figure}{5}

 \begin{figure*}
	\includegraphics[viewport= 1 20 520 720, scale=0.85]{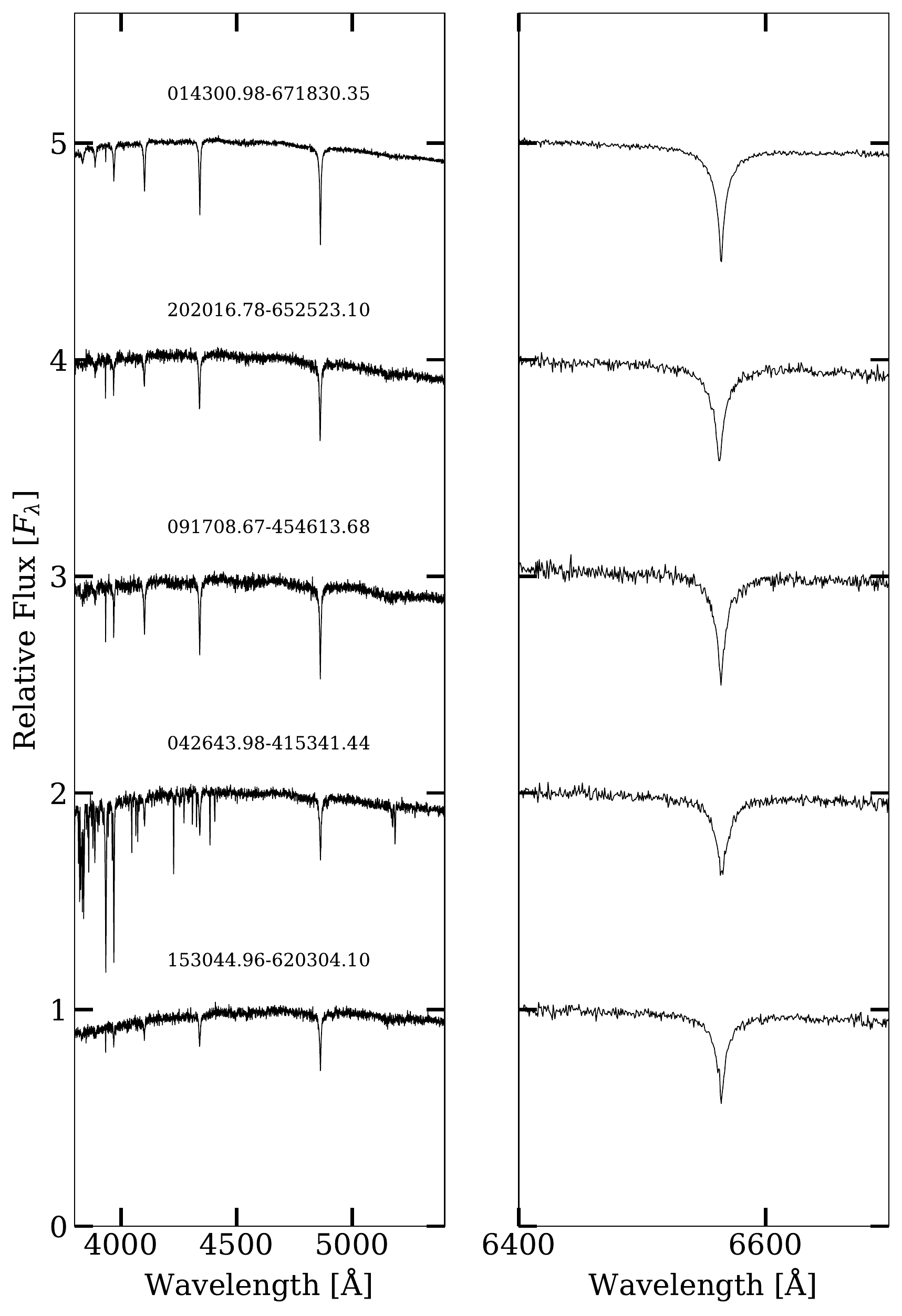}
	\vspace*{10mm}
	\caption{Spectroscopic observations of 10 DAZ white dwarfs ordered with decreasing photometric temperature (2/2). Temperature range: 6600\,K $>$ \Teff\ $>$ 5900\,K.}
        \label{fig:DAZ2}
\end{figure*}

\begin{figure*}
	\includegraphics[viewport= 1 20 520 720, scale=0.85]{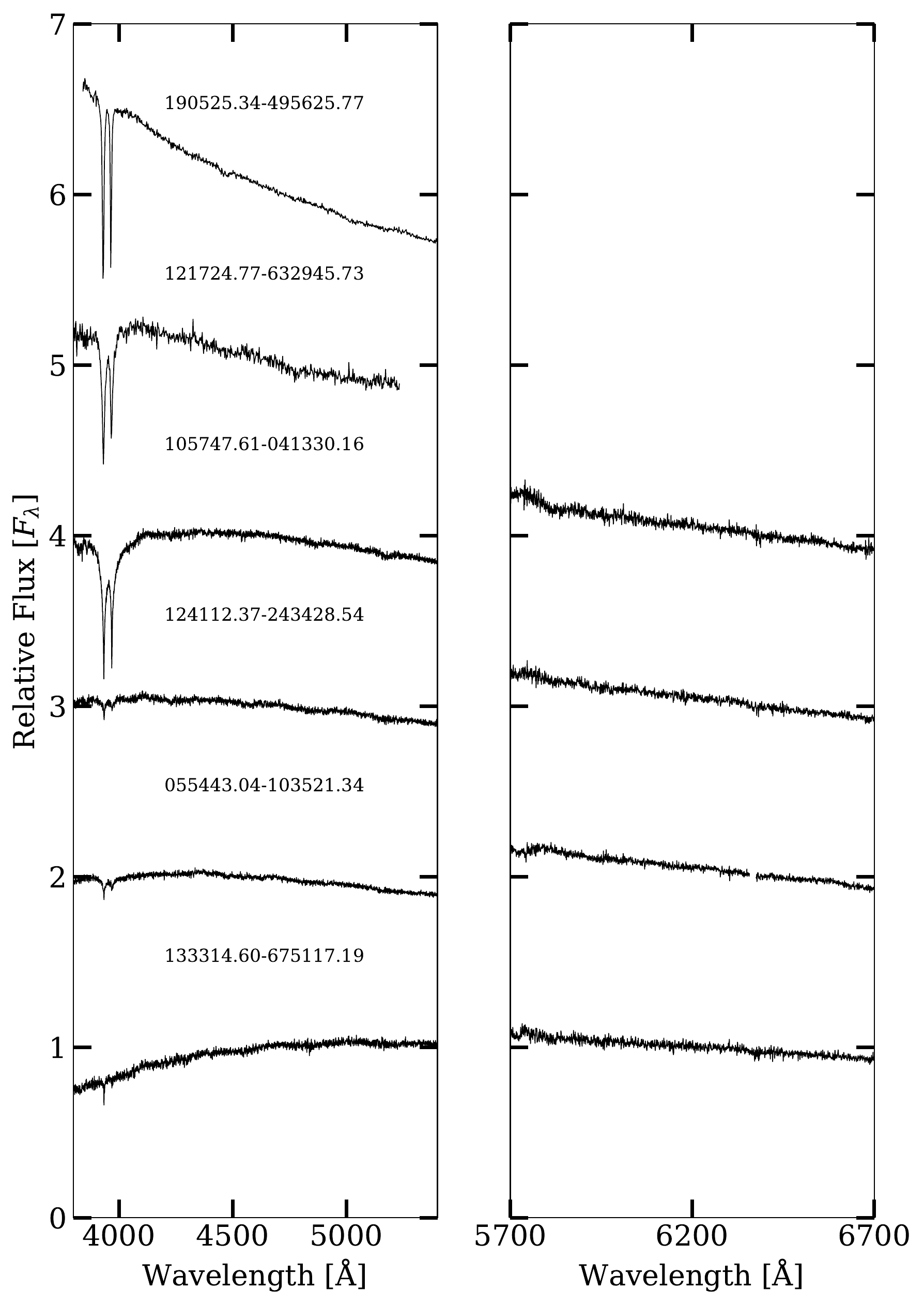}
	\vspace*{10mm}
	\caption{Spectroscopic observations of 11 DZ white dwarfs ordered with decreasing photometric temperature (1/2). Temperature range: 11\,000\,K $>$ \Teff\ $>$ 5700\,K.}
        \label{fig:DZ1}
\end{figure*}

\renewcommand{\thefigure}{A\arabic{figure}}
\setcounter{figure}{6}

\begin{figure*}
	\includegraphics[viewport= 1 20 520 720, scale=0.85]{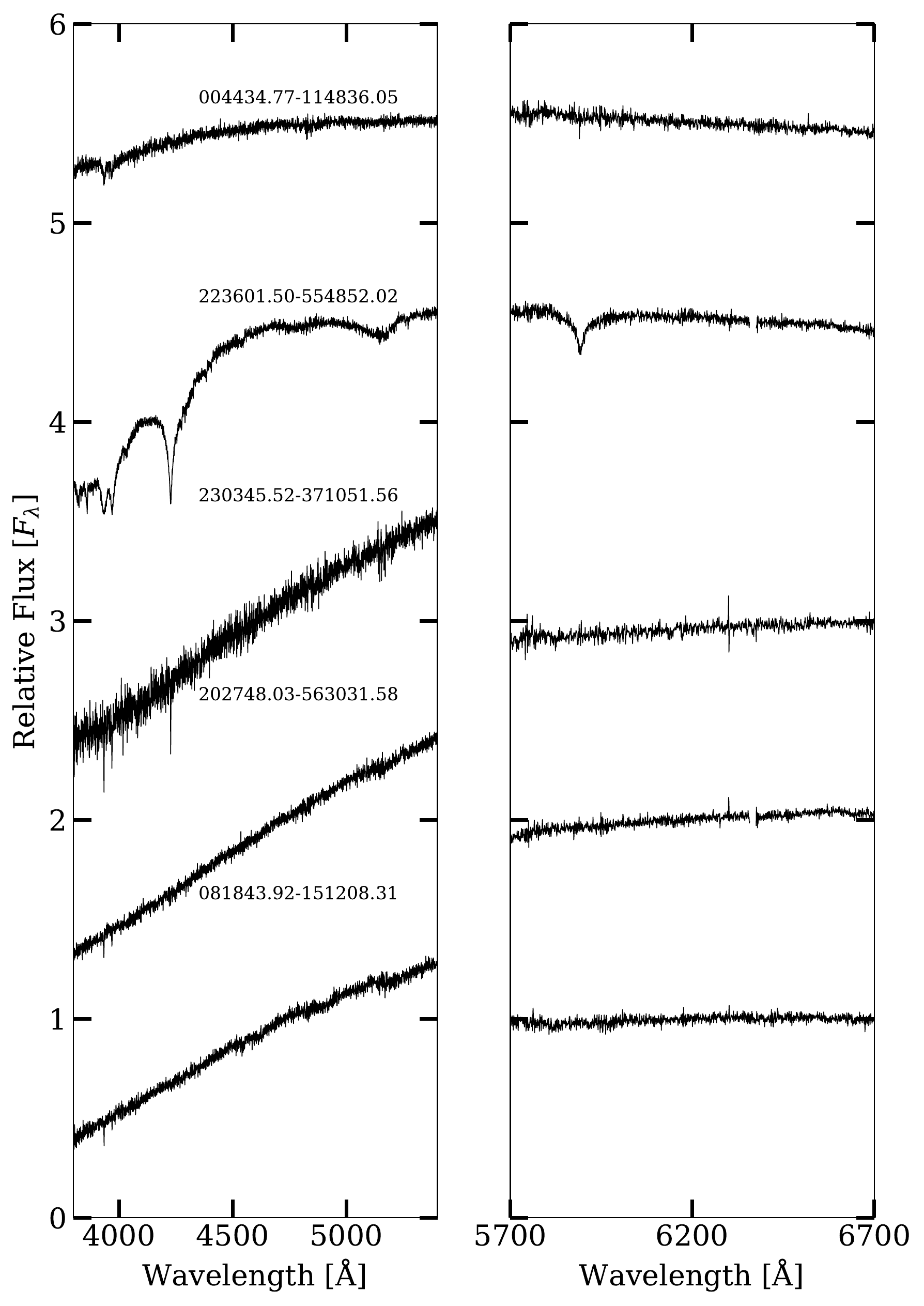}
	\vspace*{10mm}
	\caption{Spectroscopic observations of 11 DZ white dwarfs ordered with decreasing photometric temperature (2/2). Temperature range: 5700\,K $>$ \Teff\ $>$ 4000\,K.}
        \label{fig:DZ2}
\end{figure*}

 \begin{figure*}
	\includegraphics[viewport= 1 20 520 720, scale=0.85]{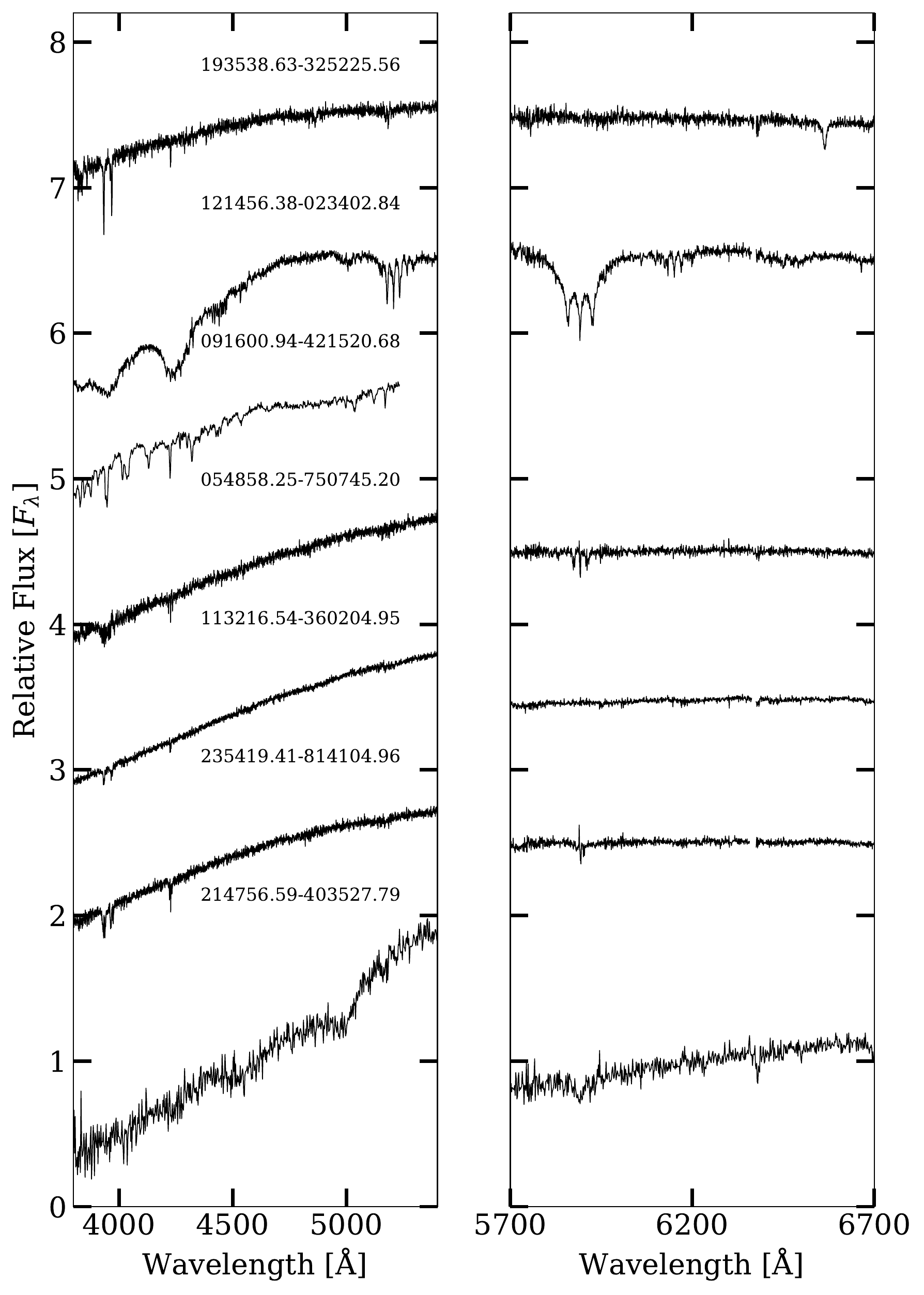}
	\vspace*{10mm}
	\caption{Spectroscopic observations of 7 DZH, DZAH, and DZQH magnetic white dwarfs ordered with decreasing photometric temperature. \textbf{WD\,J2147$-$4035} also displays carbon features (DZQH).}
        \label{fig:DZH1}
\end{figure*}

 \begin{figure*}
	\includegraphics[viewport= 1 20 520 720, scale=0.85]{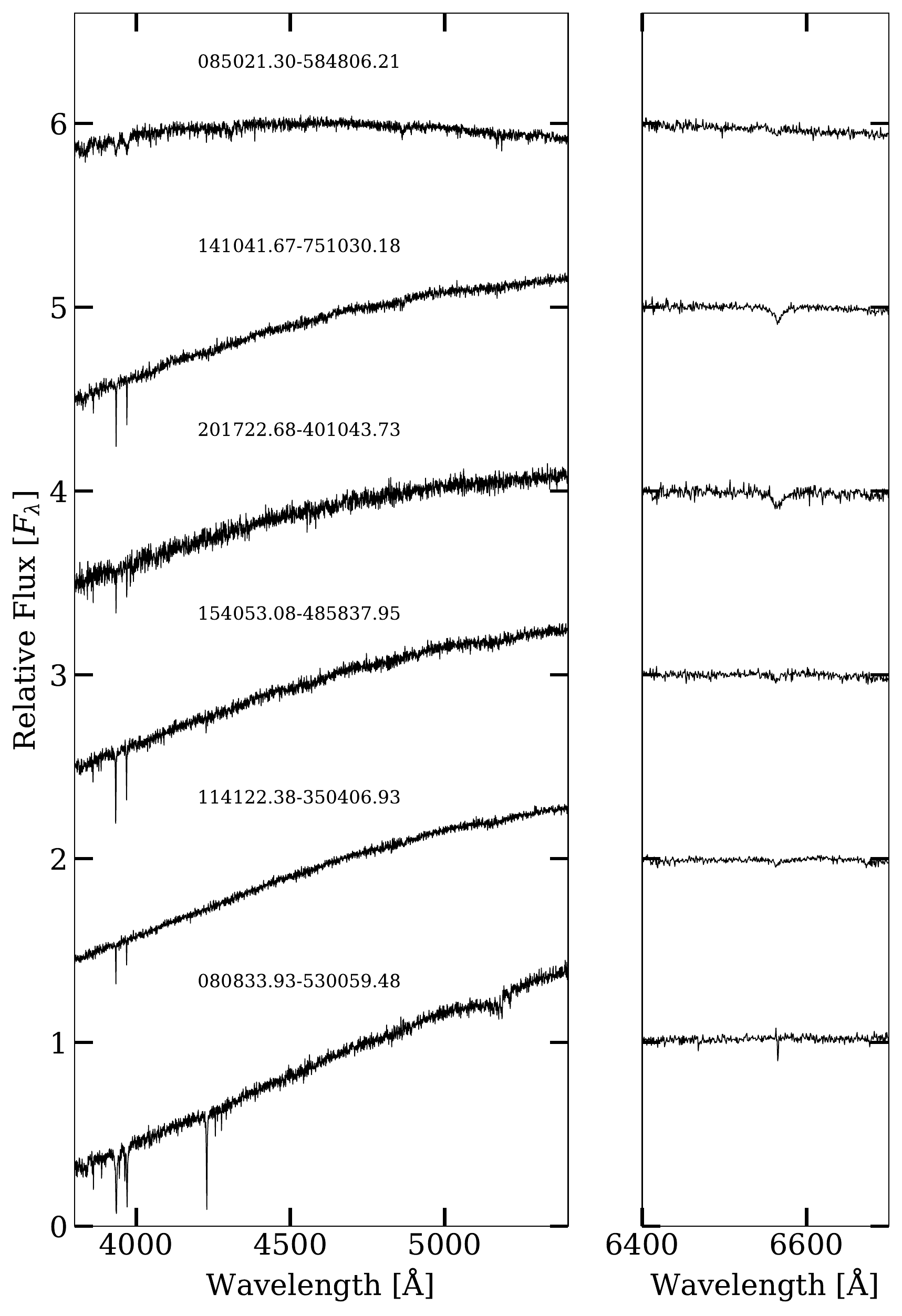}
	\vspace*{10mm}
	\caption{Spectroscopic observations of 6 DZA white dwarfs ordered with decreasing photometric temperature.}
        \label{fig:DZA1}
\end{figure*}

\begin{figure*}
	\includegraphics[viewport= 1 20 520 720, scale=0.85]{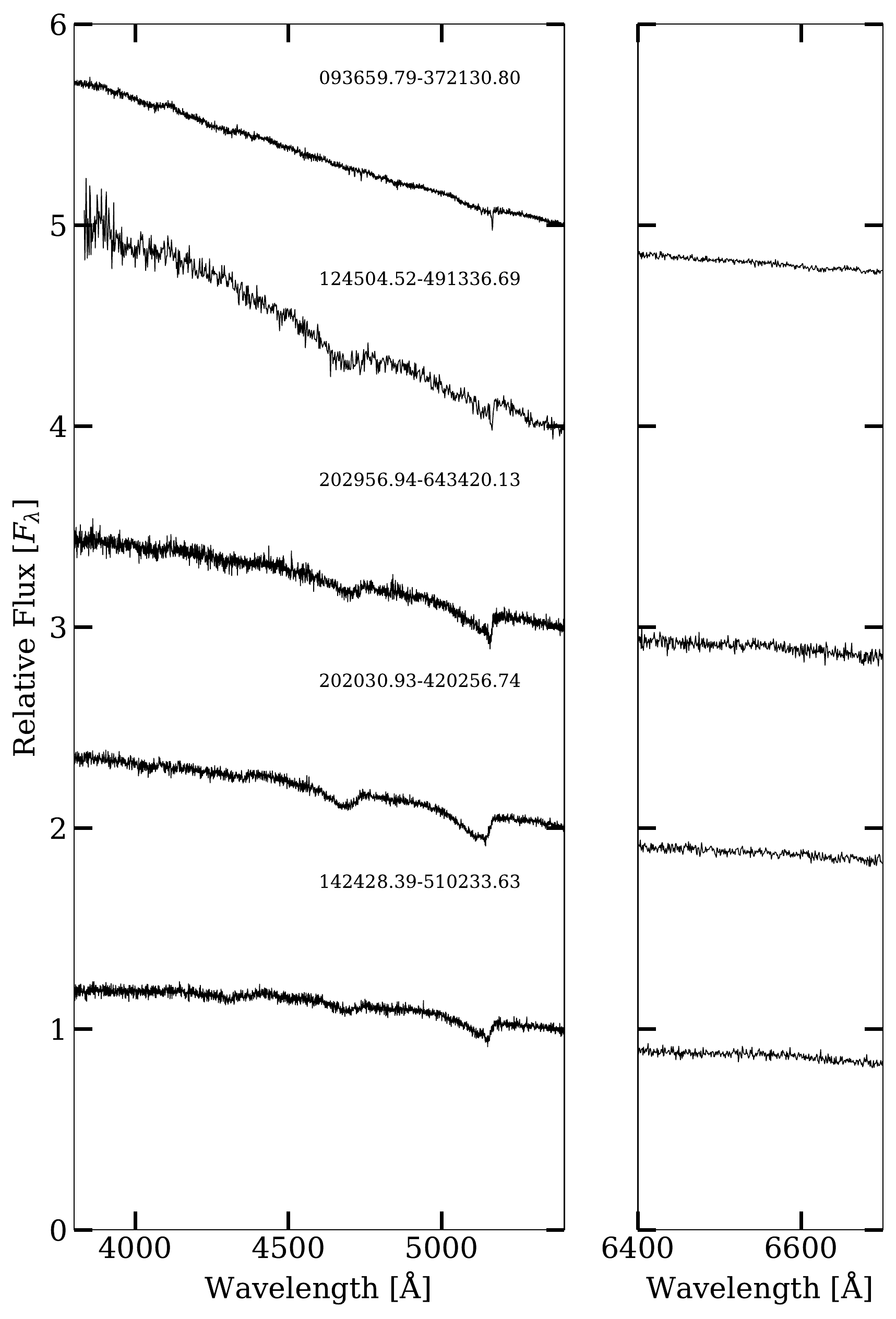}
	\vspace*{10mm}
	\caption{Spectroscopic observations of 9 DQ white dwarfs ordered with decreasing photometric temperature (1/2). Temperature range: 9200\,K $>$ \Teff\ $>$ 6700\,K.}
        \label{fig:DQ1}
\end{figure*}

\renewcommand{\thefigure}{A\arabic{figure}}
\setcounter{figure}{9}

 \begin{figure*}
	\includegraphics[viewport= 1 20 520 720, scale=0.85]{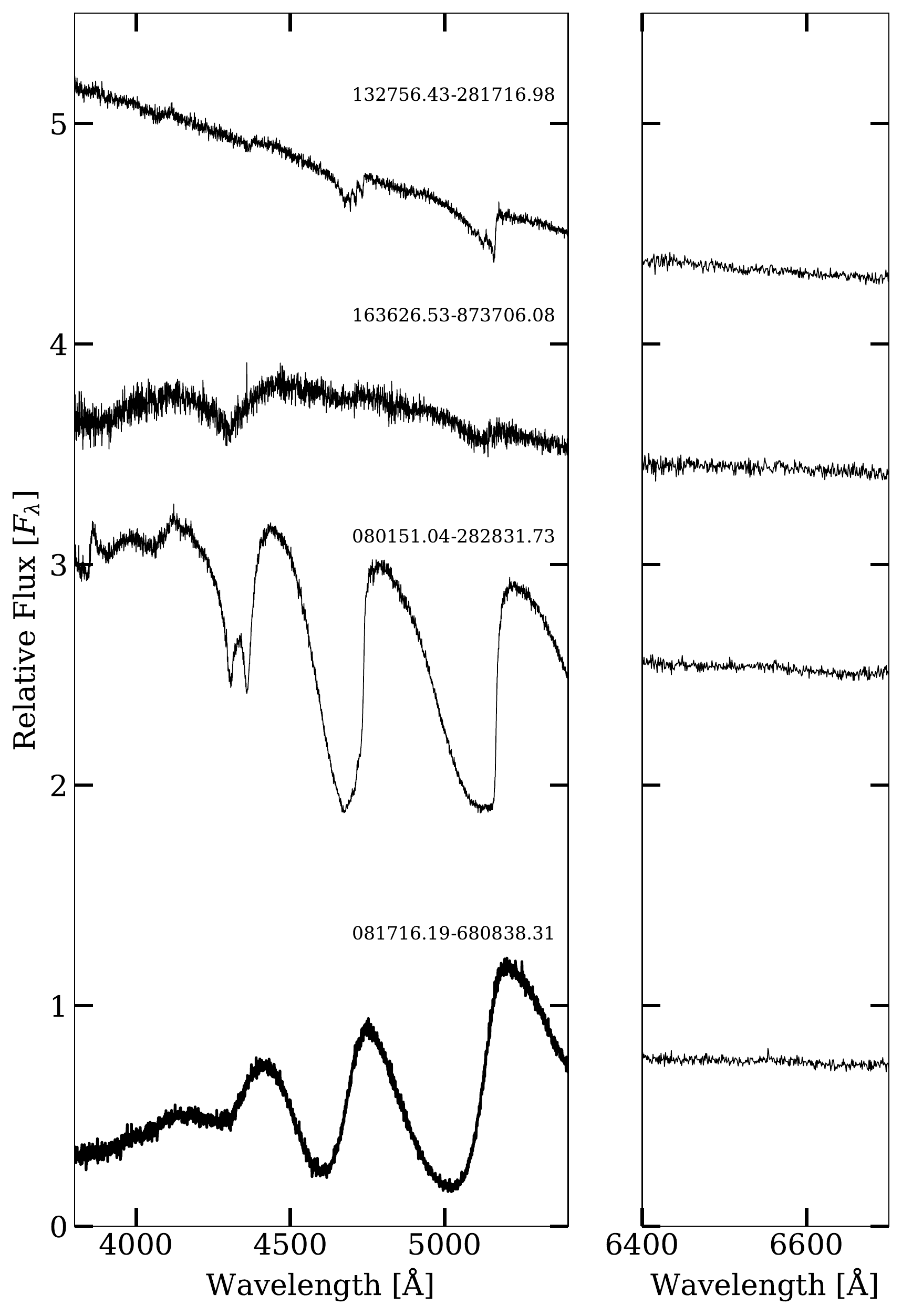}
	\vspace*{10mm}
	\caption{Spectroscopic observations of 9 DQ white dwarfs ordered with decreasing photometric temperature (2/2) The bottom two are classified as peculiar DQ (DQpec) white dwarfs. Temperature range: 6700\,K $>$ \Teff\ $>$ 4400\,K.}
        \label{fig:DQ2}
\end{figure*}

 \begin{figure*}
	\includegraphics[viewport= 1 20 520 720, scale=0.85]{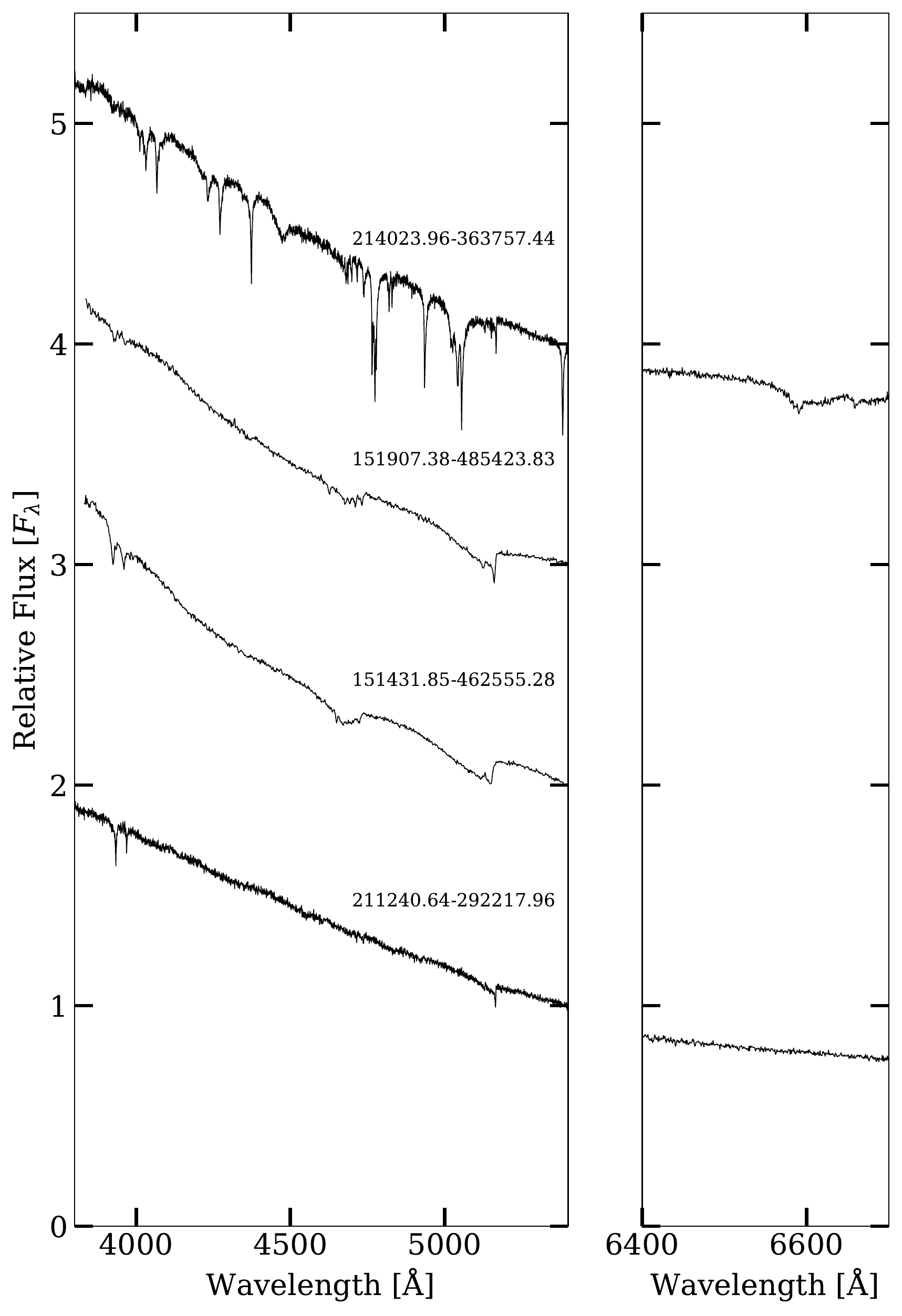}
	\vspace*{10mm}
	\caption{Spectroscopic observations of a warm DQ white dwarf (top), two DQZ white dwarfs (middle) and one DZQ white dwarf (bottom).}
        \label{fig:DQ3}
\end{figure*}

 \begin{figure*}
	\includegraphics[viewport= 1 20 520 720, scale=0.85]{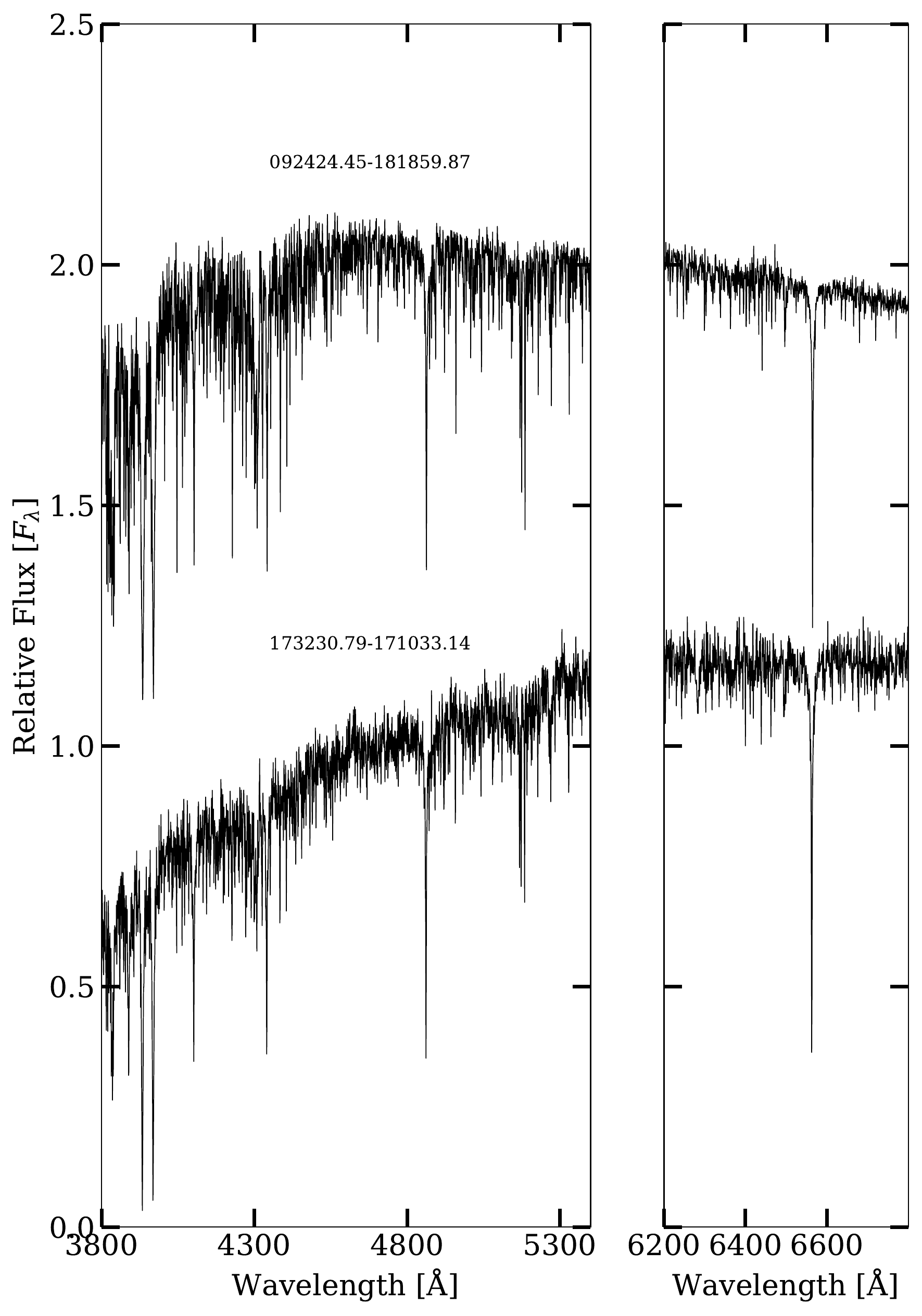}
	\vspace*{10mm}
	\caption{Spectroscopic observations of main-sequence stars that are high-probability white dwarf candidates (P$_{WD}$ $>$ 0.99) in DR3 \citet{Gentile2021}.}
        \label{fig:STAR1}
\end{figure*}








\bsp	
\label{lastpage}
\end{document}